\documentclass{emulateapj}

\newcommand{\jms}{J. Mol. Spec.}
\newcommand{\jpcrd}{J. Phys. Chem. Ref. Data}
\newcommand{\aas}{Astron. Astrophys. Suppl. Ser.}
\newcommand{\subscript}[1]{\textnormal{\scriptsize{#1}}}
\newcommand{\arcsecond}{\ensuremath{\arcsec\kern-4pt{.}\kern1pt}}
\newcommand{\tk}{$T_\subscript{K}$}
\newcommand{\trot}{$T_\subscript{rot}$}
\newcommand{\tvib}{$T_\subscript{vib}$}
\newcommand{\tev}{$T_\subscript{ev}$}
\newcommand{\tdust}{$T_\subscript{d}$}
\newcommand{\tdin}{$T_{\subscript{d}1}$}

\newcommand{\rdin}{R$_{\subscript{d}1}$}
\newcommand{\rdout}{R$_{\subscript{d}2}$}

\begin{document}

\title{A Detailed Analysis of the Dust Formation Zone of IRC+10216\\
Derived from Mid-IR Bands of C$_2$H$_2$ and HCN}
\shorttitle{The Dust Formation Zone of IRC+10216}
\author{J. P. Fonfr\'{\i}a}
\author{J. Cernicharo}
\affil{Dpto. Astrof\'{\i}sica Molecular e Infrarroja,
Instituto de la Estructura de la Materia (IEM),
Consejo Superior de Investigaciones Cient\'{\i}ficas (CSIC),
\mbox{C/ Serrano 121,} 28006, Madrid (Spain)}
\email{jpablo.fonfria,cerni@damir.iem.csic.es}
\and
\author{M. J. Richter\altaffilmark{1}}
\affil{Physics Dept. - UC Davis, One Shields Ave., Davis,
CA 95616 (USA)}
\email{richter@physics.ucdavis.edu}
\and
\author{J. H. Lacy\altaffilmark{1}}
\affil{Astronomy Dept., University of Texas, Austin,
TX 78712 (USA)}
\email{lacy@shrub.as.utexas.edu}

\altaffiltext{1}{Visiting Astronomer at the Infrared Telescope Facility,
which is operated by the University of Hawaii under
contract from the National Aeronautics and Space
Administration.}

\shortauthors{Fonfr\'{\i}a et al.}

\begin{abstract}
A spectral survey of IRC+10216 has been carried out in the range 11 to 
14~$\mu$m with a spectral resolution of about 4~km~s$^{-1}$. We have 
identified a forest of lines in six bands of
C$_2$H$_2$ involving the vibrational states
from the ground to $3\nu_5$ and in two bands of HCN, 
involving the vibrational states from the ground
up to $2\nu_2$. 
Some of these transitions are observed
also in H$^{13}$CCH and H$^{13}$CN.
We have estimated the kinetic, vibrational, 
and rotational temperatures, and the abundances and
column densities of C$_2$H$_2$ 
and HCN between 1 and 300 R$_*$ ($\simeq 1.5\times 10^{16}$~cm)
by fitting about 300 of these ro-vibrational lines.
The envelope can be divided into three regions
with approximate boundaries at 0\arcsecond019 
(the stellar photosphere), 0\arcsecond1 (the 
inner dust formation zone),
and 0\arcsecond4 (outer dust formation zone). 
Most of the lines might require a
large microturbulence broadening. 
The derived abundances of C$_2$H$_2$ and HCN 
increase by factors of 10 and 4, respectively, from the 
innermost envelope outwards. 
The derived column densities for both C$_2$H$_2$ and HCN are 
$\simeq 1.6\times 10^{19}$~cm$^{-2}$.
Vibrational states up to 3000~K above ground are populated, suggesting
pumping by near-infrared radiation from the star and innermost envelope.
Low rotational
levels can be considered under LTE while those with $J>20-30$ are not thermalized. 
A few lines require special analysis to
deal with effects like overlap with lines of other molecules.
\end{abstract}

\maketitle

\section{Introduction}
\label{sec:intro}

IRC+10216 is a carbon AGB star surrounded by a circumstellar envelope (CSE) 
of gas and dust. Since it is the nearest AGB star of this type and the
strongest infrared object in the sky, it has become
the paradigm for this kind of sources
(see \S\ref{sec:prevwork} for more information about previous 
work). The central star itself cannot
be observed directly due to the 
large dust optical depth but the observational data from the CSE suggest
that it is a physically and chemically rich environment.
Although this source has been
studied many times since its 
discovery by \citet{neugebauer_1969}, most of these physical and chemical 
processes remain poorly understood across the envelope 
\citep*[see, for example,][]{agundez_2006}.

The physical conditions in the innermost part of the
CSE maintain chemical thermodynamical 
equilibrium. However, the gas could depart
from the latter due to
the action of periodic shock waves arising from
stellar pulsation
\citep{cherchneff_1992,agundez_2006}.
At the temperatures prevailing in these zones, most atoms are 
integrated into molecular species, primarily H$_2$. The chemistry is so rich 
that more than 50\% of the molecules 
known in space were first discovered in this source
\citep*[see, for example,][and references therein]
{morris_1975,betz_1981,cernicharo_2000,cernicharo_2004}.
Some of these molecules, such as SiO \citep{schoier_2006}, SiS, CS
\citep{lucas_1995}, and metal-bearing species
\citep{cernicharo_1987,cernicharo_2000}, are refractory and are formed in 
the inner envelope. As soon as the temperature of the gas is below a 
critical value these species start to condense and form dust grains.
Other species, mainly radicals, display emission from the external shell 
of the envelope ($A_\subscript{V}\simeq 1$~mag) 
\citep{dayal_1993,dayal_1995,lucas_1995,cernicharo_1996a,lindqvist_2000}
where the Galactic UV field starts to photodissociate the stable molecules 
formed in the inner CSE and neutral-radical reactions produce long carbon 
chain radicals \citep{cernicharo_1996a,guelin_1997} 
and also 
the interesting gas-phase oxygen chemistry \citep{agundez_2006}.

The most abundant molecular species, after H$_2$, are CO, C$_2$H$_2$, and HCN.
CO has an abundance of $\simeq 8\times 10^{-4}$ from 1~R$_*$ to beyond 
1000~R$_*$ \citep{lafont_1982, knapp_1985, agundez_2006}. C$_2$H$_2$, 
is predicted by thermodynamical chemical equilibrium models 
\citep{tejero_1991} to be the most abundant molecule after CO and H$_2$. 
Its abundance has been derived to be $\simeq 8\times 10^{-5}$ in the zone 
$1-40$~R$_*$ from infrared observations \citep{keady_1993, cernicharo_1999}. 
The abundance of HCN $\simeq 3-4\times 10^{-5}$ has been derived from mid-, 
far-infrared, and radio observations \citep{keady_1993, cernicharo_1996b,
cernicharo_1999}.

Radiation pressure by stellar photons accelerates the dust grains formed 
near the star. For the densities prevailing in these dusty regions
of the CSE, the coupling between gas and dust is high and hence the gas
is also accelerated \citep{gilman_1972,kwok_1975}.
\citet{keady_1988} have found that the terminal velocity of the gas is about 
14~km~s$^{-1}$ \citep*[see also][]{cernicharo_2000} and is reached before 20~R$_*$.
These authors also found that the velocity field seems to have more than one 
acceleration regime occurring at different places of the inner CSE. These 
zones might be related to the condensation temperature of different 
refractory molecular species.

The main goal of this work is to learn more about the physical conditions 
and chemical composition of the CSE of IRC+10216 from modeling the line 
profiles of C$_2$H$_2$ and HCN observed in the mid-infrared with the 
high-resolution spectrograph TEXES \citep{lacy_2002}.
These observations are presented in
\S\ref{sec:obs}.
The models we have used to fit the observational data are described in 
\S\ref{sec:modeling} and Appendices~\ref{sec:LineFrequencies} and 
\ref{sec:LineIntensities}, where we discuss the line identification and 
give the dipole moment for the different observed vibrational transitions.
The continuum emission is analyzed in \S\ref{sec:continuum}.
The results concerning C$_2$H$_2$ and its isotopologues are presented in 
\S\ref{sec:acetylene} and \S\ref{sec:h13cch}, and those related to HCN 
and H$^{13}$CN in \S\ref{sec:hcn}. The results obtained for C$_2$H$_2$ and HCN, 
their uncertainties, and the sensitivity of the model to the physical 
conditions of the CSE are discussed and analyzed in \S\ref{sec:sens.disc}.
Finally, the results are summarized in \S\ref{sec:conclusions}.

\section{Observations}
\label{sec:obs}

We observed IRC+10216 with TEXES, the Texas Echelon-cross-Echelle 
Spectrograph \citep{lacy_2002}, at the NASA Infrared Telescope Facility
on 2002 Dec 12 (UT), corresponding to an IR stellar phase 
$\phi_\subscript{IR}\simeq 0.08$ \citep*[following][]{monnier_1998}.
We used the TEXES high resolution echelon grating with a first order 
grating as the cross-disperser.  In this mode, we 
obtained a
spectral coverage of roughly $0.25$~$\mu$m per
setting.  To cover the entire range shown 
here, 11.6 to 13.9~$\mu$m, we required 10 separate settings
and stored the data in 10 different files.

At these wavelengths, the echelon orders are larger than our detector, 
meaning that there are gaps in the spectrum between orders.  The gaps increase in
size toward longer wavelengths.  Telluric features become stronger and
more frequent toward 13.9~$\mu$m.  
In regions where
the telluric atmosphere was nearly opaque ($<$5\%\ transmission)
we discarded the data, resulting in additional gaps.

When using the first order grating as the cross disperser, the TEXES slit
must be very small to prevent orders from overlapping.  For these observations
the slit was 2\arcsecond2 long.  With the short slit, we 
nod IRC+10216 off the slit for sky subtraction.  Sky conditions were 
good enough that we do not believe this introduced any significant
systematics.  

The data were reduced with the standard TEXES pipeline \citep{lacy_2002}.
The pipeline corrects optical distortions, combines nod pairs, removes
spikes, flatfields, performs correction for telluric absorption, establishes
a frequency scale using telluric features, and extracts a spectrum.
We normalized each spectrum before analysis using a fourth-degree
polynomial to estimate the baseline.  

Typically, observations of asteroids provide the best telluric corrections.  
However, IRC+10216 is so much brighter than the available asteroids that we 
would have limited our signal-to-noise ratio by using asteroid measurements.
Instead, we used
a blackbody-sky difference spectrum to correct for the atmosphere.
In the ideal case of the atmosphere, telescope, and blackbody being at
a uniform temperature, the difference spectrum indicates the absorption
from the sky.  In our experience, this procedure for atmospheric correction
works fairly well except for telluric water vapor, which can change on a 
short time scale and from one line of sight to another.

\begin{figure*}
\centering
\includegraphics[height=0.95\textheight]{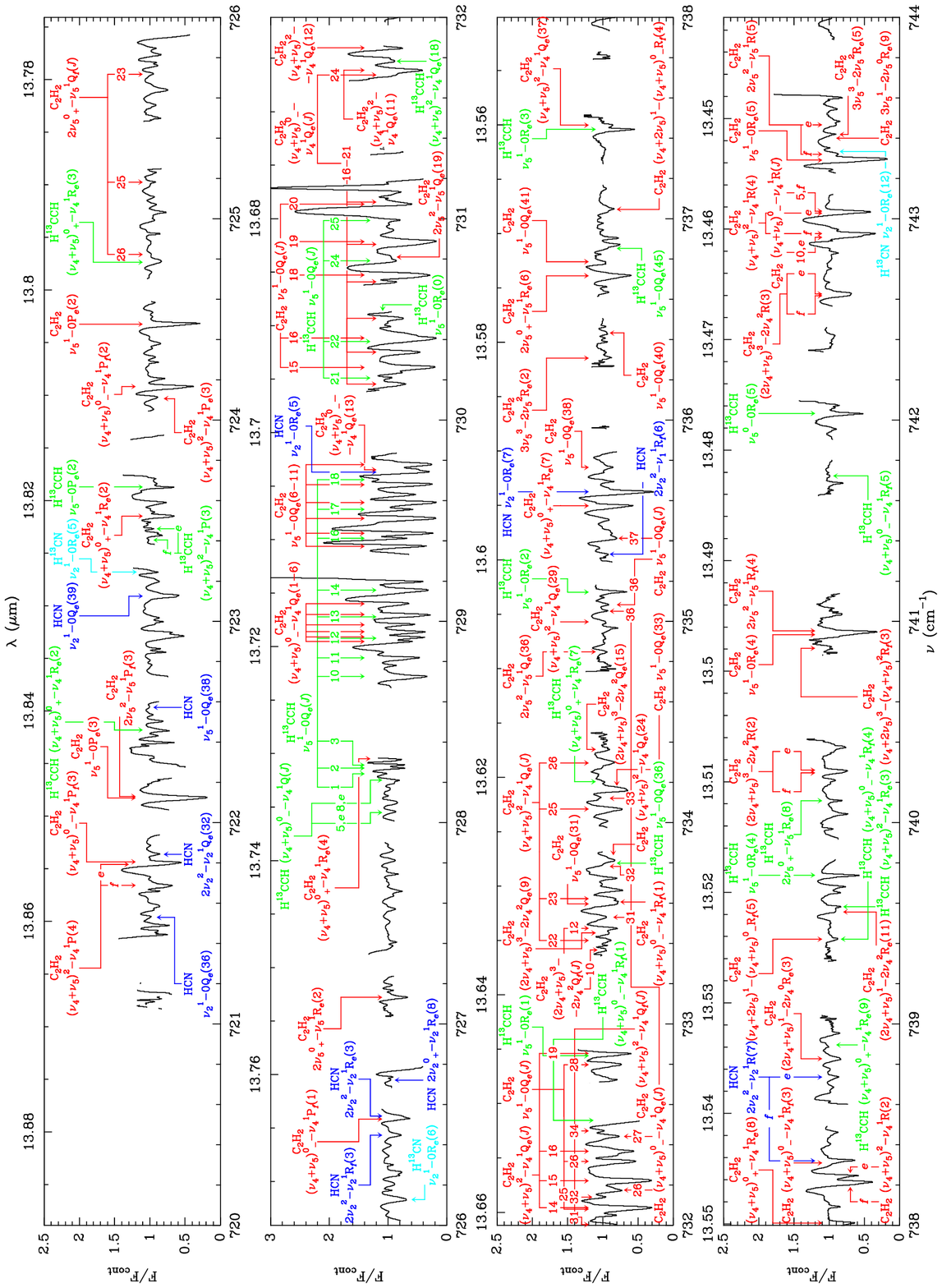}
\vspace*{-2mm}
\caption{The observed
mid-infrared spectrum of IRC+10216 in the range
720-744~cm$^{-1}$ with resolution $\simeq 10^{-2}$ cm$^{-1}$
($\simeq 3-4$ km s$^{-1}$). The most important transitions
of C$_2$H$_2$ (red), HCN (blue), H$^{13}$CCH (green), and H$^{13}$CN (light blue)
are indicated. The sensitivity is high enough to detect molecular species 
with abundances as low as $10^{-8}-10^{-7}$
relative to H$_2$.
Gaps occur between grating orders and in regions of high atmospheric opacity.
Many of the unlabeled lines remain unidentified. }
\label{fig:f1}
\end{figure*}

\begin{figure*}
\centering
\includegraphics[height=0.95\textheight]{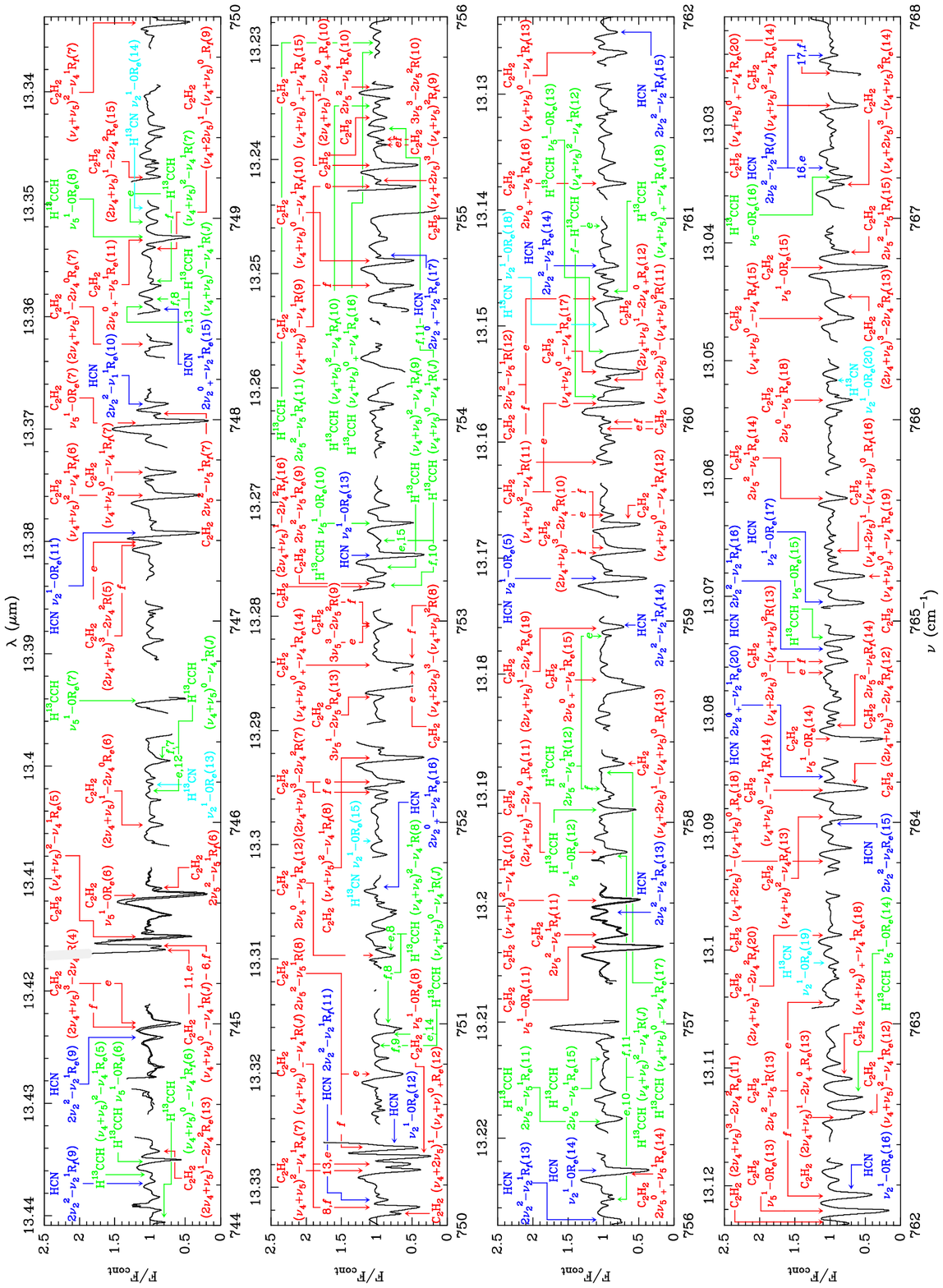}
\caption{The observed
mid-infrared spectrum of IRC+10216 in the range
744-768~cm$^{-1}$ with resolution $\simeq 10^{-2}$ cm$^{-1}$
($\simeq 3-4$ km s$^{-1}$). The most important transitions
of C$_2$H$_2$ (red), HCN (blue), H$^{13}$CCH (green), and H$^{13}$CN (light blue)
are indicated.}
\label{fig:f2}
\end{figure*}

\begin{figure*}
\centering
\includegraphics[height=0.95\textheight]{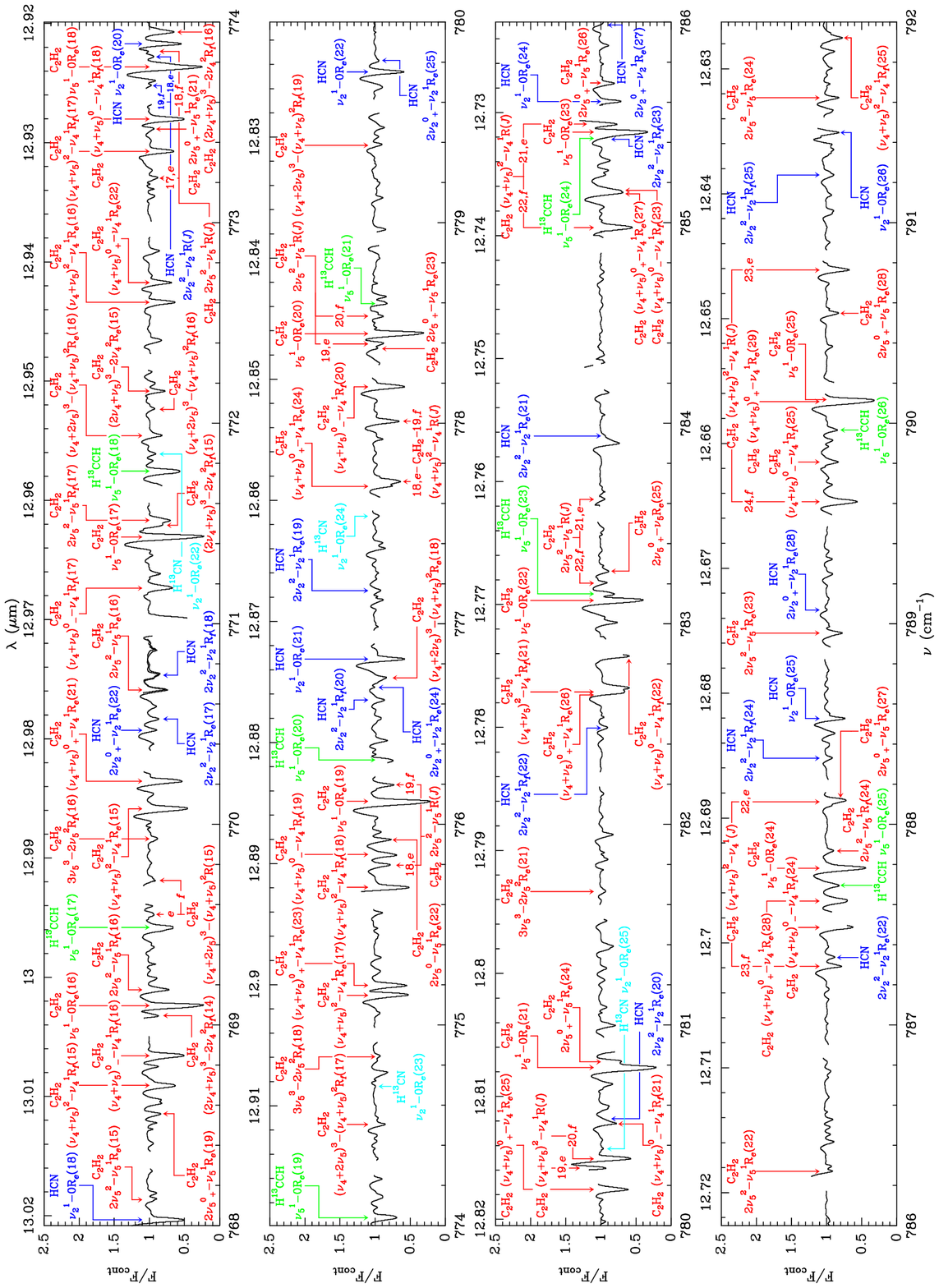}
\caption{The observed
mid-infrared spectrum of IRC+10216 in the range
768-792~cm$^{-1}$ with resolution $\simeq 10^{-2}$ cm$^{-1}$
($\simeq 3-4$ km s$^{-1}$). The most important transitions
of C$_2$H$_2$ (red), HCN (blue), H$^{13}$CCH (green), and H$^{13}$CN (light blue)
are indicated.}
\label{fig:f3}
\end{figure*}

\begin{figure*}
\centering
\includegraphics[height=0.95\textheight]{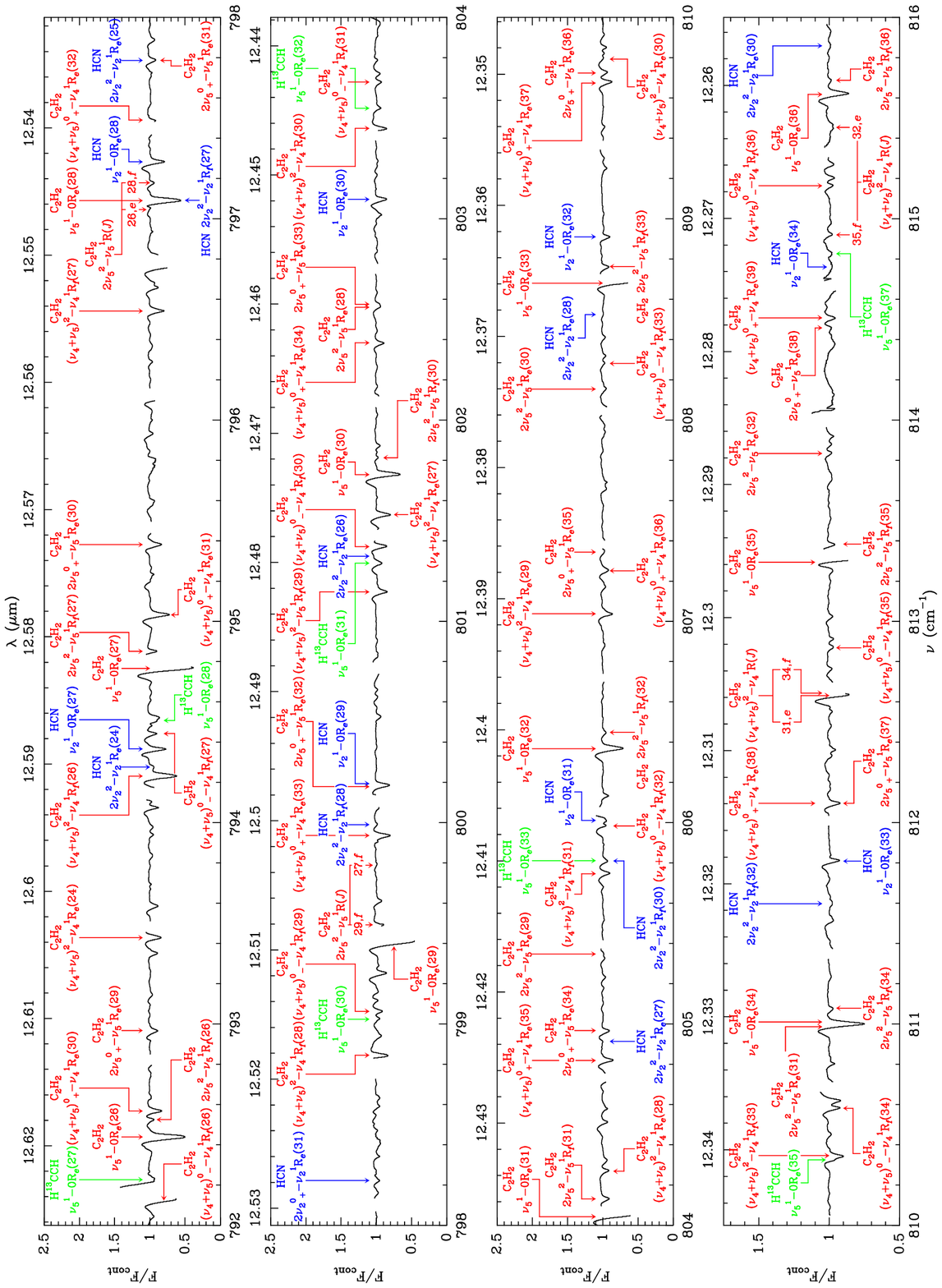}
\caption{The observed mid-infrared 
spectrum of IRC+10216 in the range
792-816~cm$^{-1}$ with resolution $\simeq 10^{-2}$ cm$^{-1}$
($\simeq 3-4$ km s$^{-1}$). The most important transitions
of C$_2$H$_2$ (red), HCN (blue), H$^{13}$CCH (green), and H$^{13}$CN (light blue)
are indicated.
The $y$-axis scale has been changed beyond 810~cm$^{-1}$ to show
the molecular features more clearly.}
\label{fig:f4}
\end{figure*}

\begin{figure*}
\centering
\includegraphics[height=0.95\textheight]{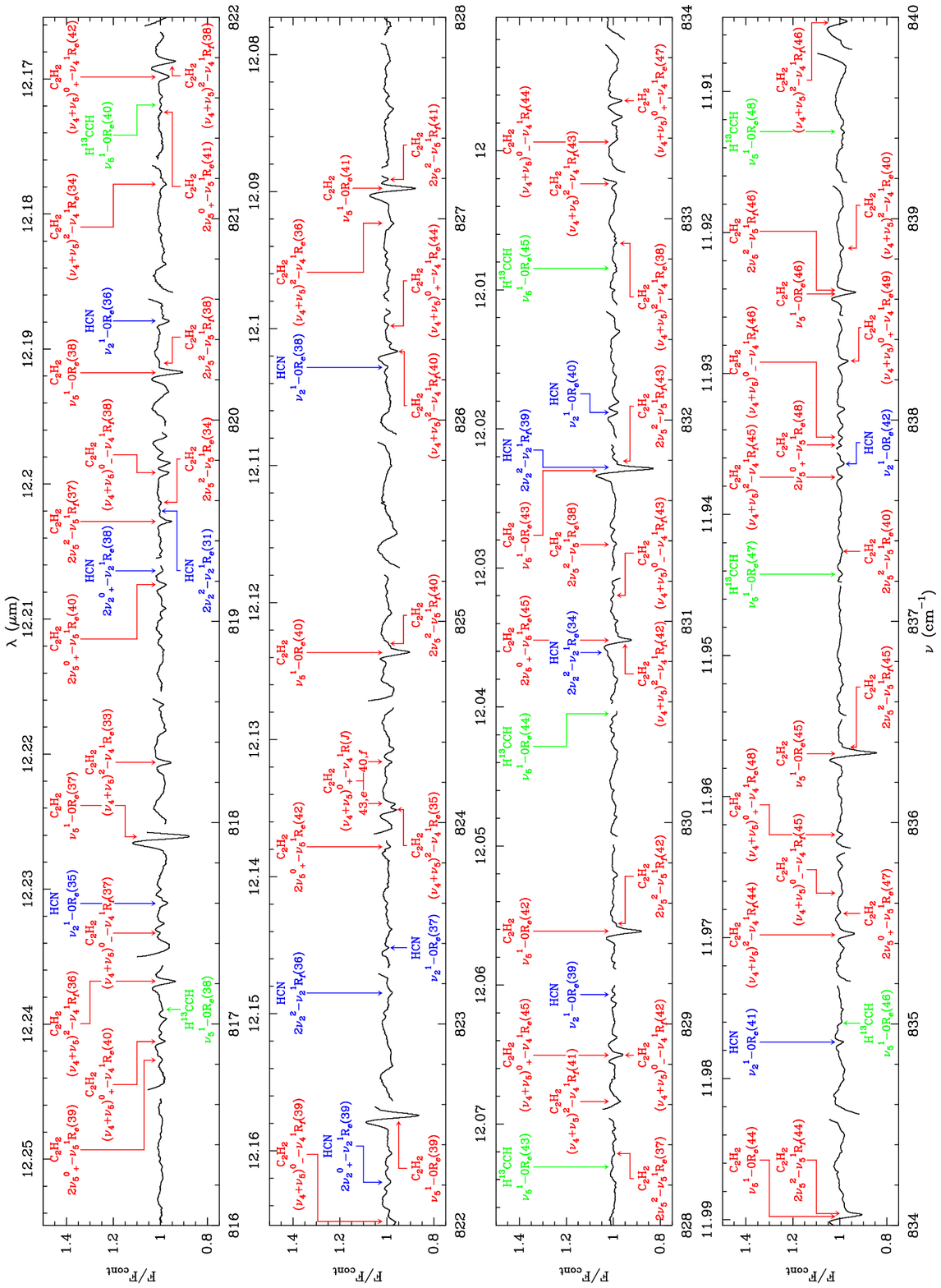}
\caption{The observed
mid-infrared spectrum of IRC+10216 in the range
816-840~cm$^{-1}$ with resolution $\simeq 10^{-2}$ cm$^{-1}$
($\simeq 3-4$ km s$^{-1}$). The most important transitions
of C$_2$H$_2$ (red), HCN (blue), H$^{13}$CCH (green), and H$^{13}$CN (light blue)
are indicated.
The $y$-axis scale has been changed with respect to 
Figures~\ref{fig:f1}-\ref{fig:f3}
to show the molecular features more clearly.}
\label{fig:f5}
\end{figure*}

\begin{figure*}
\centering
\includegraphics[height=0.95\textheight]{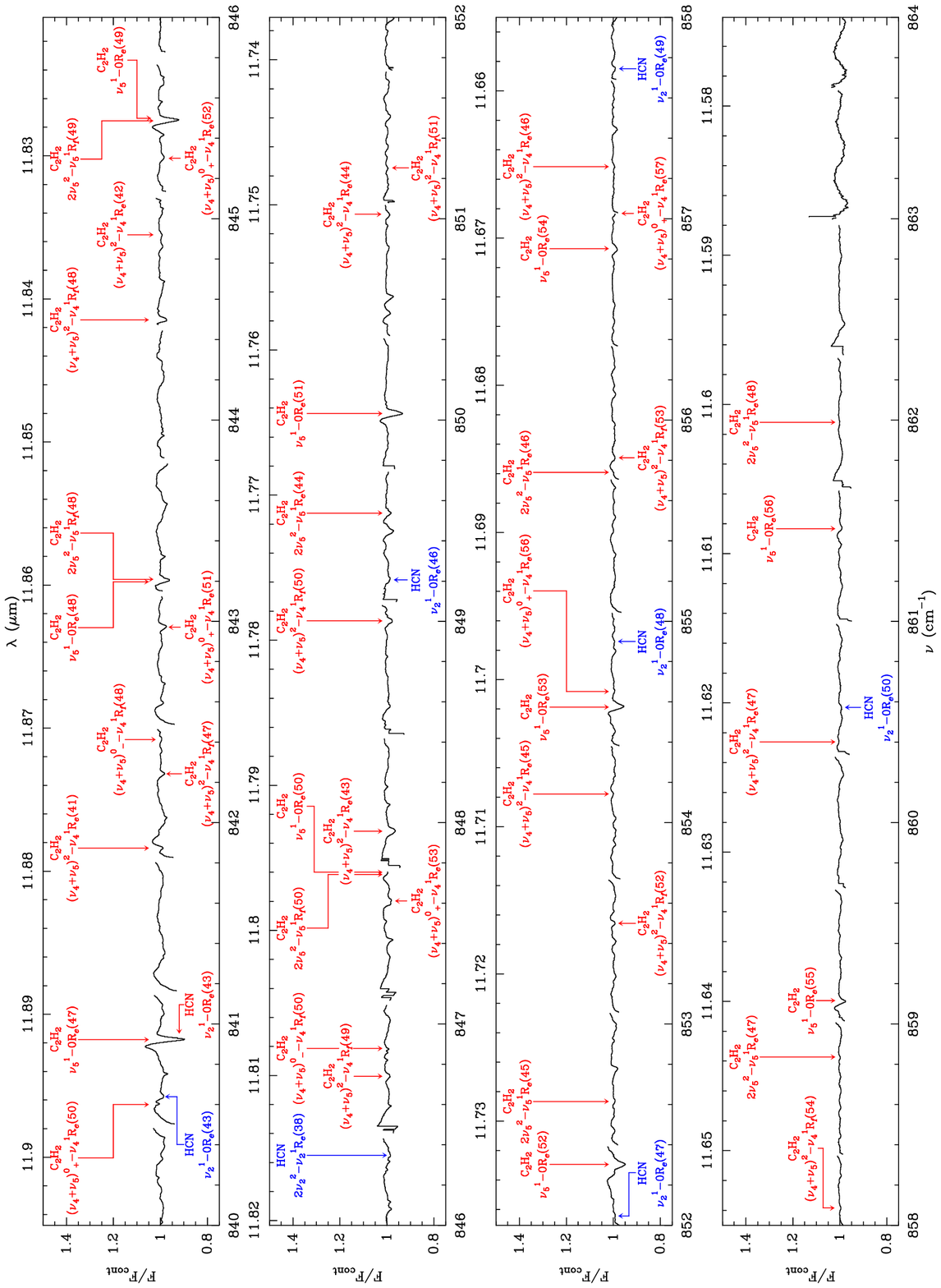}
\caption{The observed
mid-infrared spectrum of IRC+10216 in the range
840-864~cm$^{-1}$ with resolution $\simeq 10^{-2}$ cm$^{-1}$
($\simeq 3-4$ km s$^{-1}$). The most important transitions
of C$_2$H$_2$ (red), HCN (blue), H$^{13}$CCH (green), and H$^{13}$CN (light blue)
are indicated.
The $y$-axis scale has been changed with respect to 
Figures~\ref{fig:f1}-\ref{fig:f3}
to show the molecular features more clearly.}
\label{fig:f6}
\end{figure*}

The observed spectrum covers the wavenumber range 720 to 864~cm$^{-1}$.
Taking into account the molecules detected in this source 
\citep{keady_1993,cernicharo_2000}, there are several candidates that 
could contribute to the observed features: C$_2$H$_2$, HCN, and SiS, including 
their isotopologues. All of them have vibrational bands in the observed 
wavelength range (see
Figures~\ref{fig:f1}$-$\ref{fig:f6}).
Although many lines of SiS and other molecules have been identified in
the Figures, they will be analyzed and studied in a forthcoming paper.

The spectra was corrected from the source movement
by identifying and modeling 
most of the C$_2$H$_2$ $\nu_5$R$_e$ and HCN $\nu_2$R$_e$
lines in the spectrum, and calculating
the mean shift between the observed frequencies and those
published from laboratory work \citep{HITRAN}. This shift is
$\simeq 45$~km~s$^{-1}$. However, because of small uncertainties in the observational
process, 
the data in a few files needed for an extra correction smaller than 2~km~s$^{-1}$.
Moreover, we had to correct the central frequency of some C$_2$H$_2$ and HCN bands 
taken from the \citet{HITRAN_database} by blue-shifting them 
in less than 1.5~km~s$^{-1}$. A sample of the 
velocity-corrected data is shown in Table~\ref{tab:spectrum}.

\begin{deluxetable}{cccc}
\tabletypesize{\footnotesize}
\tablecolumns{4}
\tablewidth{0pt}
\tablecaption{Observed Spectrum towards IRC+10216\label{tab:spectrum}}
\tablehead{\colhead{Freq. (cm$^{-1}$)} & 
\colhead{Normalized Flux} & \colhead{Atmospheric Trans.} & \colhead{Model}}\\
\startdata
721.579 & 1.023 & 0.1810 & 1.000\\
721.581 & 1.089 & 0.1877 & 1.000\\
721.583 & 1.126 & 0.1940 & 1.000\\
721.586 & 1.121 & 0.2009 & 1.000\\
721.588 & 1.111 & 0.2068 & 1.000\\
721.590 & 1.079 & 0.2127 & 1.000\\
721.592 & 1.019 & 0.2166 & 1.000\\
721.594 & 0.9778 & 0.2159 & 1.000\\
721.597 & 0.8599 & 0.2144 & 1.000\\
721.599 & 0.7520 & 0.2155 & 1.000
\enddata
\tablecomments{Observed spectrum towards IRC+10216 ranging from 721 to 864~cm$^{-1}$.
Column 1 contains the observed frequencies in cm$^{-1}$
corrected from velocity shifts due to proper motions of the source,
frequency calibration and uncertainties related to the band center of laboratory
determined frequencies. 
The observed flux, having removed the baseline, can be found in Column 2. Column
3 accounts for an estimation of the atmospheric transmission during the observations
at the considered frequencies. The model results have been included in Column 4
to allow comparisons. See the text for details.
\textit{[The complete version of this table is in the electronic edition of
the Journal. The printed edition contains only a sample.]}}
\end{deluxetable}

In order to fit the continuum, we have used ISO/SWS observations 
carried out on 1996 May 31 (UT), that corresponds to 
$\phi_\subscript{IR}\simeq 0.34$.

\begin{deluxetable*}{c@{~}c@{~}ccc|c@{~}c@{~}ccc}
\tabletypesize{\footnotesize}
\tablecolumns{10}
\tablewidth{0pt}
\tablecaption{Identified and Modeled Lines of C$_2$H$_2$, H$^{13}$CCH, HCN, and H$^{13}$CN\label{tab:lines}}
\tablehead{\multicolumn{3}{c}{Ro-Vibrational Band} & \multicolumn{2}{c}{\# Detected Lines} &
\multicolumn{3}{c}{Ro-Vibrational Band} & \multicolumn{2}{c}{\# Detected Lines}\\
\colhead{\makebox[17mm]{Upper level}} & \colhead{\makebox[17mm]{Lower level}} &
\colhead{Branch} & \colhead{C$_2$H$_2$} & \colhead{H$^{13}$CCH} &
\colhead{\makebox[17mm]{Upper level}} & \colhead{\makebox[17mm]{Lower level}} &
\colhead{Branch} & \colhead{C$_2$H$_2$} & \colhead{H$^{13}$CCH}}
\startdata
$\nu_5(\pi_u)$            & G.S.$(\sigma_g^+)$ & $R_e$ & 50~(30) & 39~(26) & $2\nu_5(\delta_g)$    & $\nu_5(\pi_u)$          & $R_f$ & 40~(24) & $\phantom{0}$2~($\phantom{0}$0) \\
                         &                    & $Q_e$ & 21~($\phantom{0}$0) &  17~($\phantom{0}$0) &                      &                        & $P_f$ & $\phantom{0}$1~($\phantom{0}$0) & $\phantom{0}$0~($\phantom{0}$0) \\
                         &                    & $P_e$ & $\phantom{0}$2~($\phantom{0}$0) &  $\phantom{0}$1~($\phantom{0}$0) & $2\nu_4+\nu_5(\pi_u)$ & $2\nu_4(\sigma_g^+)$    & $R_e$ &  $\phantom{0}$7~($\phantom{0}$0) & $\phantom{0}$0~($\phantom{0}$0)\\
$\nu_4+\nu_5(\sigma_u^+)$ & $\nu_4(\pi_g)$      & $R_e$ & 46~(15) & 11~($\phantom{0}$9) & $2\nu_4+\nu_5(\pi_u)$ & $2\nu_4(\delta_g)$       & $R_e$ &  $\phantom{0}$4~($\phantom{0}$0) & $\phantom{0}$0~($\phantom{0}$0)\\
$\nu_4+\nu_5(\sigma_u^-)$ & $\nu_4(\pi_g)$      & $R_f$ & 42~(25) & $\phantom{0}$9~($\phantom{0}$4) &                        &                       & $R_f$ & $\phantom{0}$2~($\phantom{0}$0)  & $\phantom{0}$0~($\phantom{0}$0)\\ 
                         &                     & $Q_e$ & 20~($\phantom{0}$0) & $\phantom{0}$2~($\phantom{0}$0) & $2\nu_4+\nu_5(\phi_u)$  & $2\nu_4(\delta_g)$    &  $R_e$  & $\phantom{0}$7~($\phantom{0}$0) & $\phantom{0}$0~($\phantom{0}$0) \\
                         &                     & $P_f$ & $\phantom{0}$3~($\phantom{0}$0) & $\phantom{0}$0~($\phantom{0}$0) &                       &                        & $Q_e$  & $\phantom{0}$2~($\phantom{0}$0) & $\phantom{0}$0~($\phantom{0}$0) \\
$\nu_4+\nu_5(\delta_u)$  & $\nu_4(\pi_g)$      & $R_e$ & 37~(25) & $\phantom{0}$9~($\phantom{0}$7)  &                       &                        & $R_f$  & 11~($\phantom{0}$0) & $\phantom{0}$0~($\phantom{0}$0) \\
                         &                    & $Q_f$ & $\phantom{0}$3~($\phantom{0}$0) & $\phantom{0}$0~($\phantom{0}$0)  &                       &                        & $Q_f$  & $\phantom{0}$2~($\phantom{0}$0) & $\phantom{0}$0~($\phantom{0}$0) \\
                         &                    & $P_e$ & $\phantom{0}$2~($\phantom{0}$0) & $\phantom{0}$1~($\phantom{0}$0)  & $\nu_4+2\nu_5(\pi_g)$ & $\nu_4+\nu_5(\sigma_u^+)$ & $R_e$ & $\phantom{0}$2~($\phantom{0}$0) & $\phantom{0}$0~($\phantom{0}$0)\\
                        &                     & $R_f$ & 44~(27) & $\phantom{0}$6~($\phantom{0}$5) & $\nu_4+2\nu_5(\pi_g)$  & $\nu_4+\nu_5(\sigma_u^-)$ & $R_f$ & $\phantom{0}$5~($\phantom{0}$0) & $\phantom{0}$0~($\phantom{0}$0)\\
                         &                    & $Q_e$ & 13~($\phantom{0}$0) & $\phantom{0}$1~($\phantom{0}$0)  & $\nu_4+2\nu_5(\phi_g)$ & $\nu_4+\nu_5(\delta_u)$ & $R_e$ & $\phantom{0}$7~($\phantom{0}$0) & $\phantom{0}$0~($\phantom{0}$0)\\
                         &                    & $P_f$ & $\phantom{0}$1~($\phantom{0}$0) & $\phantom{0}$1~($\phantom{0}$0)  &                        &                        & $R_f$ & $\phantom{0}$8~($\phantom{0}$0) & $\phantom{0}$0~($\phantom{0}$0)\\
$2\nu_5(\sigma_g^+)$     & $\nu_5(\pi_u)$      & $R_e$ & 34~(18) & $\phantom{0}$2~($\phantom{0}$2) & $3\nu_5(\pi_u)$         & $2\nu_5(\sigma_g^+)$     & $R_e$ & $\phantom{0}$2~($\phantom{0}$0) & $\phantom{0}$0~($\phantom{0}$0)\\
                        &                     & $Q_f$ & $\phantom{0}$3~($\phantom{0}$0) & $\phantom{0}$0~($\phantom{0}$0) & $3\nu_5(\phi_u)$       & $2\nu_5(\delta_g)$      & $R_e$ & $\phantom{0}$5~($\phantom{0}$0) & $\phantom{0}$0~($\phantom{0}$0)\\
$2\nu_5(\delta_g)$       & $\nu_5(\pi_u)$      & $R_e$ &  31~(16) & $\phantom{0}$5~($\phantom{0}$5) &                       &                         & $R_f$  & $\phantom{0}$4~($\phantom{0}$0) & $\phantom{0}$0~($\phantom{0}$0)\\
                         &                    & $Q_e$ & $\phantom{0}$2~($\phantom{0}$0) & $\phantom{0}$0~($\phantom{0}$0) &                       &                         &       &      & \\[4pt]
\hline\\[-4pt]
\colhead{\makebox[17mm]{Upper level}} & \colhead{\makebox[17mm]{Lower level}} &
\colhead{Branch} & \colhead{HCN} & \colhead{H$^{13}$CN} &
\colhead{\makebox[17mm]{Upper level}} & \colhead{\makebox[17mm]{Lower level}} &
\colhead{Branch} & \colhead{HCN} & \colhead{H$^{13}$CN}\\[4pt]
\hline
 $\nu_2(\pi)$       & G.S.$(\sigma^+)$   & $R_e$ & 37~(19) & $\phantom{0}$7~($\phantom{0}$5) & $2\nu_2(\delta)$ & $\nu_2(\pi)$ & $R_e$ & 23~(14) & $\phantom{0}$0~($\phantom{0}$0) \\
                   &                    & $Q_e$  & $\phantom{0}$3~($\phantom{0}$0) & $\phantom{0}$0~($\phantom{0}$0) &                  &              & $R_f$ & 22~(17) & $\phantom{0}$0~($\phantom{0}$0) \\
 $2\nu_2(\sigma^+)$ & $\nu_2(\pi)$       & $R_e$ & 11~(10) & $\phantom{0}$0~($\phantom{0}$0) &                  &              &  $Q_e$ & $\phantom{0}$1~($\phantom{0}$0) & $\phantom{0}$0~($\phantom{0}$0) \\[-6pt]
\enddata
\tablecomments{The numbers without parentheses are
the detected lines while the numbers within parentheses
are the modeled
lines. Many lines for each band are contaminated by blending with other
features, overlapping with telluric features, and other effects such
as echelon order boundaries. 
The detected lines involving high-$J$ and high energy ro-vibrational levels
will be published in
a forthcoming paper. Those chosen to be fitted are the 
best quality lines for each band. The spectroscopic notation 
used is described in
Appendix~\ref{sec:LineFrequencies}.}
\end{deluxetable*}

\section{The Model}
\label{sec:modeling}

\subsection{The Physical Structure of IRC+10216 from Previous 
Work}
\label{sec:prevwork}

IRC+10216 is ejecting matter (gas and dust) at a rate of 
$1-2 \times 10^{-5}$~M$_\odot$~yr$^{-1}$ \citep{keady_1988,cernicharo_1996b, cernicharo_1999}.
The derived dust ejection rate is between 
$\simeq 2-4\times 10^{-7}$~M$_\odot$~yr$^{-1}$ \citep{rigdway_1988,menshchikov_2001}. 
The star is pulsating with a period of $636\pm 3$ days 
\citep{rigdway_1988, dyck_1991, jones_1989}. The stellar effective temperature, 
$T_{\subscript{eff}}$, quoted in the literature varies from author to author: 
$2330\pm 350$~K \citep{rigdway_1988}, $2200\pm 150$~K \citep{ivezic_and_elitzur_1996}, 
1915~K at phase 0.16 and 2105~K at phase 0.27 \citep{bergeat_2001}, and 2800 and 2500~K
for maximum and minimum brightness respectively \citep{menshchikov_2001}.

The distance of IRC+10216 is poorly established; values vary between 120 and 300~pc
\citep{doty_1997, keady_1988, weigelt_2002, bergeat_2001, loup_1993, 
cernicharo_2000, herbig_1970}. From interferometric and lunar occultation data,
\citet{rigdway_1988} have obtained a stellar angular radius of 
$0\arcsecond019\pm 0\arcsecond003$, \citet{keady_1988} assumed 
$0\arcsecond023$, the value derived by \citet{monnier_2000a} is 
0\arcsecond022, and \citet{menshchikov_2001} suggest an angular
stellar radius ranging from 0\arcsecond014 to 0\arcsecond018
over the whole period of pulsation (corresponding to $\simeq 815$~R$_\odot$ 
for a distance of 200~pc derived from observations at 1.65, 2.2, 3.15, 
and 4.95~$\mu$m; $\simeq 970$~R$_\odot$ with a distance of 200~pc;
$\simeq 635$~R$_\odot$ for a distance of 135~pc from 8-12~$\mu$m observations;
and a radius ranging from 390 to 500~R$_\odot$ with an assumed distance of 130~pc,
from measurements between 0.6~$\mu$m and 6~mm, respectively).

Infrared observations during a lunar occultation \citep{rigdway_1988}
showed that the dusty inner envelope of IRC+10216 is asymmetric.
There is a large amount of dust in the equatorial plane and two bright
lobes along the poles where the dust density and resulting extinction are 
lower. Interferometric observations by \citet{weigelt_1998} revealed  at least 
four clumps in the lobes. Later observations showed that  these clumps evolve with 
time \citep{weigelt_2002,tuthill_2005} with timescales of $\simeq 1$ year 
\citep{menshchikov_2002}. The large scale density profile of the dust reveals
nearly concentric shells corresponding to increased ejection of matter over 
periods of $200-800$ years lasting $20-40$ years each 
\citep{mauron_and_huggins_1999,mauron_and_huggins_2000,murakawa_2002}. The 
latter ejection episodes are compatible with results by 
\citet{menshchikov_2001} that suggest that the star has experienced at least 
two episodes of high mass loss over the last 1000 years.

From the work of  \citet{keady_1988}, dust grains are formed by amorphous carbon (AC)
with some inclusions of SiC and perhaps other components containing Mg or S. 
SiC condenses in the photosphere, forming the seeds of dust grains (the 
condensation temperature of SiC is $\simeq 2000$~K -- \citealt{menshchikov_2001}).
\citet{ivezic_and_elitzur_1996} have calculated the relative amounts of the main 
components of the dust grains to be $\simeq 95$\% amorphous carbon, $3-8$\% 
SiC, and less than 10\% MgS. Other authors have suggested that the molecular bands 
detected in the continuum can be produced by molecules more complex than MgS 
\citep*[e.g.,][]{menshchikov_2001}.

The gas is accelerated to $\simeq 2$~km~s$^{-1}$ near the photosphere
(innermost envelope). At some distance from the star the carbonate material 
condenses and the gas is accelerated again until it reaches a velocity of
$\simeq 11$~km~s$^{-1}$. \citet{rigdway_1988} have obtained an inner radius for 
the condensation of carbonate material of 5~R$_*$. At $\simeq 11$~R$_*$ 
refractory Mg and/or S-bearing molecular species condense and the gas reaches 
an expansion velocity of $\simeq 14$~km~s$^{-1}$. \citet{keady_1988} propose 
a turbulence in the innermost CSE of $\simeq 5$~km~s$^{-1}$ and a 
terminal turbulence velocity of $\simeq 1.0$~km~s$^{-1}$. Lower values for the
terminal turbulence velocity can be found in the literature, e.g., 0.9~km~s$^{-1}$ 
\citep{huggins_1986} and 0.65~km~s$^{-1}$ \citep{skinner_1999}.

\subsection{Model Description}
\label{sec:modeldesc}

A good approach to the spectral line radiative transfer problem in a CSE is
given by the LVG method. It is very fast and produces good results but 
assumes that the linewidth of a line is negligible compared 
to the velocity 
gradient between very close emitting regions of the CSE. Hence, they cannot
be used to accurately model warm regions like the innermost CSE 
($T_\subscript{K}\simeq 1000-2500$~K) where the linewidth can be larger than 
the expansion velocity gradient.

The most exact model we can develop should be a non-local model which
involves both the numerical resolution of the statistical equilibrium and 
radiative transfer equations.
Studies at radio wavelengths have 
successfully applied this
method to CSEs
(ALI codes by, e.g., \citealt{justtanont_2005} and
Monte Carlo codes by, for example, \citealt{gonzalezalfonso_1997,crosas_1997,schoier_2001}), 
although 
they have needed a significant amount of CPU time even when only a few 
vibrational and rotational levels have been considered. 
Nevertheless, the most important reason that led us to reject the use of
a non-LTE code is the lack of ro-vibrational collisional rates quoted 
in the literature for C$_2$H$_2$ and HCN.
We have identified ro-vibrational lines of C$_2$H$_2$ and HCN
involving vibrational levels with
energies up to 2200~cm$^{-1}$ that are created close to the star. 
In the innermost CSE, the kinetic temperature is high enough to 
significantly populate these
vibrational levels having energies between 3000 and 4000~cm$^{-1}$ 
or more. For example, C$_2$H$_2$ has about 50 vibrational levels below 
4000~cm$^{-1}$. Therefore, a thorough model of the envelope should 
account for many vibrational and rotational states. 
The large number of ro-vibrational 
levels to be considered, makes the problem computationally impractical
with these kind of methods. We can significantly reduce the computing time
by avoiding
exact solutions of the statistical equilibrium equations. 
We do this by assuming a 
temperature dependence with radius (supported by previous work
or physical considerations and by allowing
several parameters free in order to fit the observed lines. 
A reasonably good estimation to the excitation temperatures could be 
achieved taking advantage of the 
large number of observed ro-vibrational lines, allowing us to estimate also the populations
of many ro-vibrational levels below 2500~cm$^{-1}$. This handmade processing technique
accounts for all the physical phenomena having an influence in the molecular level 
populations at the nearby environment of 
each position in the envelope, for example, the collisional rates.

Concerning the geometry of the envelope, modeling the important deviations 
from spherical symmetry in the innermost CSE requires a large number of 
parameters which have limited impact in our fits
due to the low angular resolution of our observations.
Nonetheless, the likely complex velocity field of the gas might introduce some 
small features in a well defined region of the emission component of the 
the high-excitation line profiles.
Unfortunately, the signal-to-noise ratio is not large enough
to unmistakeably identify the signs of the inner structure.
Moreover, the envelope 
approaches spherical geometry
at large scales. Therefore we have ignored the
complex inner dust structure and assumed spherical symmetry for the
whole CSE.

These approximations allow us to solve the problem with one dimensional
calculations. We choose a two dimensional coordinate system with the
$y$ axis parallel to the line of sight pointing away from the Earth 
and the $x$ axis perpendicular to $y$ in an arbitrary direction.
Then, we study the evolution of just one ray of light 
parallel to $y$ for each $x$ and build the final spectrum using the 
symmetry of the envelope. The emerging intensity at each $x$ is calculated by 
summing the emission of each region 
backwards from the star, 
so that the 
radiation emitted by a region
is affected by the optical 
depth of those in the foreground. 
Due to its importance in the mid-IR, dust is included in the calculations and
radiatively coupled to the gas.
The last step is to multiply the emerging intensity
by the point spread function (PSF)
of the telescope, take into account the slit dimensions of
the spectrometer and to convolve the resultant flux with the frequency 
response of the detector.
By virtue of simplicity, 
we have adopted a gaussian profile for both the PSF and the detector response
with the HPBW of the telescope and the width of each channel
of the detector as FWHM, respectively.

We have adopted the velocity field structure proposed by \citet{keady_1988}
which consists of three different zones:
Region I starts at the stellar photosphere and ends at the distance
of the first dust formation layer, \rdin; Region II starts at that 
radius and extends to the position of the second dust formation layer, 
\rdout; finally, the rest of the envelope is Region III.
We have initially assumed that the two dust formation regions
are located at 5 and 15~R$_*$ and have a thickness of
1~R$_*$ \citep{keady_1988}. We allow these values to vary in the model
to get the best fit for the dust emission (continuum of the star) and the 
molecular features. We have divided the envelope into a large number of 
concentric shells to follow the rapid variation of physical conditions.
Using a logarithmic step for the radius increment, most shells are placed 
in the inner and middle regions (Regions I and II), where the temperature and
density gradients are more important.

We have tried to find some simple laws for the variation of molecular 
abundances, density, and gas kinetic and dust grain black-body temperatures
as a function of $r$ that best reproduce the observed continuum and line 
intensities and profiles. We have assumed that all the temperatures (dust, 
kinetic, vibrational, and rotational) follow a continuous radial dependence 
$r^{-\alpha}$, where $\alpha$ could be different for each region of the envelope
and for each ro-vibrational level ($\alpha\to\alpha_{vJ}$), depending on the 
considered temperature. In the case of the dust temperature, we have assumed 
that $\alpha$ remains constant over the whole dusty CSE. 
The parameter $\alpha$ (for Regions I 
and II, and for each temperature except the dust temperature) is completely 
determined through the input of the corresponding temperature at R$_*$ and 
\rdin{} for Region I, and \rdin{} and \rdout{} for Region II. In Region I
$\textnormal{T}_x\left(\textnormal{R}_*\right)/
\textnormal{T}_x\left(\textnormal{\rdin{}}\right)=
\left(\textnormal{R$_*$/\rdin{}}\right)^{-\alpha_{x,\subscript{I}}}$
and in Region II $\textnormal{T}_x\left(\textnormal{\rdin{}}\right)/
\textnormal{T}_x\left(\textnormal{\rdout{}}\right)=
\left(\textnormal{\rdin{}/\rdout{}}\right)^{-\alpha_{x,\subscript{II}}}$,
where $x$ can be dust, kinetic, vibrational, or rotational. However, $\alpha$ 
in Region III cannot be obtained accurately from a modeling of different lines. 
In consequence, we have assumed $\alpha=1.0$ for all the temperatures in Region III 
\citep{doty_1997}. As we will see below, this hypothesis is 
compatible with
the fits. On the other hand, the gas density is assumed to satisfy
the continuity equation:
\begin{equation}
\dot{M} = 4\pi r^2 v_\subscript{exp} \mu m_{\subscript{H}_2} n_{\subscript{H}_2}(r)
\end{equation}
where $\mu=\sum m_i x_i/m_{\subscript{H}_2}$, $m_i$ is the mass of the 
$i^\subscript{th}$ most abundant species and $x_i$ its abundance with respect to H$_2$.
Since the most abundant species after H$_2$ are He and CO (with $x\simeq 0.2$ 
-- solar abundance, \citealt{cox_2000} -- and 
$\simeq 8\times 10^{-4}$, respectively) then, $\mu\simeq 1.4$.

For dust grains, we have assumed a static density profile following a $r^{-2}$ variation
law:
\begin{equation}
n_\subscript{d}(r) =
\frac{\tau_{\lambda_0}}{a_{\lambda_0}}\frac{1}{R_{d1}}\left(\frac{R_{d1}}{r}\right)^2
\end{equation}
where $\tau_\lambda$ and $a_\lambda$ are the optical depth and the absorption
of a dust grain at wavelength $\lambda$ respectively, $\lambda_0$ is 
a fixed wavelength and $\tau_{\lambda_0}$ is an input parameter of the 
model. The wavelength $\lambda_0$ that we have adopted is $11~\mu$m.
The dust opacity is derived from the optical properties of amorphous
carbon (AC) and silicon carbide (SiC) at each wavelength and the emission is
computed from the dust opacity and temperature in each volume element.

The emission and absorption in an elementary integration step is calculated
from the adopted velocity field, the corresponding vibrational and
rotational temperatures, the H$_2$ density and the molecular abundances.
For the gas, the adopted line frequencies used for the line identification
are given in Appendix~\ref{sec:LineFrequencies}, while all the data
relevant to line intensities (opacity, dipole moment, and partition function)
are in Appendix~\ref{sec:LineIntensities}.

\begin{figure}
\centering
\includegraphics[angle=-90,width=0.475\textwidth]{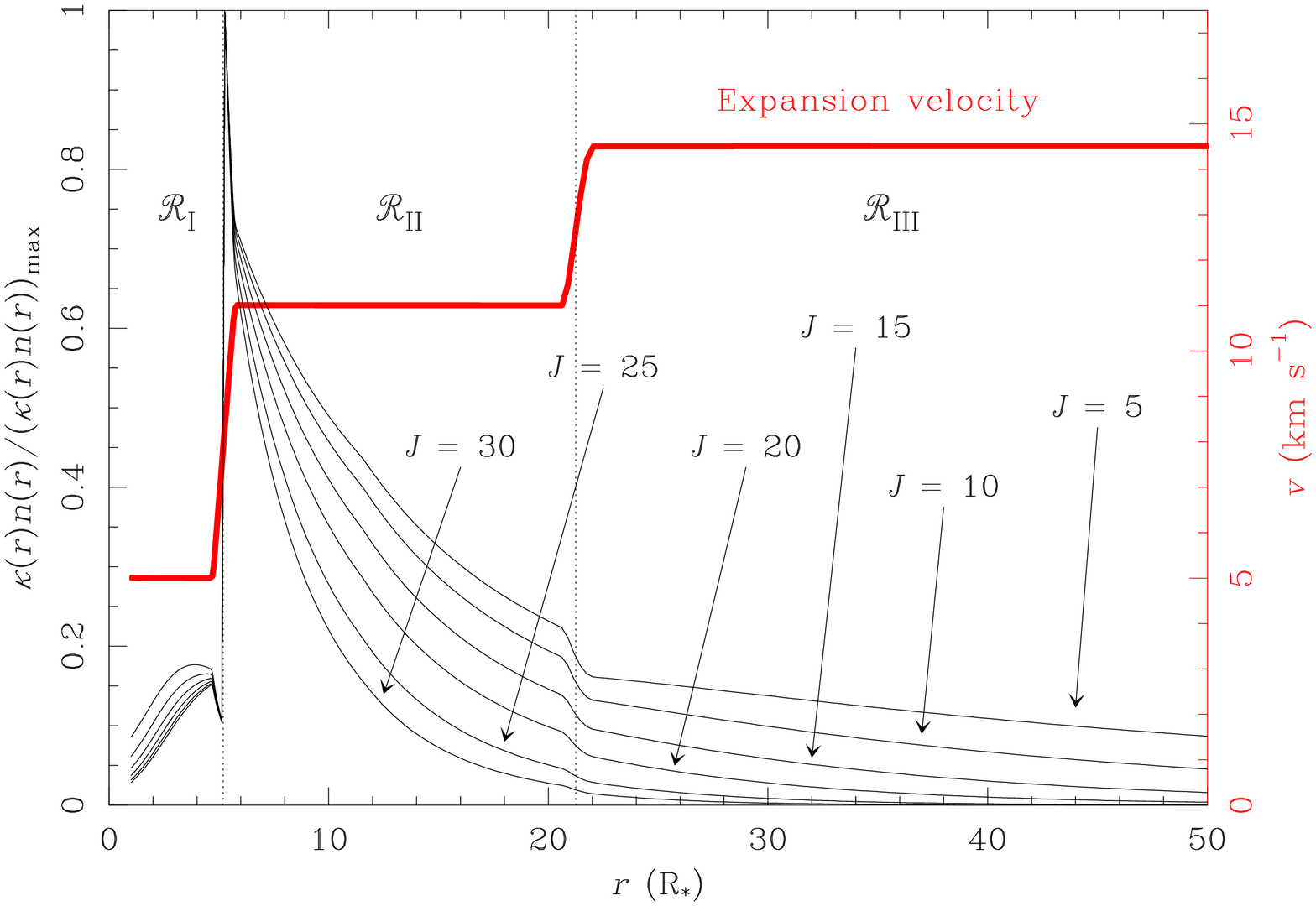}
\caption{Optical depth per unit length
integrated over frequency 
for several lines of the $\nu_5(\pi_u)$R$_e$ branch of C$_2$H$_2$.
The optical depth for low-$J$ lines is still large at 50~R$_*$.
However, high-$J$ lines ($J> 20$) will be mainly formed
in the inner and middle regions (Region I, 
$\mathcal{R}_\subscript{I}$, and Region II, $\mathcal{R}_\subscript{II}$).
Hence, we can obtain information about the outer envelope (Region III, 
$\mathcal{R}_\subscript{III}$)
from low-$J$ lines while high-$J$ lines are sensitive to the physical 
and chemical conditions near the star. Choosing the correct line can 
help improve the accuracy of the determination of the parameters in a 
particular region (see Table~\ref{tab:errors}).
The thick red line represents the velocity field. The sharp increase in 
opacity at $\simeq 5$~R$_*$ corresponds to the change in C$_2$H$_2$ abundance 
in Region I found in this work.}
\label{fig:op}
\end{figure}

\notetoeditor{The figure f7.eps is a colored figure. It should appear just
in the on-line version. In the printed version, the figure should be f7bw.eps.}

The line profile is assumed to be gaussian with a thermal linewidth over 
the external CSE because the dust and the gas are expected to expand at their
terminal velocity in a steady flow. In contrast, the linewidth next to the 
photosphere has not been well determined by previous observations
of IRC+10216. The matter ejection near the star could increase the microturbulent 
velocity adding a non-thermal contribution to the linewidth. In our code, 
this non-thermal contribution has been implemented as $\Delta v_{m,1} e^{-(r-1)/\ell}$, 
where $\Delta v_{m,1}$ is the non-thermal microturbulent velocity over
the photosphere and $\ell$ is a characteristic length ($r$ and $\ell$
are measured in R$_*$). Since the lines affected by temperature and/or 
microturbulence present gaussian profiles, the resultant linewidth is
$(\Delta v)^2=(\Delta v_\subscript{temperature})^2+(\Delta v_\subscript{microturbulence})^2$.
In order to obtain physically significant parameters from the code,
the parameter we have used is the total linewidth in the photosphere, $\Delta v_1$.
High energy ro-vibrational lines, which are formed at the photosphere, could 
carry some information on these parameters.

As many observed lines are modified by many weak features, telluric effects,
and overlaps with other lines, the fits derived by the minimization
of $\chi^2$ would be worse than those obtained here. In addition, the model depends 
on many parameters and minimizing the $\chi^2$ function for all the selected lines 
is computationally unrealistic. 
In fact, we were able to obtain reasonably good eye fits 
which are in agreement with those derived by minimizing the $\chi^2$ function
when diminishing the number of parameters (as a first
approximation).

However, in the case of the uncertainties of the parameters (see \S\ref{sec:sens.disc}),
we have used a numerical method. Given an observed line, 
several parameters could similarly affect the synthetic profile 
masking their effects and hindering an accurate determination of
all their values that reasonably fit the line.
Thus, we could calculate the maximum value adopted by 
a given parameter, by considering it as a function of the others (having imposed
a restriction on the synthetic profile with respect to the observed one; 
for example, setting the function $\chi^2$ to be equal to a given number or 
forcing the synthetic profile to differ from the observed one in less than a given 
quantity), and by following the gradient of this function. It is also possible to
find the minimum value of that parameter by following the field opposite to
the gradient. This is a good method to be used 
for a large number of parameters and for a few lines, since it is possible to 
approach the maximum (minimum) 
in several iterations avoiding unnecessary calculations and
spending a reasonable amount of CPU time.

In our models we have adopted a stellar temperature of 2330~K, a distance of
180~pc, an angular radius for the star of 0\arcsecond019 (corresponding to a stellar
radius of 735~R$_\odot$) and a mass loss rate of $2.1\times 10^{-5}$~M$_\odot$~yr$^{-1}$.

The region between 1 and 100~R$_*$ is shielded against Galactic UV photons by
the outer dust envelope. Hence, we have considered the radiation field coming
only from the star and the dust component of the CSE.

Although the envelope can extend to more than 1000~R$_*$, the dust in the 
outermost part of the CSE does not contribute significantly to
the continuum flux in the mid-infrared (an envelope with a maximum radius
of 600~R$_*$ divided in 300 shells is enough to find a good fit for
the continuum).

The vibrational and rotational temperatures decrease so fast that only the
emission and absorption of the low-$J$ levels of the ground vibrational state of
C$_2$H$_2$ and HCN could be affected by the choice of the external radius in 
our models (e.g., at 100~R$_*$ the kinetic temperature is $\simeq 85$~K and the 
most populated level of C$_2$H$_2$ is $J\simeq 4-5$, see 
Figure~\ref{fig:op}).
The outer radius of the CSE for fitting the lines has been fixed to 300~R$_*$
since there is no significant modification due to the CSE region between 300 
and 600~R$_*$. A total of 100 layers were modeled.

The absorbing feature at terminal velocities is mainly produced in Region III
for low-$J$ ro-vibrational lines. Optically thin absorbing lines are produced 
essentially in Region II, related to the $(-10,10)$~km~s$^{-1}$ velocity range 
in the profiles, setting the absorption maxima at velocities around $-8$ and 
$-10$~km~s$^{-1}$. The innermost region of the CSE (Region I), and the fact 
that the dust continuum is largely formed outside of this radius, has
little effect in low-$J$ line profiles due to its small angular
size compared with the rest of the CSE. The kinetic temperature and density 
are high enough to assure LTE throughout the envelope for $J\le 20$ within 
vibrational levels up to $2\nu_5$ for C$_2$H$_2$ and up to $2\nu_2$ for HCN.
The validity of this assumption has been checked in our models and found 
to be acceptable. However, high energy ro-vibrational levels are populated 
almost completely in Region I, allowing us to derive physical parameters 
for the regions closest to the star. 
Consequently, the sensitivity of the
observed line profiles to the physical conditions of the gas and the large 
variation of these parameters across the envelope, limit the radial 
resolution we can achieve for the molecular abundance profiles.

\section{Continuum Emission}
\label{sec:continuum}

The dust properties affect considerably the molecular excitation due to 
radiative coupling between the dust and the gas. P-Cygni profiles arise 
in all the lines created in the inner CSE. In particular, the C$_2$H$_2$ 
and HCN lines show this kind of lineshape
(see Figures~\ref{fig:f1}--\ref{fig:f6}).
The line profile depends on the dust parameters such as the dust temperature, 
composition, size of the dust grains, and absorption and scattering cross-sections.

For the wavelength range under consideration (MIR, $\lambda\simeq 11-14~\mu$m),
the size of the grains has a small effect on the continuum.
This parameter can be neglected at larger wavelengths (FIR and radio) but is much more
important in the NIR range \citep{ivezic_and_elitzur_1996}, i.e.,
$\lambda\le 1-2~\mu$m. We have assumed that dust grains are spheres with 
a constant radius not larger than 0.1~$\mu$m and calculated their opacity 
using Mie Theory \citep*[e.g.,][]{hoyle}. According to the theory, the 
scattering cross-section is small compared to the absorption cross-section for 
dust grains with diameter small compared to the wavelength of radiation. Since 
we are considering $\lambda\simeq 11-14~\mu$m radiation, we ignore scattering.
The composition of the dust grains are guessed to be amorphous carbon (AC) and
silicon carbide (SiC) (see \S\ref{sec:prevwork}). The complex refractive 
index of AC as a function of $\lambda$ has been taken from 
\citet{rouleau_1991}. Laboratory works on SiC \citep{mutschke_1999} suggest
that SiC is probably crystalline in space. Unfortunately, the difficulties in 
measuring the optical constants for crystalline SiC have led us to adopt the 
laboratory data
for the refractive index of amorphous SiC in our models.
For the dust temperature, \tdust{}, we initially assume the $r^{-0.4}$ 
dependence,
proposed by \citet{rigdway_1988}. 
Actually, the \tdust{} dependence on $r$ is steeper in the innermost
CSE than at greater distances because the opacity of the dust grains is
larger at higher frequencies, i.e., closer to the star. 
We have 
adopted the power law shown above as an approximation.
The exponent is a very sensitive parameter
and can be determined with a high accuracy (see \S\ref{sec:sens.disc}).
Finally, concerning the dust optical depth, $\tau_\lambda$, the latter authors
derived a value $\tau_\lambda\simeq 1$ at 11~$\mu$m, \citet{ivezic_and_elitzur_1996} 
obtained $\tau_\lambda\simeq 0.32-0.40$
at 10~$\mu$m, and \citet{monnier_2000a}, obtained $\tau_\lambda\simeq 0.66$
at 11.15~$\mu$m.

In this work, we have fitted the continuum data of IRC+10216 obtained by ISO/SWS, 
adjusting the dust parameters 
as described above. The ISO 
observations used for that purpose are those quoted by 
\citet{cernicharo_1996b, cernicharo_1999}.
We have assumed that the difference between the stellar phases of the observations
(see \S\ref{sec:obs}) with ISO/SWS, and IRTF/TEXES, does not introduce 
any important effect on dust and molecular properties, and that the derived \tdust{} 
from the ISO data is similar to those prevailing at the moment of the TEXES 
observations.

We have fitted the continuum between 7 and 27~$\mu$m. Fitting to shorter 
wavelength ISO data is inappropriate because scattering will be important 
and the model does not include scattering. The brightness of the central star 
has a small impact on the observed continuum. The maximum contribution to the 
continuum from the star is at 5.31~$\mu$m ($F_\nu\simeq 1075$~Jy) and the emission 
at 10~$\mu$m is $\simeq~1.73$ times lower. A small modification of the
stellar temperature does not significantly affect
the flux emerging from the source.
Note, however, that a large change in the stellar temperature could modify
the physical and chemical properties of the envelope and, hence, of the emergent
flux. In any case, it seems obvious that the observed mid- and far-infrared 
emission in IRC+10216 comes mainly from the dusty envelope.

We obtain from our analysis that the inner and outer dust formation shells are 
located at radii $5.2^{+0.6}_{-0.5}$~R$_*$ 
and $21\pm 3$~R$_*$ (0\arcsecond1 and 0\arcsecond4),
respectively (see Figure~\ref{fig:op}, Table~\ref{tab:errors},
and \S\ref{sec:sens.disc}). A value, R$_{\subscript{d}1}=5.2$~R$_*$ is derived from 
fitting only the continuum. However, \rdout{} can be determined with more or less 
precision through fitting several molecular lines, while the continuum of 
the star displays little information about this outer dust formation shell. 
Studying how changes to the dust density, $n_\subscript{d}$, affect \rdout{} led us 
to very inaccurate results because the variation of $n_\subscript{d}$ over 
this shell only slightly affects the continuum.

Several studies have determined \rdin{} since the discovery of IRC+10216. 
\citet{keady_1988} proposed that R$_{\subscript{d}1}\simeq 3$~R$_*$ equivalent
to $\simeq 0\arcsecond057$, with a stellar radius of $\simeq 970$~R$_\odot$ and
a distance to the star of 200~pc. Later, \citet{rigdway_1988} found
R$_{\subscript{d}1}\simeq 5$~R$_*$ and an angular stellar radius of $\simeq 0\arcsecond019$, 
implying an angular radius of the inner dust formation zone, $\alpha_{\subscript{d}1}$, of 
$\simeq 0\arcsecond095$. \citet{monnier_2000a} found
$\alpha_{\subscript{d}1}\simeq 0\arcsecond15$ at a distance of 135~pc and
$\alpha_*\simeq 0\arcsecond022$, meaning that
R$_{\subscript{d}1}\simeq 6.8$~R$_\odot$, 
and \citet{ivezic_and_elitzur_1996} determined $\alpha_{\subscript{d}1}$ to be
$\simeq 0\arcsecond22$ at maximum luminosity and $\simeq 0\arcsecond15$ at 
minimum luminosity. Consequently, our result is in good agreement with the 
values quoted in the literature. For \rdout{}, the angular radius of the outer 
dust formation zone, \citet{keady_1988} found
R$_{\subscript{d}2}\simeq 14$~R$_*$ 
($\alpha_{\subscript{d}2}\simeq 0\arcsecond32$). We derived 
$\alpha_{\subscript{d}2}\simeq 0\arcsecond40$, a larger value 
than found by \citet{keady_1988} and \citet{keady_1993}.

The optical depth derived with our model at 11~$\mu$m is $0.7$, similar to that 
proposed by \citet{monnier_2000a}, $\tau(\lambda=11.15~\mu\textnormal{m})=0.66$, 
and larger than the value proposed by \citet{ivezic_and_elitzur_1996}, who 
derived $\tau(\lambda=10~\mu\textnormal{m})\simeq 0.32-0.40$.
On the other hand, the value for the temperature of the innermost dust 
formation shell, \tdin, derived in this work is $850\pm 25$~K
(see \S\ref{sec:sens.disc} and Table~\ref{tab:errors} for a discussion
on the errors), lower than that obtained by \citet{rigdway_1988} 
($T_{\subscript{d}1}=1040\pm 100$~K) and \citet{groenewegen_1997} 
($T_{\subscript{d}1}=1075\pm 50$~K), similar to \citet{monnier_2000a} 
($T_{\subscript{d}1}=860$~K), and higher than \citet{ivezic_and_elitzur_1996} 
($T_{\subscript{d}1}=750\pm 50$~K). The exponent of the \tdust{} law is derived to 
be 0.39, quite similar to that proposed by \citet{keady_1988}.

The best fit to the dust composition is 95\% amorphous carbon and 5\% amorphous 
silicon carbide. Modifying the proportion of SiC changes the ratio of the 
predicted flux at 11~$\mu$m, with respect to the rest of the continuum. Nevertheless, 
this change in composition could be balanced by modifying the optical depth and 
dust temperature at \rdin. Fitting the whole continuum is necessary to get more 
realistic percentages.

Taking into account the different observing periods and the wavelength
coverage in these different data sets, we consider that our estimate
of \tdust{} and opacity in \rdin{} are representative of the physical 
conditions in IRC+10216.

\section{C$_2$H$_2$, Modeling}
\label{sec:acetylene}

\begin{figure*}
\centering
\includegraphics[scale=0.8]{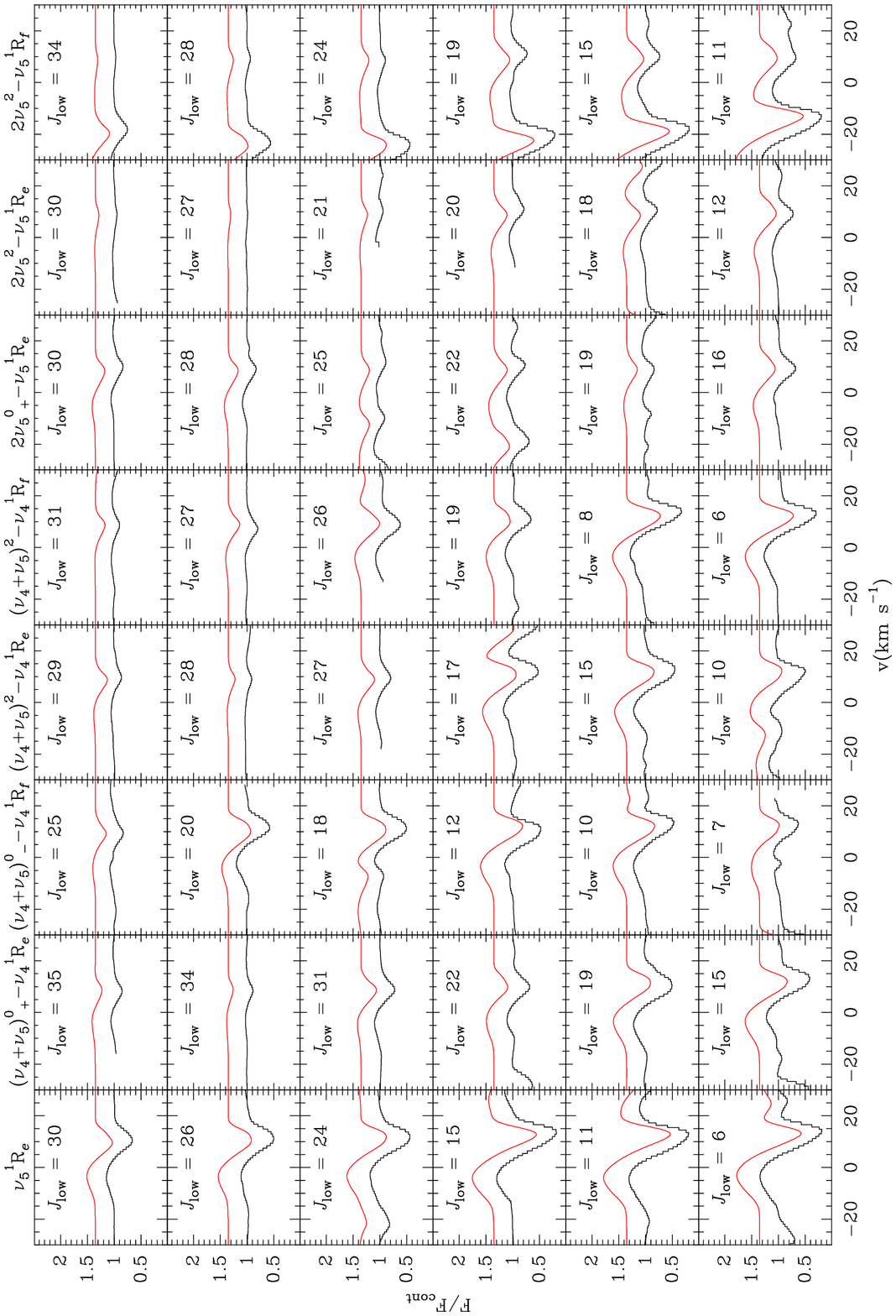}
\caption{Fits of some lines of C$_2$H$_2$. The black lines are the observed
spectra while the red lines are the fits. The data have been corrected from
observational frequency deviations adding/suppressing the mean difference
between the observational and the laboratory frequency of many lines (see
\S\ref{sec:obs}).
The existing discrepancies are small and can be assumed as blends with other 
lines and/or observational errors (the spectroscopic notation 
is described in Appendix~\ref{sec:LineFrequencies}).}
\label{fig:fitsc2h2}
\end{figure*}

In order to obtain the abundance of C$_2$H$_2$ and the physical parameters 
of the CSE, we have selected several sets of C$_2$H$_2$ lines that could be 
sensitive to the derived parameters (see Table~\ref{tab:lines} and 
Figure~\ref{fig:op}).
The assumed gas velocity profiles
and computed line profiles
are in very good agreement (see Figure~\ref{fig:fitsc2h2}). The C$_2$H$_2$ 
abundance and physical conditions from the fits are listed in the following 
subsections.

\notetoeditor{We would like this figure to be placed near the latter
paragraph. It should fill a whole page in landscape mode and occupying
as much space as possible. f8bw.eps is the black-and-white version of 
f8.eps. We would like f8.eps to appear just in the on-line version of the 
paper. We would like that the sentence ``The black lines are the observed
spectra while the RED lines are the fits'' appears in the on-line version of
the paper while ``The black lines are the observed spectra while the GRAY
lines are the fits'' goes with the black-and-white figure.}

\subsection{Velocity Profile}
\label{sec:vp}

The simultaneous modeling of different $J$ ro-vibrational transitions of the same
band requires a three-region velocity field \citep{keady_1988}. As can be seen
in Figure~\ref{fig:op}, lines with higher energy are formed closer
to the photosphere. Consequently, the effect of the expansion on the lineshape
changes with $J$. In many cases,
a variation of the velocity in one of these regions modifies the lineshapes 
significantly. For example, the gas velocities in Regions II and III control 
the position of the intensity minimum, so that lines formed closer to the
star represent the bulk absorption at lower velocities than the lines formed
at larger distances. 

On the other hand, the model is very insensitive to variations in the 
thickness
of the acceleration shells (see \S\ref{sec:sens.disc}).
The linewidths in these regions are
similar to the velocity-increasing
in the acceleration zones, and hence hide
the acceleration effects on the profiles.
The small spatial extent of the innermost region and the limited angular
resolution of our data, prohibit a better determination of the physical and 
chemical conditions in Region I.
We have found that the best fit to the lines can be achieved with the
following velocity profile (see Figure~\ref{fig:op}):
\begin{equation}
v_e(r) = \left\{
\begin{array}{ll}
5 & \textnormal{$1\le r/\textnormal{R$_*$} <4.7$}\\
5-11^* & \textnormal{$4.7\le r/\textnormal{R$_*$} <5.7$}\\
11 & \textnormal{$5.7\le r/\textnormal{R$_*$} <20.75$}\\
11-14.5^* & \textnormal{$20.75\le r/\textnormal{R$_*$} <21.75$}\\
14.5 & \textnormal{$21.75\le r/\textnormal{R$_*$}$}
\end{array}
\right.
\end{equation}
where $^*$ means \textit{constant velocity gradient}, $dv_e/dr$
between the inner and outer radii of the corresponding zone.

\subsection{Linewidth}
\label{sec:linewidth}

When microturbulence is included in the calculations, 
the emission from Region I spreads and the intensity of 
the lines reaches its maximum at higher velocities (towards lower 
frequencies), compared with the fits without microturbulence.
Unfortunately, fits for low- and intermediate-$J$ ro-vibrational 
levels are relatively insensitive to the total linewidth in the photosphere, 
$\Delta v_1$, 
and the characteristic length, $\ell$. 
A substantial change in the fitted profiles requires unrealistic 
values for these parameters. Reasonable values, for example, are
$\Delta v_1=5$~km~s$^{-1}$ and $\ell=1.5$~R$_*$. 
The value of $\ell$ is acceptable due to the physical 
phenomena thought to occur near the photosphere, i.e., a moderate pulsation
of the central star and the appearance of
sound waves \citep{bowen_1988, pijpers_1989a, pijpers_1989b}.
However, the fits obtained with these parameters are not more
accurate than those calculated without microturbulence.
In fact, when choosing $\Delta v_1=30$~km~s$^{-1}$ and 
$\ell=1.5$~R$_*$, the emission can be fitted more accurately. We will
discuss this further in \S\ref{sec:sens.disc}.

\subsection{Kinetic Temperature, \tk{}}
\label{sec:tkc2h2}

C$_2$H$_2$ low and intermediate energy levels are highly populated in each 
vibrational level, and many of the ro-vibrational transitions in 
which they are involved are optically thick and therefore hide information 
(e.g., low-$J$ transitions in the 
fundamentals in Regions I and II). 
However, there are still many lines in LTE that are not optically 
thick in the hot bands over the whole envelope 
and in the fundamental band in Regions II and III (see Figure \ref{fig:op}).
We can determine the kinetic temperature profile by fitting simultaneously 
several low- and intermediate-$J$ lines from the fundamental band and 
relatively low energy hot bands. The relationship between the intensities 
of adjacent ro-vibrational lines in LTE within a given band allows us to 
derive \tk{} immediately. In doing so,
it is necessary to derive 
the vibrational temperatures and the molecular abundances at the same time as \tk{}. 
We will discuss this determination in \S\ref{sec:vibc2h2}.

We have fixed the kinetic temperature at $r=1$~R$_*$ equal to the 
temperature of the central star. For Regions I and II, 
\tk{} at \rdin{} and \rdout{} are free parameters 
that completely determine
the value of the exponent. 
The temperature exponent obtained is $0.58$ for both Regions I and II,
while the fits support the assumption of $\alpha=1.0$ over Region III.

The kinetic temperature profile derived from the best fit to
all C$_2$H$_2$ lines is:
\begin{equation}
T_\subscript{K}(r) = \left\{
\begin{array}{rl}
2330~(r/\textnormal{R$_*$})^{-0.58} & \textnormal{Region I}\\
900~(r/\textnormal{\rdin})^{-0.58} & \textnormal{Region II}\\
400~(r/\textnormal{\rdout})^{-1.00} & \textnormal{Region III}
\end{array}
\right.
\end{equation}

\subsection{Rotational Temperatures, \trot{}}
\label{sec:trc2h2}

An accurate estimate of the rotational temperatures requires a sophisticated 
model including molecular parameters, such as collisional rates between 
ro-vibrational levels, that are currently unavailable. Therefore, we have used 
our data to derive a first estimate of rotational temperatures for all 
ro-vibrational transitions in the three zones defined above. The procedure to 
derive \trot{} profiles is as follows:
\begin{enumerate}
\item \tk{} and the vibrational temperatures are derived by
modeling the ro-vibrational transitions that involve both low and intermediate energy 
rotational levels (see \S\ref{sec:tkc2h2}).
\item Initially, all the high energy ro-vibrational levels are considered to be
in LTE.
\item We vary \trot{} for all the high-$J$ levels in each vibrational state to fit
the observed line profiles. \trot{} is considered to depend on $J$ as a 2$^\subscript{nd}$ 
degree polynomial.
\item In the next iteration, step 3 is repeated for the next ro-vibrational transition 
of the band under consideration.
\end{enumerate}
At each step, all line intensities are recalculated because a change in the rotational 
temperature of one of them implies a variation in line opacities. As the transitions 
involve higher energy levels, the variation in the opacities of the fitted transitions 
becomes lower and lower. The process is repeated until convergence is reached for
each band. The remaining lines to fit belong to ro-vibrational transitions involving 
vibrational levels with higher energies. Running the code two or three times is 
usually enough to fit the line correctly with a CPU time of less than 10~s per
line. Unfortunately, fitting lines of hot bands sometimes requires repeating the 
fitting process for lines involving lower energy vibrational levels since the upper level 
of an optically thick transition can coincide with
the lower level of an optically thin one.

\begin{figure}[!hbt]
\centering
\includegraphics[angle=-90,width=0.475\textwidth]{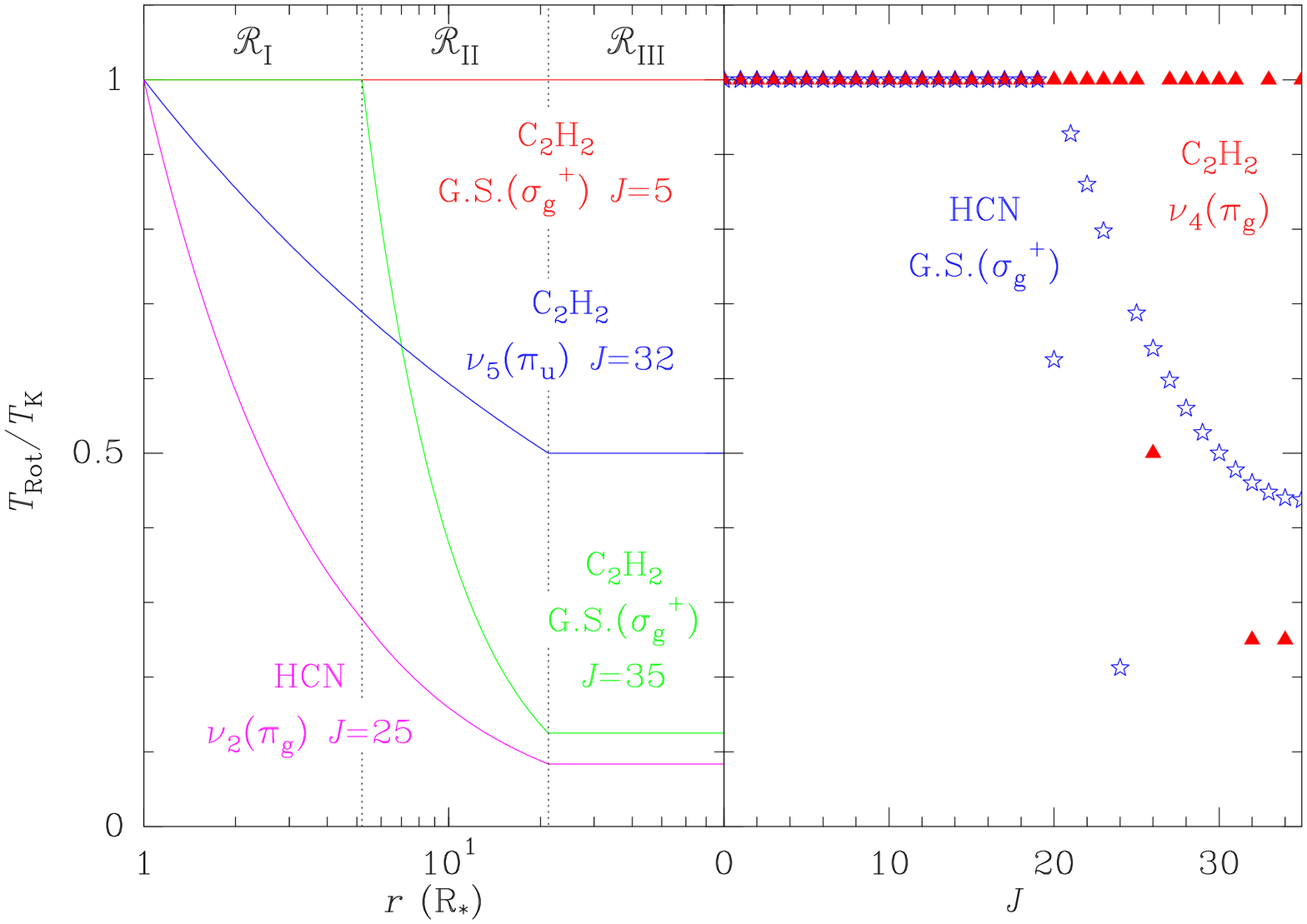}
\caption{Rotational temperatures
for some C$_2$H$_2$ and HCN levels depending
on the radius (left) and rotational temperatures
for some C$_2$H$_2$ and HCN
vibrational levels at $r=$\rdout{} depending on $J$ (right). The
rotational temperature is less than or equal to the kinetic
temperature throughout the CSE (left).
Fitting the ro-vibrational transitions sometimes needs the
introduction of \textit{ad-hoc} rotational temperatures whose
dependence usually decreases smoothly with $J$ (right).}
\label{fig:trot}
\end{figure}

As discussed above, \trot{} may be considered equal to the kinetic temperature in 
the innermost zones for low- and intermediate-$J$ levels in lower energy vibrational 
states. However, for high-$J$ levels in high energy vibrational states, 
we have found that \trot{} is below \tk{}, even near the photosphere.
In addition, for some specific transitions, it has been necessary to
include \textit{ad-hoc} rotational temperatures in order to match
the data (see Table~\ref{tab:adhocrottemp}). It is common, for
two or three
important \textit{ad-hoc} rotational temperatures to be found 
in a single vibrational
level below $J\simeq 30-35$. These \textit{anomalous} rotational temperatures 
are unrelated to frequencies having larger telluric interference,
and there are no apparent instrumental effects. 
Moreover, high-$J$ lines are 
in an optically thinner
part of the atmosphere than low-$J$ ones, and the latter seem to be 
under LTE with high accuracy (see Figure~\ref{fig:trot}). 
Therefore, these effects could reflect real physical processes such as collisional
rates, radiative selective pumping effects, or overlaps of
these lines with other spectral features \citep*[see][]{fonfria_2006}.

\begin{deluxetable}{c@{~}c@{~}c|c@{~}c}
\tabletypesize{\footnotesize}
\tablecolumns{5}
\tablewidth{0pt}
\tablecaption{\textit{Ad-hoc} Rotational Temperatures of C$_2$H$_2$ and HCN\label{tab:adhocrottemp}}
\tablehead{\multicolumn{3}{c}{C$_2$H$_2$} & \multicolumn{2}{c}{HCN}\\
\colhead{G.S.$(\sigma_g^+)$} & \colhead{$\nu_4(\pi_g)$} & \colhead{$\nu_5(\pi_u)$} & 
\colhead{G.S.$(\sigma^+)$} & \colhead{$\nu_2(\pi)$}}
\startdata
35(900,50) & 26(620,200) & 20(900,200) & 20(900,250) & 25(250,35)\\
36(900,60) & 32(425,100) & 26(900,150) & 24(640,\phantom{0}85)  & \\
           & 34(425,100) & 32(620,200) &             & 
\enddata
\tablecomments{Ad-hoc rotational temperatures of C$_2$H$_2$ and HCN
for the lowest vibrational levels.
The numbers are shown according to the nomenclature 
$J$(\trot(\rdin),\trot(\rdout)), where $J$ is the rotational level requiring
an \textit{ad-hoc} \trot. 
The temperatures are expressed in K
(see the text for an explanation for the radial dependence).
The rest of the rotational temperatures have a
smooth dependence on $J$. We have only considered those
rotational levels involved
in ro-vibrational levels up to $J=35-40$ (depending on the band). 
Maybe new \textit{ad-hoc} \trot{} should be included for higher-$J$ levels.}
\end{deluxetable}

\notetoeditor{The figure f9.eps is colored. We would like it to be included
only in the on-line version of the paper.}

We have found that \trot{} for levels below $J\simeq 20-30$ follows \tk{}
(depending on the vibrational level). However, for high-$J$ lines, \trot{} is 
systematically lower than \tk{}. This fact seems to indicate that the
physical conditions over the inner CSE are not appropriate to
maintain high-$J$ levels under LTE, as we assumed in \S\ref{sec:modeldesc}.

\subsection{Vibrational Temperatures}
\label{sec:vibc2h2}

The vibrational temperatures were determined from a fitting of line
intensities, rather than by non-LTE radiative transfer 
calculations, but we
can explain the derived populations with radiative pumping.

\begin{figure}[!htb]
\centering
\includegraphics[width=0.475\textwidth]{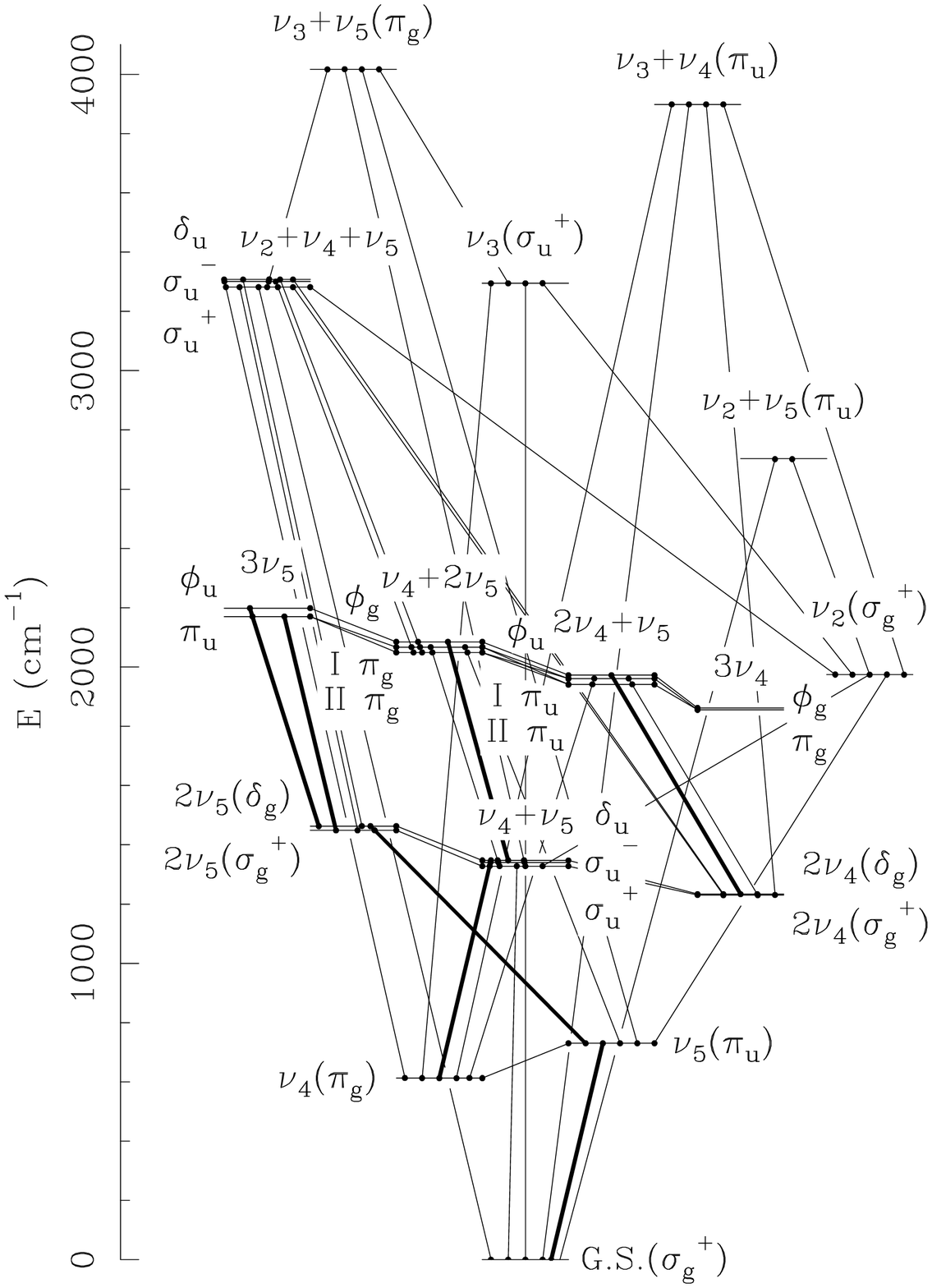}
\caption{Energy of the vibrational levels of C$_2$H$_2$. The
allowed radiative transitions are shown (only the perpendicular transitions
with $\Delta v_4=\pm 1$ and $\Delta v_5=\pm 1$, and the most
intense parallel ones). This
scheme also represents the $^{13}$C$_2$H$_2$ vibrational level structure.
The vibrational levels for H$^{13}$CCH are similar but they do not
express $g$ or $u$. Some infrared inactive modes
of C$_2$H$_2$ are weakly infrared active for H$^{13}$CCH (i.e., the $\nu_4$ mode
described by \citealt{dilonardo_2002}).
The thin lines are allowed transitions while the thick ones are
observed transitions.}
\label{fig:vlc2h2}
\end{figure}

Some excited levels of C$_2$H$_2$ are connected to the ground state only 
through collisions and radiative cascades ($\nu_4$, Raman active), while 
others are connected both by collisions and direct absorption of photons ($\nu_5$, 
IR active). Near the star, collisions with H$_2$ and He could play a role in the
pumping of the low energy vibrational levels. Moreover, selection rules for 
C$_2$H$_2$ only allow
radiative transitions from \textit{gerade} to \textit{ungerade} 
vibrational states and vice-versa (Figure~\ref{fig:vlc2h2}). 

\notetoeditor{This figure should be fitted to one column.}

Consequently, the pumping of low energy vibrational levels could occur
through absorption of near-IR photons emitted by both the star and the dust. 
These photons will pump C$_2$H$_2$ from the ground state to the strongest stretching 
mode, $\nu_3(\sigma_u^+)$, and its combination bands, $\nu_3+\nu_4(\pi_u)$,
$\nu_3+\nu_5(\pi_g)$, and $\nu_2+\nu_4+\nu_5(\sigma_u^+)$
(the notation is discussed in Appendix~\ref{sec:LineFrequencies})
followed by radiative decay to the bending modes \citep{cernicharo_1999}.
These infrared transitions are strong ($\nu_3(\sigma_u^+)$),
medium ($\nu_3+\nu_4(\pi_u)$, $\nu_3+\nu_5(\pi_g)$), and very strong 
($\nu_2+\nu_4+\nu_5(\sigma_u^+)$), as indicated by \citet{mandin_2005} and 
\citet*[][p.~290]{herzberg_ii}.

\begin{figure}[!htb]
\centering
\includegraphics[width=0.475\textwidth]{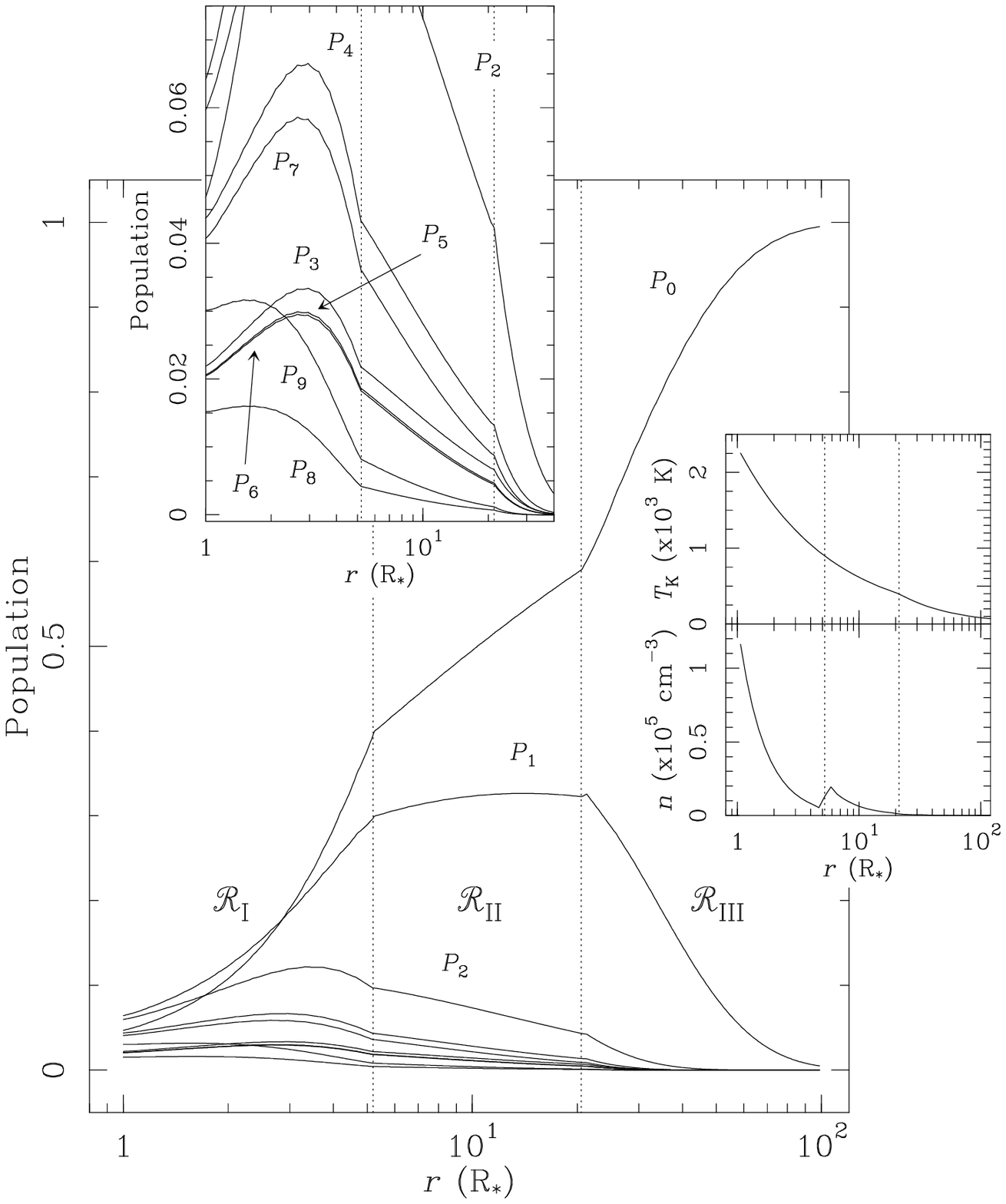}
\caption{The population, $P$, of the vibrational levels of C$_2$H$_2$. They are 
labeled as follows: $P_0\equiv P[\textnormal{G.S.}(\sigma_g^+)]$,
$P_1\equiv P[\nu_4(\pi_g)]$, $P_2\equiv P[\nu_5(\pi_u)]$,
$P_3\equiv P[2\nu_4(\sigma_g^+)]$, $P_4\equiv P[2\nu_4(\delta_g)]$, 
$P_5\equiv P[\nu_4+\nu_5(\sigma_u^+)]$,
$P_6\equiv P[\nu_4+\nu_5(\sigma_u^-)]$,
$P_7\equiv P[\nu_4+\nu_5(\delta_u)]$,
$P_8\equiv P[2\nu_5(\sigma_g^+)]$, and
$P_9\equiv P[2\nu_5(\delta_g)]$.
The $\pi$ and $\delta$ levels are split into two sublevels
with opposite parities, $e$ and $f$ (see Appendix~\ref{sec:LineFrequencies}).
In the Figure they have been merged into a single level with degeneracy 2.
Only vibrational levels up to 2$\nu_5(\delta_g)$ have been plotted although
higher energy levels have been considered in the calculations.
The main box contains the vibrational levels, while the upper inset focuses on
the high energy ones. The right boxes contain C$_2$H$_2$ density (lower) and 
\tk{} (upper).}
\label{fig:popc2h2}
\end{figure}

\notetoeditor{The figure must appear before \label{tab:tvc2h2} and should
be fitted to one column.}

Radiative pumping through $\nu_3(\sigma_u^+)$ and 
its combination bands, is only effective in the innermost region.
Farther out, absorption of mid-IR photons coming from the star and the dust 
pump C$_2$H$_2$ from the ground state to $\nu_5$, and from $\nu_4$ and $\nu_5$ to their 
combination band, $\nu_4+\nu_5$, and other vibrational levels, as overtones.
For $r>20$~R$_*$, the available near-infrared photons are strongly
reduced due to dust absorption. However, pumping through
$\nu_5(\pi_u)$ is efficient over the whole CSE. Hence, we could expect 
vibrational temperatures, \tvib, to be out of LTE as the distance 
from the star increases.

\begin{figure}[!htb]
\includegraphics[angle=-90,width=0.475\textwidth]{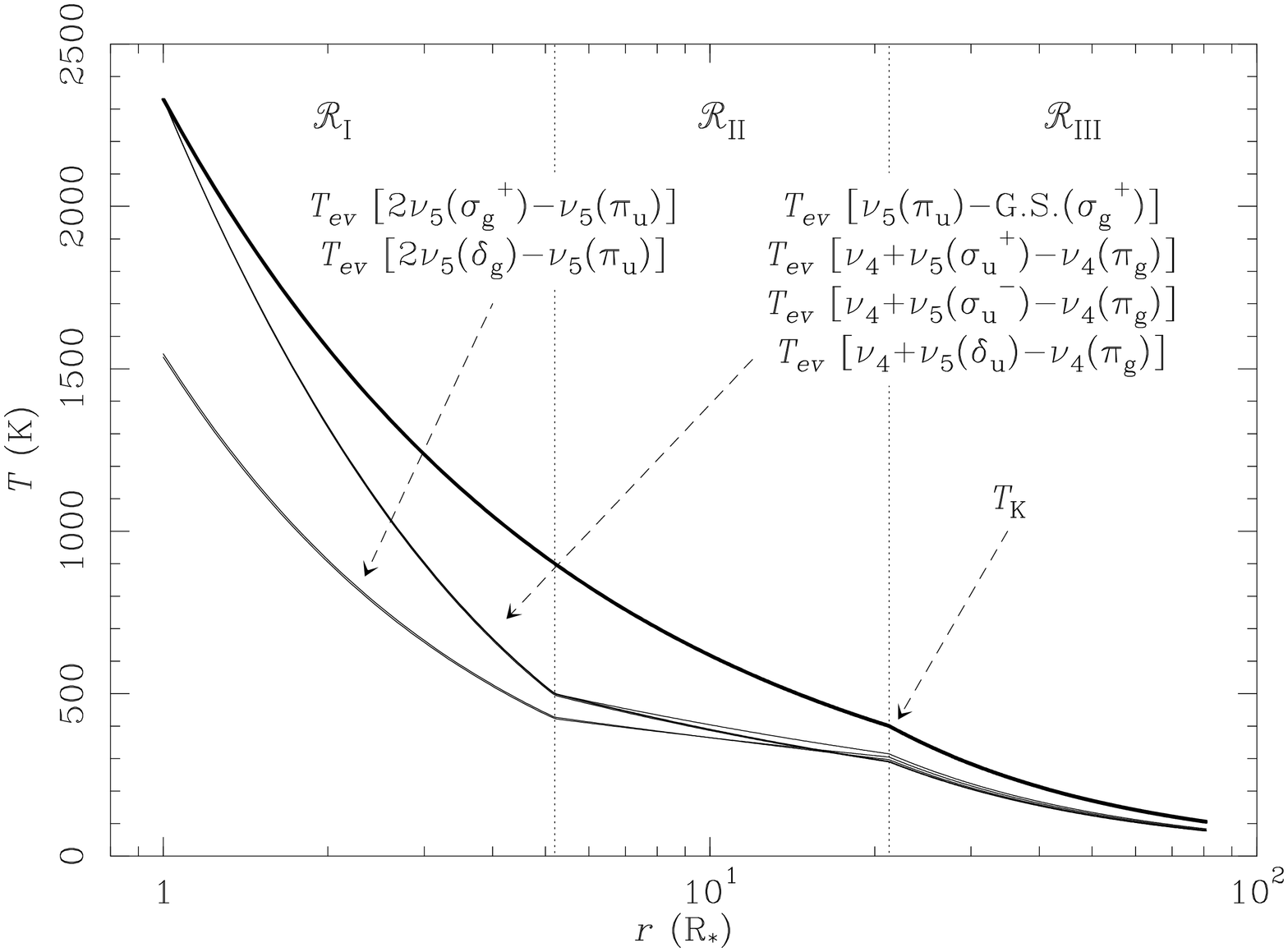}
\caption{Kinetic and vibrational excitation temperatures for the observed 
transitions of C$_2$H$_2$. \tev{} are merged into two different groups.
In spite of the similarity of the dipole moments of the transitions,
$T_\subscript{ev}[2\nu_5(\sigma_g^+)\leftrightarrow\nu_5(\pi_u)]$ and
$T_\subscript{ev}[2\nu_5(\delta_g)\leftrightarrow\nu_5(\pi_u)]$ are quite
different from $T_\subscript{ev}[\nu_4+\nu_5(\sigma_u^\pm)\leftrightarrow \nu_4(\pi_g)]$,
$T_\subscript{ev}[\nu_4+\nu_5(\delta_u)\leftrightarrow
\nu_4(\pi_g)]$, and $T_\subscript{ev}[\nu_5(\pi_u)\leftrightarrow\textnormal{G.S.}
(\sigma_g^+)]$. All the considered vibrational levels are out of LTE over the whole 
CSE. Note that $2\nu_5 (\sigma_g^+)$ and $2\nu_5(\delta_g)$ are out of LTE even near 
the star.}
\label{fig:tvc2h2}
\end{figure}

To estimate \tvib{}, we have selected some low-$J$ lines from each 
observed vibrational band. As the vibrational excitation temperature, \tev, 
depends on \tvib{} and controls the population of the upper vibrational level with
respect to the lower one, the vibrational temperatures are coupled to
the abundance of the molecular species under consideration. Hence,
an iterative procedure involving \tk{}, \tev{} (involving \tvib{}),
and the abundance are required to derive (also
fitting the observed 
ro-vibrational transitions) the populations of all the low- and intermediate-$J$ 
rotational levels within vibrational levels with low and moderate energy,
which are under LTE. So, as we will shows in \S\ref{sec:xc2h2}, the first
step in this procedure is to estimate the abundances over the whole envelope. The 
resulting \tvib{} for C$_2$H$_2$ are shown in Table~\ref{tab:tvc2h2},
having been defined \tvib{} with reference
to the lower vibrational state
(e.g., $T_\subscript{vib}(\textnormal{G.S.})\equiv{}
T_\subscript{vib}[\nu_4(\delta_g)-\textnormal{G.S.}(\sigma_g^+)]$ for C$_2$H$_2$).
Figures~\ref{fig:popc2h2} and \ref{fig:tvc2h2} show 
\tev{} and the derived population, $P_i$, of the observed vibrational levels, 
respectively (see the caption of Figure~\ref{fig:popc2h2} for the definitions 
of $P_i$).

\begin{deluxetable}{c@{}c@{}c@{}c@{}c@{}c@{}c@{}c}
\tabletypesize{\footnotesize}
\tablecolumns{8}
\tablewidth{0pt}
\tablecaption{Vibrational temperatures of C$_2$H$_2$\label{tab:tvc2h2}}
\tablehead{\colhead{\parbox{0.75cm}{\centering Vib.\\Level}} & \colhead{$P_i$} & \colhead{$T_1$ (K)} & \colhead{$\alpha_1$} & 
\colhead{$T_2$ (K)} & \colhead{$\alpha_2$} & \colhead{$T_3$ (K)} & \colhead{$\alpha_3$}}
\startdata
G.S.                & $P_0$      & 2330 & 0.58 & 900 & 0.19 & 685 & 1.00 \\
$\nu_4(\pi_g)$      & $P_1$      & 2330 & 1.66 & 150 & 0.43 & 82  & 1.00 \\
$\nu_5(\pi_u)$       & $P_2$     & 2330 & 0.58 & 900 & 0.26 & 625 & 1.00 \\
$2\nu_4(\sigma^+_g)$  & $P_3$     & 2330 & 0.58 & 900 & 0.58 & 400 & 1.00 \\
$2\nu_4(\delta_g)$     & $P_4$    & 2330 & 0.58 & 900 & 0.58 & 400 & 1.00 \\
$\nu_4+\nu_5(\sigma^+_u)$ & $P_5$ & 2330 & 0.58 & 900 & 0.58 & 400 & 1.00 \\
$\nu_4+\nu_5(\sigma^-_u)$ & $P_6$ & 2330 & 0.58 & 900 & 0.58 & 400 & 1.00 \\
$\nu_4+\nu_5(\delta_u)$   & $P_7$ & 500 &  0.98 & 100 & 0.16 & 80 & 1.00 \\
$2\nu_5(\sigma^+_g)$      & $P_8$ & 2330 & 0.58 & 900 & 1.40 & 125 & 1.00 \\
\tk                      & --- & 2330 & 0.58 & 900 & 0.58 & 400 & 1.00
\enddata
\tablecomments{Vibrational temperatures of C$_2$H$_2$.
$T_1$ is the vibrational temperature very near the stellar surface, $T_2$ is at 
the inner dust formation shell ($r=5.2$~R$_*$), and $T_3$ is at the outer
dust formation shell ($r=21.2$~R$_*$). $\alpha_1$, $\alpha_2$, and $\alpha_3$
are the exponents of the temperature power law ($r^{-\alpha}$) in Regions I, II,
and III, respectively. Note that the vibrational temperature of level
$\nu_4+\nu_5(\delta_u)$ is very low even close to the stellar surface.
\tk{} has been included in the last row of the Table to allow quick 
comparisons. The vibrational
temperatures refer to the lower vibrational level.}
\end{deluxetable}

The analysis of the derived vibrational temperatures and populations
(compared with those under LTE) indicates that:
\begin{itemize}
\item In the innermost envelope (Region I), the population of the ground
state is smaller than for levels $\nu_4$ and $\nu_5$ because
they are doubly degenerate ($e$ and $f$ parities) while the ground state is an $e$ 
vibrational level. \tk{} and the strong radiation field near the star are able 
to pump many molecules to high energy levels (see Figure~\ref{fig:popc2h2}; 
2$\nu_5(\delta_g)$ for example is at $\simeq 2100$~K).
\item $\nu_4(\pi_g)$ is a metastable vibrational level 
(see Figure~\ref{fig:vlc2h2}). It is not radiatively
connected to the ground state through electric dipole transitions.
Collisions and/or radiative cascades are its main pumping mechanisms.
\item Almost all vibrational levels, except $\nu_5(\pi_u)$
and $2\nu_5(\sigma_g^+)$, can be considered to be in LTE in Region I. $\nu_5(\pi_u)$ is out 
of LTE because of the strength of the radiative connection to the ground state 
(the dipole moment of the transition is $\mu\simeq 0.31$~D, \citealt{jacquemart_2001}).
The population of $2\nu_5(\sigma_g^+)$ departs from LTE over the whole CSE,
even near the stellar surface ($P_8/P_7<\left. P_8/P_7\right|_\subscript{LTE}$),
because it is connected radiatively with $\nu_5(\pi_u)$ (the dipole moment 
of the transition $2\nu_5(\sigma_g^+)\leftrightarrow\nu_5(\pi_u)$
is similar to that of $\nu_5(\pi_u)\leftrightarrow{}$G.S.$(\sigma_g^+)$).
\item The radial dependence of the population of all the vibrational levels 
changes in Region II with respect to Region I.
The slope of $P_0$ and $P_1$ falls while it increases for the other levels. 
The absorption of 3~$\mu$m radiation, which pumps C$_2$H$_2$
from the ground to high energy levels, changes the slope of
the population of the ground state with respect to that under LTE throughout
Regions I and II.
This radiation also affects the $\nu_4(\pi_g)$ level
which starts to be efficiently pumped from the radiative cascade from the stretching modes. 
The ratio $P_1/P_0$ is larger in Regions II and III than under LTE. However, as
the transition $\nu_5(\pi_u)\leftrightarrow{}$G.S.$(\sigma_g^+)$
is strong, the ratio $P_2/P_1$ is lower than expected under LTE.
All the vibrational levels above $\nu_5$ are underpopulated in Region II.
Collisions are less efficient as the volume density
in this region is low and all vibrational pumping is radiatively
dominated.
\item $2\nu_4(\sigma_g^+)$ and $2\nu_4(\delta_g)$ are metastable states
like $\nu_4(\pi_g)$. On the other hand, $\nu_5(\pi_u)$ is
strongly radiatively connected to the ground state.
Consequently, $P_3/P_2$ is larger than
it would be under LTE for $r>$R$_{\subscript{d}1}$.
\item $P_6/P_5$ is larger in Region II than expected under LTE. 
The populations of the vibrational levels $\nu_4+\nu_5(\sigma_u^+)$, 
$\nu_4+\nu_5(\sigma_u^-)$, and $\nu_4+\nu_5(\delta_u)$ are similarly increased
due to radiative cascades from higher energy levels. However, the
strong radiative transition $\nu_4+\nu_5(\sigma_u^+)\leftrightarrow{}$G.S.$(\sigma_g^+)$
\citep*[][p.~290]{herzberg_ii} increases the ratio $P_6/P_5$ due to stimulated de-excitation 
of molecules from $\nu_4+\nu_5(\sigma_u^+)$ by 7.5~$\mu$m radiation.
This mechanism is further strengthened due to the forbidden 
transition $\nu_4+\nu_5(\sigma_u^-)
\leftrightarrow{}$G.S.$(\sigma_g^+)$.
\item $P_4/P_3$, $P_5/P_4$ and $P_7/P_6$ are consistent with LTE
as these levels are separated by an energy equivalent to
4.5, 136.1 and 10.9~K, respectively.
As levels $\nu_4+\nu_5(\sigma_u^-)$ and $\nu_4+\nu_5(\delta_u)$ are not radiatively 
connected to the ground state, the collisions keep $P_7\simeq 2P_6$ (see 
Figure~\ref{fig:popc2h2}). There may be a slight departure from LTE for the 
ratio $P_5/P_4$ in Region III ($T_\subscript{K}(100~\textnormal{R}_*)\simeq 85$~K), 
though it cannot be derived from the observational data.
\item For distances larger than 20~R$_*$ (Region III), only the lower
energy levels are significantly populated. The most intense transition
in the outer envelope is $\nu_5(\pi_u)\leftrightarrow{}$G.S.$(\sigma_g^+)$. 
Further away from the star, it is not possible to find any emission from
C$_2$H$_2$.
\item As for Region II, $P_1/P_0$ in Region III is larger than it would be under LTE, 
but it decreases rapidly, as in the LTE case. This comes about from absorption of 
mid-infrared radiation in Region III. This radiation excites molecules from the state 
$\nu_4$ towards $\nu_4+\nu_5$ and, then, the molecule de-excites to the ground state.
The 3~$\mu$m pumping mechanism, important in Region II, loses its effectiveness
at distances larger than \rdout{} due to the decreasing intensity of the near-IR 
radiation. In addition, since the gas density and \tk{} are low, collisions are  
ineffective in de-exciting molecules from level $\nu_4(\pi_g)$. Furthermore,
no molecules are in that state in the outer CSE.
\item $P_9/P_8$ is consistent with LTE in Region I but shows a 
departure from LTE in Regions II and III. $2\nu_5(\sigma_g^+)$
is radiatively connected to the ground state
through $2\nu_5(\sigma_g^+)\leftrightarrow\nu_2+\nu_4+\nu_5(\sigma_u^+)
\leftrightarrow{}$G.S.$(\sigma_g^+)$ and 
$2\nu_5(\sigma_g^+)\leftrightarrow\nu_5(\pi_u)\leftrightarrow{}$G.S.$(\sigma_g^+)$.
On the contrary, $2\nu_5(\delta_g)$ is connected to the ground state
via the transition 
$2\nu_5(\delta_g)\leftrightarrow\nu_5(\pi_u)\leftrightarrow{}$G.S.$(\sigma_g^+)$
but not through $2\nu_5(\delta_g)\leftrightarrow\nu_2+\nu_4+\nu_5(\delta_u)
\leftrightarrow{}$G.S.$(\sigma_g^+)$ because
$\nu_2+\nu_4+\nu_5(\delta_u)\leftrightarrow{}$G.S.$(\sigma_g^+)$ is forbidden.
Therefore, when collisional pumping loses its effectiveness ($r>$ \rdin), $P_9/P_8$ 
becomes lower than under LTE.
\end{itemize}

The estimation of the populations in the $P_3$ and $P_4$ levels
are subject to large uncertainties because the observed lines arising from 
these levels are weak. We believe that even for $P_3$ and $P_4$ the obtained 
vibrational temperatures are good estimators (see Figure~\ref{fig:popc2h2}).
The rest of the vibrational populations are well determined based on the
observational data.

\subsection{Abundances}
\label{sec:xc2h2}

The condensation of refractory molecules on dust grains over the two
acceleration zones modifies the abundances of several molecular 
species. It is unclear to what degree C$_2$H$_2$ (and isotopologues) 
condenses onto dust
grains but our data provide a tool to check whether or not
significant changes occur from one region to another.

As we will show (see \S\ref{sec:sens.disc}), the 
abundances
in Region I has large uncertainties due to their small influence
on most observed lines compared with the rest of the CSE. The angular 
size of the innermost region is small ($\simeq 0\arcsecond2$) and, 
despite the high values of density and \tk{}, the observed flux is small 
with respect to that coming from outer shells of the envelope. 
On the other hand, the efficiency of a PSF $\simeq 1''$ 
is significant for the angular size of Region II ($\simeq 0\arcsecond8$).
Loss of light at the spectrograph entrance slit, which is 
$\simeq$ 1\arcsecond6 wide, is still relatively unimportant 
for Region II but is a large fraction of the 
radiation emitted in
Region III. In addition, \tk{} and the density are still high in Region II
(the mean \tk{} and $n$ for C$_2$H$_2$ are $\simeq 560$~K and 
220~cm$^{-3}$, respectively). Therefore, most of the emission
for optically thick lines (extending farther out than \rdin{}) and the 
bulk absorption for thin ones (produced in Region I)
are due to Region II. In Region III, \tk{} and the density are low so emission
from this region is anticipated to be small.  Furthermore, Region
III suffers the most from light loss at the slit, which pertains
only to the emission component of the thick line profiles. Consequently
Region III need only be considered for the very thick lines
(e.g. ro-vibrational lines of the vibrational transition 
$\nu_5(\pi_u)\leftrightarrow{}$G.S.$(\sigma_g^+)$), for which the 
bulk absorption is produced by this region.
This implies that the maximum information on the C$_2$H$_2$ abundance in 
Region III can be obtained from absorption in the P-Cygni profiles
of these thick lines. Unfortunately, their thickness precludes the possibility 
of getting accurate information on the abundances of C$_2$H$_2$ at angles 
larger than $1''$ ($r > 50$~R$_*$). At larger radii, the absorption component 
of the lines is almost completely saturated. At distances larger than 300~R$_*$, 
there is no emission from C$_2$H$_2$ because all the molecules remain in the 
ground state (see Figure~\ref{fig:popc2h2}) and pure rotational transitions 
are forbidden due to the lack of electric dipole moment.

The C$_2$H$_2$ abundances derived from the best fits for each region are:
\begin{equation}
x_{\subscript{C}_2\subscript{H}_2}(r) = \left\{
\begin{array}{ll}
7.5\times 10^{-6} & \textnormal{Region I}\\
8.0\times 10^{-5} & \textnormal{Region II}\\
8.0\times 10^{-5} & \textnormal{Region III}
\end{array}
\right.
\end{equation}
with an assumed initial value of $8\times 10^{-5}$ for the CSE.
The derived column density is $\simeq 1.6\times 10^{19}$~cm$^{-2}$.

Note that the abundance in Region I is an order of magnitude lower
than in the rest of the CSE.
A higher abundance in that region implies diminishing its size, 
incompatible with previous results (see \S\ref{sec:continuum}). 
We will discuss this topic further in \S\ref{sec:sens.disc}.
The abundances of C$_2$H$_2$ in Regions II and III  are similar to those
obtained in previous studies 
$8\times 10^{-5}$ and $5\times 10^{-5}$ from \citet{keady_1993} and 
\citet{cernicharo_1999} respectively.

In addition, the increment in the C$_2$H$_2$ abundance between Regions I and II
is compatible with that yielded by LTE calculations (M. Ag\'undez, private
communication) and with the results of the chemistry in 
a pulsating star published by \citet{cherchneff_2006}.
The variation in the LTE abundance profile between both Regions
ranges from 15\% to 99\% (with respect to the abundance in Region II), depending on
the physical conditions of the very inner envelope \citep{agundez_2006}.
The action of shocks on chemical abundances at radii smaller than 5~R$_*$
produces abundance profiles which increase in 99\% the abundance in Region II
\citep{cherchneff_2006}. Both models could explain the increment of 90\%
derived in this work. We will also discuss this topic in \S\ref{sec:sens.disc}.

The uncertainties on the abundances (see Table~\ref{tab:errors}) are 
compatible with a small condensation of C$_2$H$_2$ (and isotopologues) 
on dust grains and the 
consequent existence of an acceleration regime between Regions II and III.
Unfortunately, we cannot assure the existence of these processes with the
data derived.

\section{H$^{13}$CCH \& $^{13}$C$_2$H$_2$ Modeling}
\label{sec:h13cch}

As in the case of C$_2$H$_2$, we can derive the abundance, vibrational
temperatures and constrain the parameters related to the physical 
conditions of the CSE, through fitting as many H$^{13}$CCH lines as possible.
The fitting procedure is the same as for C$_2$H$_2$ (see \S\ref{sec:acetylene}).
We have used the physical conditions determined previously with C$_2$H$_2$ 
as initial inputs for the fits of the H$^{13}$CCH lines.

\subsection{Abundances}
\label{sec:xh13cch}

The [C$_2$H$_2$]/[H$^{13}$CCH] ratio is easily determined taking into account 
that the abundance of H$^{13}$CCH must be proportional to that of C$_2$H$_2$:
\begin{equation}
\frac{\textnormal{[C$_2$H$_2$]}}{\textnormal{[H$^{13}$CCH]}} \simeq 
\frac{1}{0.049} \simeq 20.5
\end{equation}
where the fitted lines are shown in Table~\ref{tab:lines}.
This value is very similar to [C$_2$H$_2$]/[H$^{13}$CCH]$\simeq 22$, proposed by 
\citet{cernicharo_1996b,cernicharo_1999,cernicharo_2000}. Hence, the abundance 
ratio [$^{12}$C]/[$^{13}$C] derived from our data is $\simeq 41$ in IRC+10216.

\subsection{Vibrational Temperatures}
\label{sec:tvh13cch}

As shown in Appendix~\ref{sec:LineFrequencies} and \ref{sec:LineIntensities},
the spectrum of H$^{13}$CCH presents some differences with respect to
C$_2$H$_2$ (e.g., $\nu_4(\pi)-$G.S.$(\sigma^+)$ is IR active for H$^{13}$CCH while it
remains forbidden for C$_2$H$_2$). Consequently, although
H$^{13}$CCH is very similar to C$_2$H$_2$, its vibrational temperatures
are different (see Table~\ref{tab:tvh13cch} and 
Figure~\ref{fig:tvh13cch}).
The observed H$^{13}$CCH transitions also support the near-IR pumping mechanism 
proposed in \S\ref{sec:vibc2h2}. For H$^{13}$CCH, more transitions are 
allowed than for C$_2$H$_2$ due to combined levels involving both
stretching and 
bending modes. These differences can be understood in terms of the
radiative pumping paths for both isotopologues (see below).
\begin{itemize}
\item In Region I, the population of $2\nu_5(\sigma^+)$ is lower than in LTE 
compared with that of $\nu_4+\nu_5(\delta)$. Radiative cascades from higher energy 
levels populate these states in a different manner than in the C$_2$H$_2$ case 
(e.g., $\nu_3+\nu_5\leftrightarrow{}2\nu_5$ is allowed for H$^{13}$CCH), 
keeping level $2\nu_5(\sigma^+)$ almost in LTE next to the photosphere 
(see \S\ref{sec:vibc2h2}). Therefore, all the vibrational levels 
can be considered in LTE close to the star (see 
Figure~\ref{fig:tvh13cch}).
\item The allowed transitions for H$^{13}$CCH connecting $2\nu_5$ to lower 
energy levels and the increase of the population of $\nu_5$ through 
$\nu_4+\nu_5\leftrightarrow\nu_5$, result in a decrease of 
$T_{\subscript{ev}}[2\nu_5\leftrightarrow\nu_5]$ in Regions II and III, 
with respect to the C$_2$H$_2$ case. 
\item In Regions II and III, $P_2/P_1$, $P_3/P_2$, and $P_8/P_7$
are lower than in LTE, as for the C$_2$H$_2$ case. However, several 
radiative transitions in H$^{13}$CCH, which are forbidden for 
C$_2$H$_2$, keep these ratios closer to LTE.

\begin{deluxetable}{c@{}c@{}c@{}c@{}c@{}c@{}c@{}c}
\tabletypesize{\footnotesize}
\tablecolumns{8}
\tablewidth{0pt}
\tablecaption{Vibrational temperatures of H$^{13}$CCH
\label{tab:tvh13cch}}
\tablehead{\colhead{\parbox{0.75cm}{\centering Vib.\\Level}} & $P_i$ & \colhead{$T_1$ (K)} &
\colhead{$\alpha_1$} & \colhead{$T_2$ (K)} & \colhead{$\alpha_2$} &
\colhead{$T_3$ (K)} & \colhead{$\alpha_3$}}
\startdata
G.S.                     & $P_0$ & 2330 & 0.58 & 900 & 0.56 & 410 & 1.00 \\
$\nu_4(\pi)$             & $P_1$ & 2330 & 0.58 & 900 & 1.92 & 60 & 1.00 \\
$\nu_5(\pi)$             & $P_2$ & 2330 & 0.58 & 900 & 1.43 & 120 & 1.00 \\
$2\nu_4(\sigma^+)$       & $P_3$ & 2330 & 0.58 & 900 & 0.58 & 400 & 1.00 \\
$2\nu_4(\delta)$         & $P_4$ & 2330 & 0.58 & 900 & 0.58 & 400 & 1.00 \\
$\nu_4+\nu_5(\sigma^+)$  & $P_5$ & 2330 & 0.58 & 900 & 0.58 & 400 & 1.00 \\
$\nu_4+\nu_5(\sigma^-)$  & $P_6$ & 2330 & 0.58 & 900 & 0.58 & 400 & 1.00  \\
$\nu_4+\nu_5(\delta)$    & $P_7$ & 2330 & 0.93 & 500 & 2.78 & 10 & 1.00 \\
$2\nu_5(\sigma^+)$       & $P_8$ & 2330 & 0.58 & 900 & 0.58 & 400 & 1.00 \\
\tk                      & --- & 2330 & 0.58 & 900 & 0.58 & 400 & 1.00
\enddata
\tablecomments{Vibrational temperatures of H$^{13}$CCH. See 
Table~\ref{tab:tvc2h2} for details about the meaning
of the constants.}
\end{deluxetable}

\item The $\nu_4(\pi_g)$ vibrational level in C$_2$H$_2$ is only collisionally 
connected to the ground state, while it is connected both collisionally and 
radiatively in H$^{13}$CCH. This fact reduces the ratio $P_1/P_0$ and thermalizes 
the $\nu_4$ state over almost the whole CSE.
In addition, $P_2/P_1$ is in LTE in Region I, where collisions can 
thermalize the $\nu_5$ state, but it is out of LTE in Regions II and III 
once \tk{} and the density have decreased and radiative excitation has 
become the main pumping mechanism.
As the transition $\nu_4\leftrightarrow{}$G.S. is weaker than
$\nu_5\leftrightarrow{}$G.S. \citep{dilonardo_2002}, we can conclude that
pumping of molecules from the ground state to $\nu_4$ and $\nu_5$ through radiative
cascades from high-energy vibrational levels seems to be almost the same
(see Figure~\ref{fig:vlc2h2}).
\item $P_3/P_2$ is lower than in LTE in Regions II and III for
several reasons: $2\nu_4(\sigma^+)$ is radiatively connected to 
$\nu_4(\pi)$ and G.S.$(\sigma^+)$; molecules are pumped from the 
ground state to $\nu_5(\pi)$ through radiative cascades, and collisional pumping 
is not effective enough to keep the vibrational populations under LTE in
these regions.
\item As for C$_2$H$_2$, the model points out that $P_4/P_3$, $P_5/P_4$, and $P_7/P_6$
can be considered to be in LTE.

\begin{figure}[!hbt]
\includegraphics[angle=-90,width=0.475\textwidth]{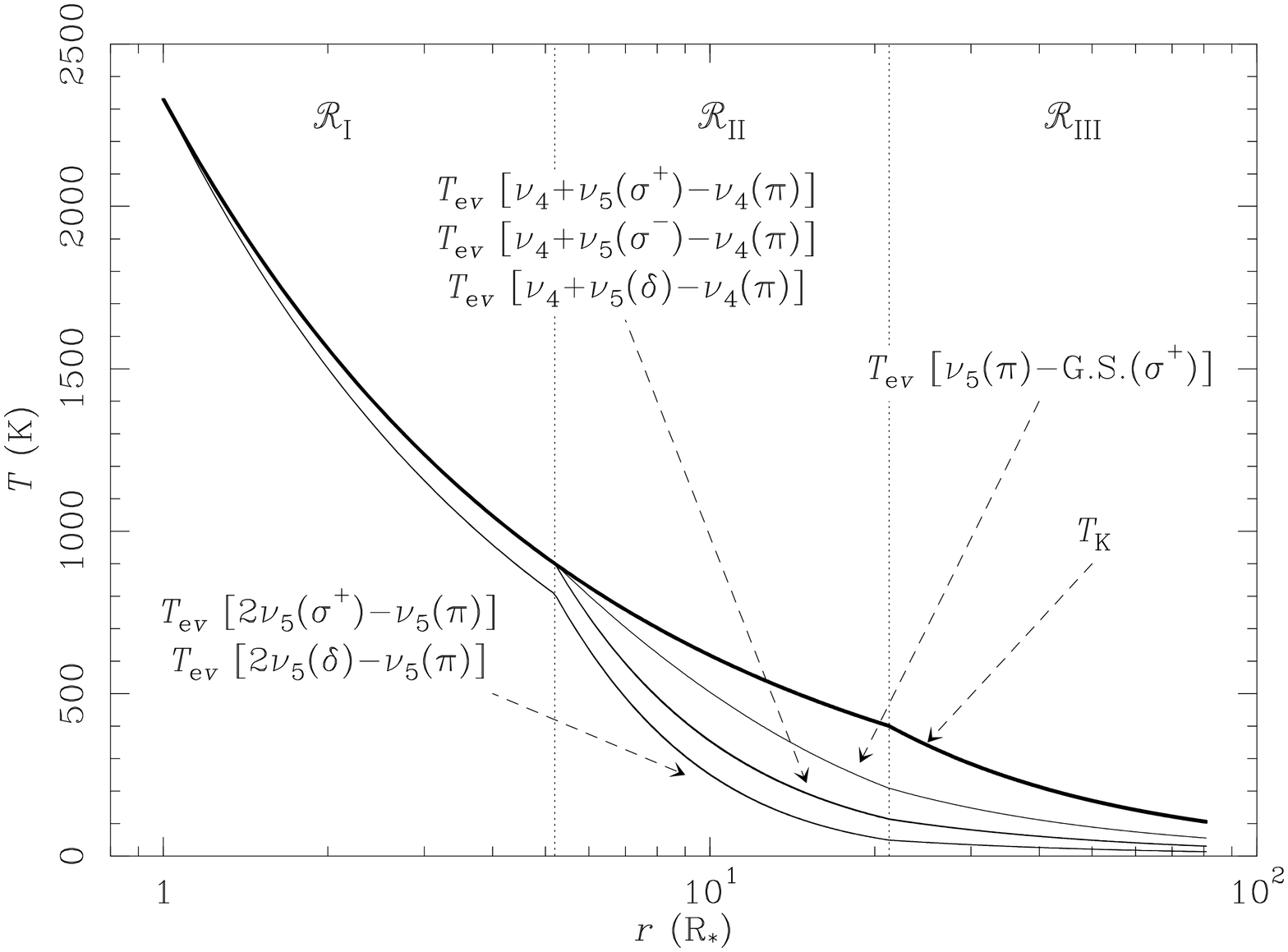}
\caption{Kinetic and vibrational excitation temperatures for the H$^{13}$CCH
observed transitions. As for C$_2$H$_2$, \tev{} are merged
into two different groups, although in this case, the transition
$\nu_5(\pi)\leftrightarrow{}$G.S.$(\sigma^+)$ stands alone with a \tev{}
not much lower than \tk{}. All the selected transitions are under LTE in  
Region I.}
\label{fig:tvh13cch}
\end{figure}

\item $T_{\subscript{ev}}[\nu_5(\pi)\leftrightarrow{}$G.S.$(\sigma^+)]<T_\subscript{K}$ 
because the transition is strong.
On the other hand, the transition $\nu_4+\nu_5\leftrightarrow\nu_5$ could depopulate
$\nu_4+\nu_5$, which is compatible with 
$T_{\subscript{ev}}[\nu_4+\nu_5\leftrightarrow\nu_4]<T_{\subscript{ev}}[\nu_5\leftrightarrow{}$G.S.], 
in contrast to the behavior of these temperatures for C$_2$H$_2$ where
they are equal.
\end{itemize}
\newpage
\subsection{Rotational Temperatures}
\label{sec:trh13cch}

Unlike C$_2$H$_2$, all H$^{13}$CCH chosen lines can be fitted quite well with the rotational 
levels in LTE. No \textit{ad-hoc} rotational temperatures are needed.

\subsection{Searching for $^{13}$C$_2$H$_2$}
\label{sec:13c2h2}

Given the large abundance of C$_2$H$_2$, 
detection of $^{13}$C$_2$H$_2$ ro-vibrational 
lines may be possible. The double isotopic substitution of the C-atoms greatly 
reduces the abundance of this species compared with C$_2$H$_2$.
As we have derived in \S\ref{sec:h13cch}, $[^{12}$C]/$[^{13}$C$]\simeq 41$. 
Consequently, $[^{13}$C$_2$H$_2$]/[C$_2$H$_2]\simeq (1/41)^2\simeq 6\times 10^{-4}$.
Unfortunately, the strongest bands of $^{13}$C$_2$H$_2$ are in a
range of the spectrum crowded with strong C$_2$H$_2$, H$^{13}$CCH,
and HCN transitions (see Figure~\ref{fig:f1}). 
We have compared the optical depth of one of the strongest $^{13}$C$_2$H$_2$ lines,
$\nu_5(\pi_u)$R$_e(3)$, with the same line for C$_2$H$_2$. The dipole moment
employed in the calculations for $^{13}$C$_2$H$_2$ has been 
set equal to that of C$_2$H$_2$ due to the lack of data on intensities 
quoted in the literature.
The optical depth of this line has been computed by running the code with
the parameters derived from the fits to the observed C$_2$H$_2$ lines
and using the double-substitution isotopic ratio given above.
The derived optical depths along the line of sight, without considering the
Doppler effect, are 630 for C$_2$H$_2$ and 0.033 for $^{13}$C$_2$H$_2$.
The low optical depth for the $^{13}$C$_2$H$_2$ line results in a maximum
absorption $<1\%$ of the continuum flux and cannot be observed 
in our data. An equal or smaller
absorption is expected for other lines of the
same band.

\section{HCN and H$^{13}$CN, Modeling}
\label{sec:hcn}

\begin{figure*}[!htb]
\centering
\includegraphics[scale=0.8,angle=-90]{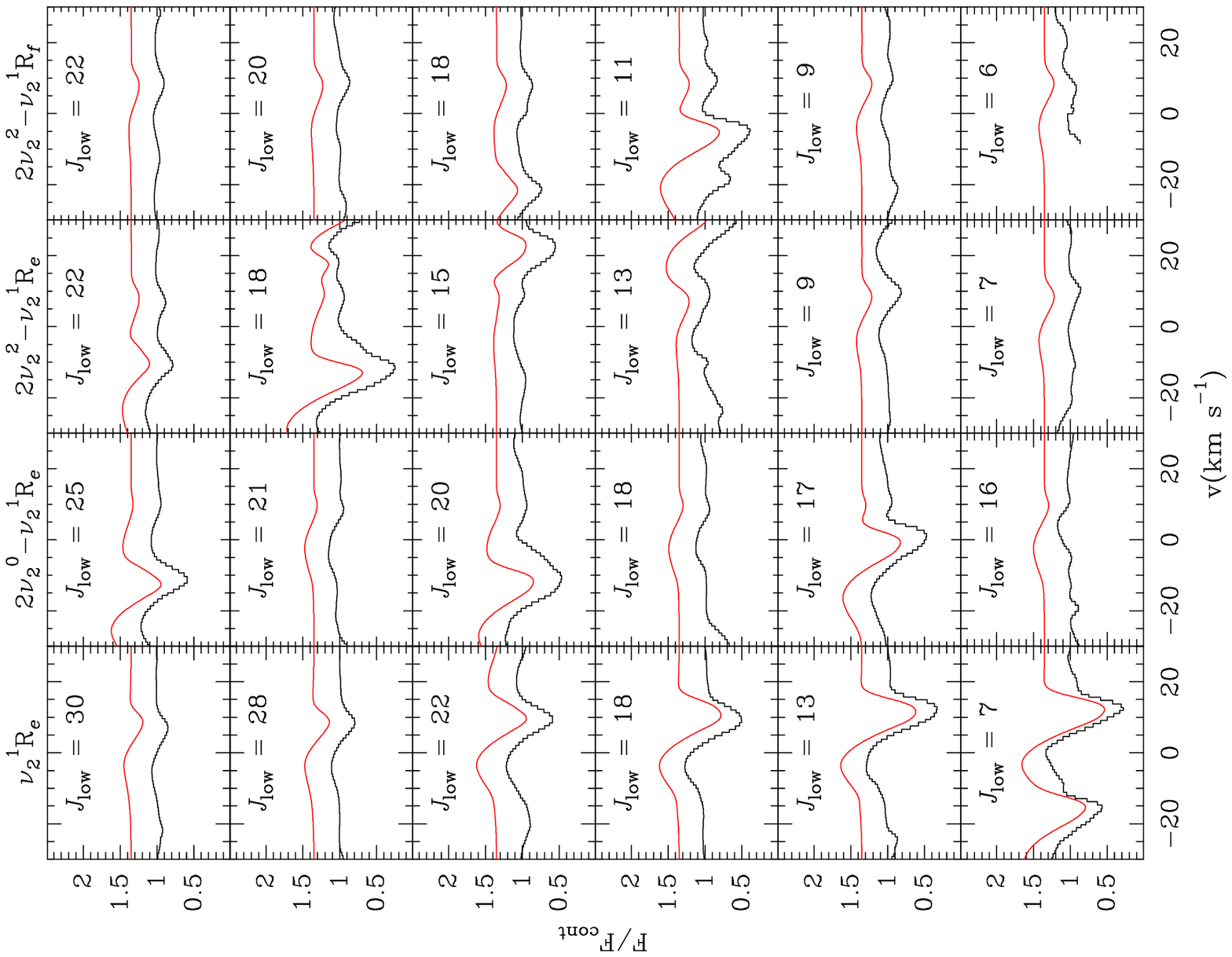}
\caption{Fits of some lines of HCN. The black lines are the observed
spectra while the red lines are the fits. The data have been corrected from
observational frequency deviations adding/suppresing the mean difference
between the observational and the laboratory frequency of many lines (see
\S\ref{sec:obs}).
The existing discrepancies are small and can be assumed as blends with other 
lines and/or observational errors (the spectroscopic notation 
is described in Appendix~\ref{sec:LineFrequencies}).}
\label{fig:fitshcn}
\end{figure*}

Modeling HCN (and H$^{13}$CN) lines adds additional constraints to
the physical conditions over the CSE for $r<100$~R$_*$. The spectra
shown in Figures~\ref{fig:f1}$-$\ref{fig:f6}
indicate the presence of many ro-vibrational lines
from the $\nu_2$ bending mode and several associated overtones.
As for C$_2$H$_2$ and H$^{13}$CCH, 
we have selected a sample of HCN (and H$^{13}$CN) lines to 
fit to the data and to derive physical and chemical conditions through
the CSE (see some fits in Figure~\ref{fig:fitshcn}). 
The observation of transitions arising from 
overtones and H$^{13}$CN allows us to study optically thinner lines 
compared to those of the fundamental bending mode.

\notetoeditor{We would like this figure to appear in the paper with the same
scale than Figure 8, which contains some observed features and their fits.
As in the case of Figure 8, we would like that ``RED lines'' appears in the on-line
paper and ``GRAY lines'' in the black-and-white one.}

The differences between H$^{13}$CN and the main isotopologue, HCN, 
lead to a significant increment in the molecular 
mass and a slight variation in the 
interatomic distances. Consequently, the wavefunctions
of each molecular state and their related energies are not equal for HCN and H$^{13}$CN.
Although some spectroscopic work on H$^{13}$CN can be found in the literature,
there is no data available about the intensities of its transitions,
while HCN has been studied more extensively.
Therefore, we have used the dipole moments of HCN for H$^{13}$CN as a reasonable
approximation for the transition strengths.

The number of lines used for the fits is shown in Table~\ref{tab:lines}.
Information about frequencies, intensities, and nomenclature for HCN and 
its isotopologues can be found in Appendixes~\ref{sec:LineFrequencies} and 
\ref{sec:LineIntensities}, and references therein.

\subsection{Vibrational Temperatures}
\label{sec:tvhcn}

\begin{deluxetable}{c@{}c@{}c@{}c@{}c@{}c@{}c@{}c}
\tabletypesize{\footnotesize}
\tablecolumns{8}
\tablewidth{0pt}
\tablecaption{Vibrational temperatures of HCN and H$^{13}$CN\label{tab:tvhcn}}
\tablehead{\colhead{\parbox{0.75cm}{\centering Vib.\\Level}} & $P_i$ & \colhead{$T_1$ (K)} &
\colhead{$\alpha_1$} & \colhead{$T_2$ (K)} & \colhead{$\alpha_2$} &
\colhead{$T_3$ (K)} & \colhead{$\alpha_3$}}
\startdata
G.S.               & $P_0$ & 2330 & 0.58 & 900 & 1.56 & 130 & 1.00 \\
$\nu_2(\pi)$       & $P_1$ & 1250 & 0.82 & 600 & 0.62 & 250 & 1.00 \\
$2\nu_2(\sigma^+)$ & $P_2$ &  800 & 1.26 & 100 & 1.24  & 17.5 & 1.00 \\
\tk                & --- & 2330 & 0.58 & 900 & 0.58 & 400 & 1.00
\enddata
\tablecomments{Vibrational temperatures of HCN and H$^{13}$CN. See 
Table~\ref{tab:tvc2h2} for details about the meaning
of the constants. The parameters of the vibrational temperatures of H$^{13}$CN 
are only those related to the ground state.}
\end{deluxetable}

As seen in previous sections, C$_2$H$_2$ does not present any permanent dipole moment 
(it is very small for H$^{13}$CCH and can be neglected to a first approximation).
Considering
rotational LTE as suitable initial input, 
there will not be any
pure rotational emission.
Fortunately, in spite of the permanent dipole moment, HCN (and H$^{13}$CN) seems to be
under LTE for low-$J$ rotational levels, making easier the determination
of the abundance and \tvib{} (see \S\ref{sec:trhcn}).

\begin{figure}[!htb]
\centering
\includegraphics[height=0.475\textheight]{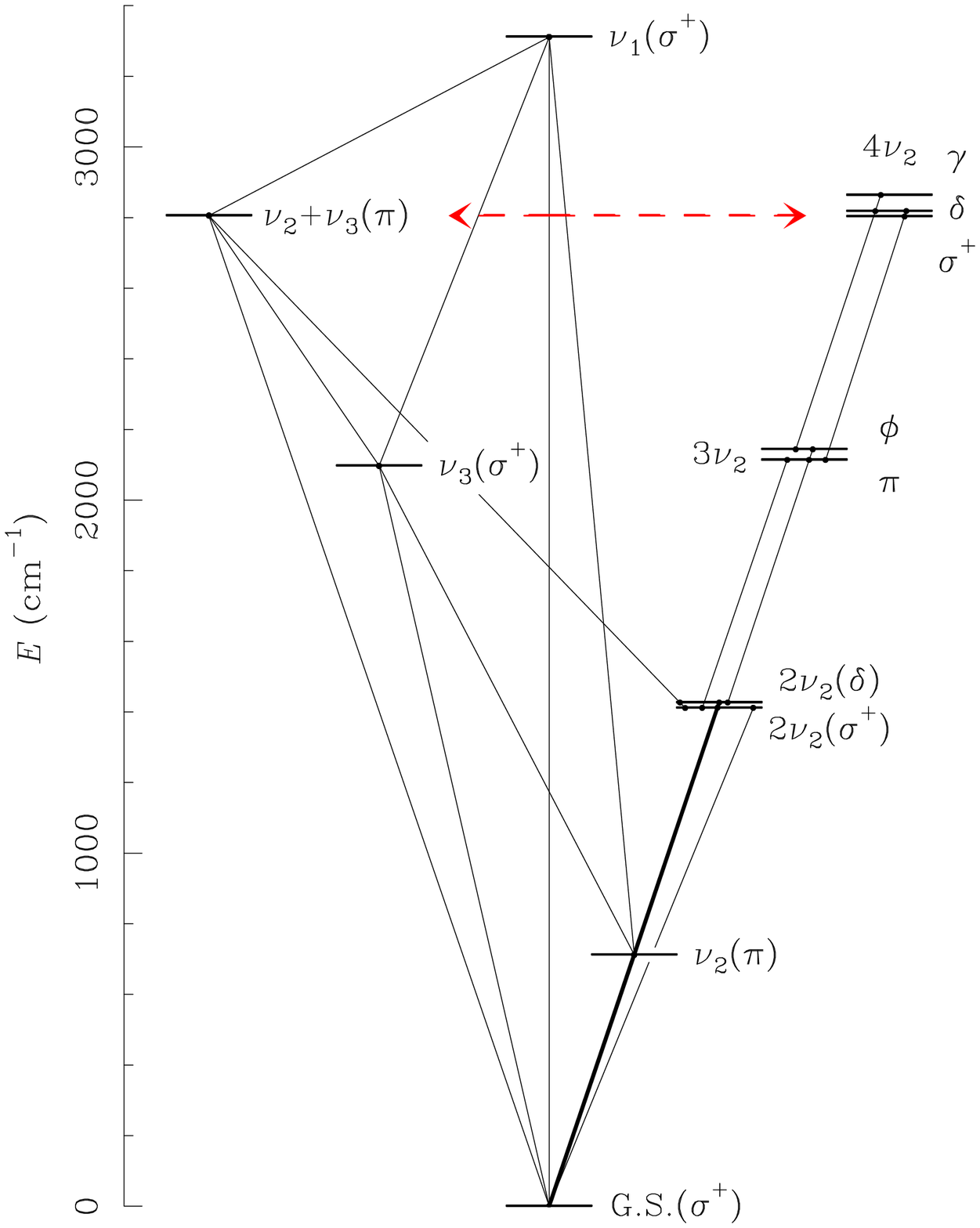}
\caption{Energy of the vibrational levels of HCN. This pattern is the
same for the H$^{13}$CN isotopologue. The transitions given
are the perpendicular ones with $\Delta v_2=\pm 1$ and the most intense 
parallel ones. The thin lines are allowed transitions. The 
thick lines are observed vibrational transitions. The red dashed
double-arrow indicates the resonance between several rotational levels
of the vibrational states
$\nu_2+\nu_3(\pi)$ and $4\nu_2(\sigma^+)$.}
\label{fig:tvhcn}
\end{figure}

\notetoeditor{The figure f15.eps is colored. We would like it to be included
only in the on-line version of the paper.}

Near \rdin{}, $T_\subscript{d}\simeq 800-850$~K and the dust black-body 
emission peaks at $\simeq 3.5~\mu$m ($\simeq 2850$~cm$^{-1}$).
This radiation pumps HCN from the ground state to the stretching 
and combination levels $\nu_3(\sigma^+)$, $\nu_2+\nu_3(\pi)$, and 
$\nu_1(\sigma^+)$, with energies 2096.85, 2807.05, and 3311.48~cm$^{-1}$
respectively (see Figure~\ref{fig:tvhcn}).
In addition, several rotational levels of 4$\nu_2(\sigma^+)$ (with energy 
2802.96~cm$^{-1}$) are strongly connected to the ground state through a 
resonance with the corresponding rotational levels of $\nu_2+\nu_3(\pi)$. 
This resonance could explain the HCN masers observed by \citet{schilke_2003} 
in the 4$\nu_2(\sigma^+)$ state. Radiative deexcitation changes the population 
of mode $\nu_2(\pi)$ and its overtones with respect to the LTE case.

\begin{figure}[!hbt]
\includegraphics[angle=-90,width=0.475\textwidth]{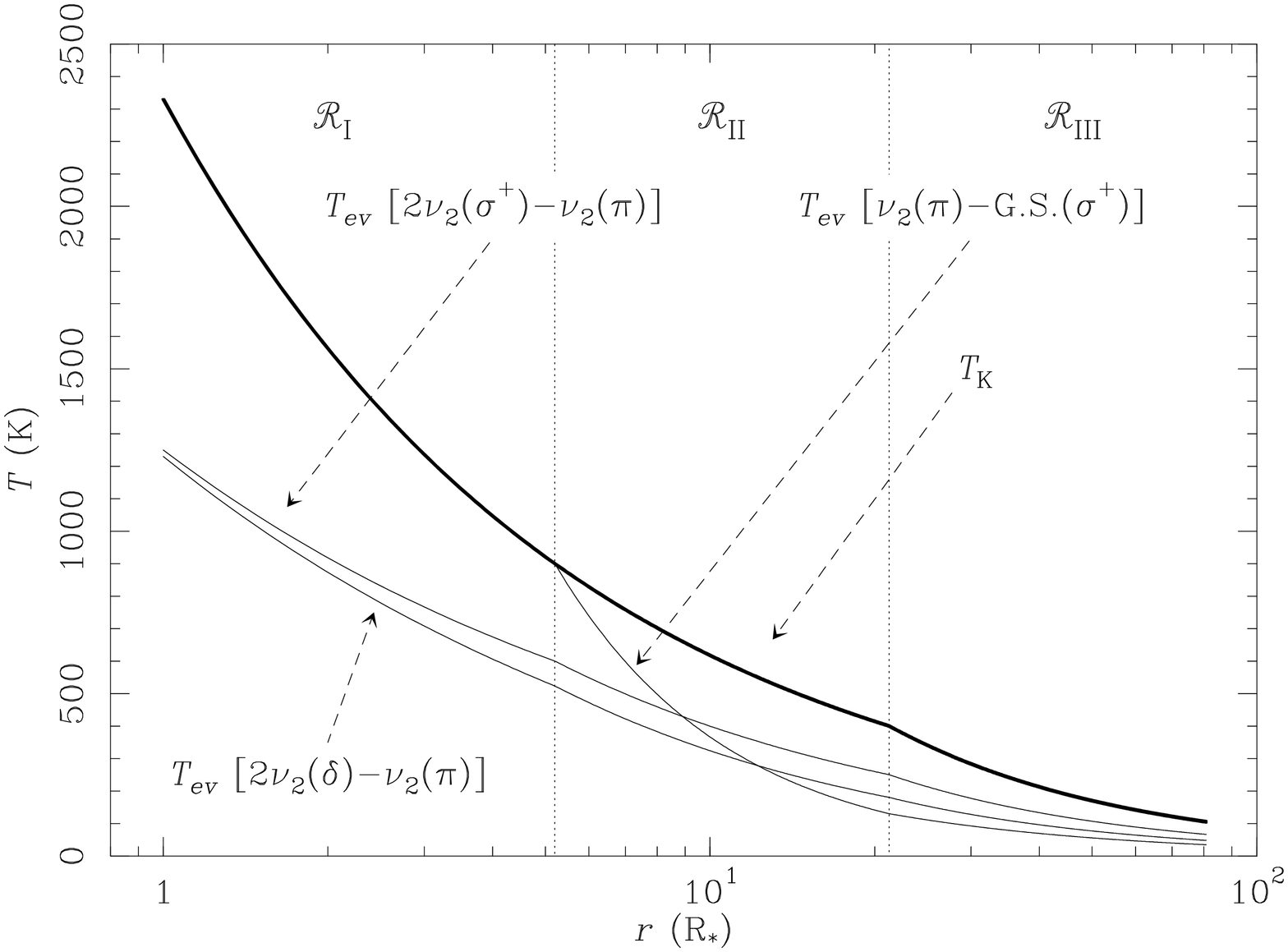}
\caption{Kinetic and vibrational excitation temperatures for 
HCN observed transitions. See text for details.}
\label{fig:tvhxcn}
\end{figure}

The vibrational temperatures derived from the fits are presented in
Table~\ref{tab:tvhcn} and plotted in
Figure~\ref{fig:tvhxcn}.
As in the case of C$_2$H$_2$, we have labeled the populations of the levels as
follows: $P_0\equiv P[$G.S.$(\sigma^+)]$, $P_1\equiv P[\nu_2(\pi)]$, 
$P_2\equiv P[2\nu_2(\sigma^+)]$, and $P_3\equiv P[2\nu_2(\delta)]$.
\begin{itemize}
\item $T_\subscript{vib}[\nu_2(\pi)\leftrightarrow{}$G.S.$(\sigma^+)]\simeq T_\subscript{K}$
in Region I, departing from \tk{} in Regions II and III.
In Region I, collisions thermalize $\nu_2(\pi)$, while at larger distances only
infrared photons play a role in the pumping.
Vibrational levels $\nu_1(\sigma^+)$ and $\nu_3(\sigma^+)$ are radiatively 
connected to the bending level $\nu_2(\pi)$. In addition, $\nu_2(\pi)$ can also be 
populated through the loop G.S.$(\sigma^+)\rightarrow\nu_2+\nu_3(\pi)\rightarrow\nu_3(\sigma^+)
\rightarrow\nu_2(\pi)$. The band $\nu_2(\pi)-$G.S.$(\sigma^+)$ is quite strong
and the deexcitation rate is high.
\item $2\nu_2(\sigma^+)$ and $2\nu_2(\delta)$ are
out of LTE over the whole CSE, as well as close to the stellar photosphere.
The dipole moment of the transitions involving these levels are so high that
they are not thermalized even in the dense warm innermost region.
Changing \tvib{} of these levels at the stellar surface from 1250~K to 2330~K 
produces a change of $20-30$\% in the emission of 
$2\nu_2(\sigma^+)\leftrightarrow\nu_2(\pi)$ ro-vibrational transitions.
Hence, our data is very sensitive to the value of \tvib{}.
$2\nu_2(\sigma^+)$ can be pumped from the ground state and from $\nu_2+\nu_3(\sigma^+)$.
However, the $2\nu_2(\delta)$ level is not connected to the ground state except 
through the radiative cascades from higher energy levels \citep*[see][]{cernicharo_1999}. 
Therefore, $P_3/2P_2$ is always lower than $2P_2/P_1$ (we add the factor 2 
to account for the difference in the degeneracy between levels
$2\nu_2(\sigma^+)$ and $2\nu_2(\delta)$, and $2\nu_2(\sigma^+)$ and $\nu_2(\pi)$).
\item $T_\subscript{vib}[\nu_2-$G.S.$]$ and $T_\subscript{vib}[2\nu_2-\nu_2]$ are equal 
to each other at $\simeq 10$~R$_*$ (see 
Figure~\ref{fig:tvhxcn}).
Outside of the region where the NIR radiation emitted by the dust arises, the pumping via 
higher energy levels from the ground state (through 7~$\mu$m radiation, which is 
the wavelength corresponding to the maximum emission of a black-body at
$T_\subscript{bb}\simeq 400$~K) becomes less important, meaning that
the population of level $2\nu_2(\sigma^+)$ increases with respect to $\nu_2(\pi)$.
\end{itemize}

\subsection{Rotational Temperatures}
\label{sec:trhcn}

The rotational structure of the ground state seems to be in LTE 
for rotational levels up to $J=20$.
It is necessary to adopt \textit{ad-hoc} rotational temperatures
for rotational levels $J=$~20 and 24, when fitting the 
$\nu_2$R$_e$ lines
(see Table~\ref{tab:adhocrottemp}). 
The line $\nu_2(\pi)-$G.S.$(\sigma^+)$R$_e(21)$ can be fitted more
accurately with an \textit{ad-hoc} \trot{} at $r=$~\rdout{} 
for $J=20$ (although 
the fit is relatively good even under LTE). 
However, this \textit{ad-hoc}
temperature is necessary to fit higher-$J$ ro-vibrational lines.
$J=24$ must be out of LTE. Under LTE, the absorption
and the emission of the line $\nu_2(\pi)-$G.S.$(\sigma^+)$R$_e(25)$ 
is 15\% and 70\% larger respectively
compared with the observed feature. In addition, considering that
this level departs from LTE with the same smooth dependence as 
adjacent levels, does not give a good fit as synthetic profile
presents 77\% more emission and 6\% more absorption than the best fit.
Using an \textit{ad-hoc} rotational temperature improves the fits
to higher-$J$ ro-vibrational lines. The fits suggest a quadratic
deviation from LTE for the rotational pattern of the ground state 
beginning at $J=20$ with a decrease per rotational level
of $96-7.8(J-20)$~K at $r=$~\rdin{} and of $46-3.2(J-20)$~K
at $r=$~\rdout{}.
From the fits of bands $\nu_2(\pi)-$G.S.$(\sigma^+)$ and 
$2\nu_2(\sigma^+)-\nu_2(\pi)$, we can assume that $\nu_2(\pi)$ is in
LTE for the rotational levels up to $J=22$ departing smoothly
from LTE for higher levels. The most important \textit{ad-hoc}
temperature corresponds to $J=25$, being necessary to fit ro-vibrational
lines involving higher-$J$ levels. We can assume that $2\nu_2(\sigma^+)$
rotational levels are in LTE for $J\le 26$, with it necessary to add an
\textit{ad-hoc} rotational temperature for $J=27$. Nevertheless, fitting 
the bands $2\nu_2(\delta)-\nu_2(\pi)$ also requires \textit{anomalous} 
non-LTE rotational temperatures at the photosphere \rdin{}, and \rdout{}
for lines $J=12$ and 26. The best fits to all the lines involving 
\textit{anomalous} rotational temperatures require 
$T_{\subscript{rot}}($\rdout$)\le 50$~K, although temperatures below 100~K are 
acceptable despite poorer fits.

\subsection{Abundances}
\label{sec:xhcn}

Low-$J$ HCN lines of band $\nu_2(\pi)-$G.S$(\sigma^+)$ are optically thick due
to the large dipole moment of the vibrational transition. Many lines are saturated
and little information can be obtained from them. In particular, the abundance
is a parameter which can not be accurately determined through fits of
low energy HCN lines. The intensity of the G.S.$(\sigma^+)$ pure 
rotational lines is directly related to the abundance of HCN while the intensity 
of ro-vibrational lines also depend on \tvib{}, which is initially unknown. The 
lack of radio observations of these Regions hinders an accurate 
determination of the HCN abundance. Hence it is better to obtain the H$^{13}$CN 
abundance by fitting the optically thin $\nu_2(\pi)-$G.S.$(\sigma^+)$ lines, and using 
the derived ratio [$^{12}$C]/[$^{13}$C]$\simeq 41$ to determine the abundance of HCN.
We found that the abundances which produces the best fits are:
\begin{equation}
x_\subscript{HCN}(r) = \left\{
\begin{array}{ll}
1.23\times 10^{-5} & \textnormal{Region I}\\
4.5\times 10^{-5} & \textnormal{Region II}\\
2.0\times 10^{-5} & \textnormal{Region III}
\end{array}
\right.
\end{equation}
which means that the column density is $1.6\times 10^{19}$~cm$^{-2}$, i.e.,
similar to that of C$_2$H$_2$. In fact, HCN seems to be more abundant than
C$_2$H$_2$ in Region I, where the gas density is rather high.
However, as in the C$_2$H$_2$ case, the inner abundance is poorly determined (see 
\S\ref{sec:sens.disc}) and the middle and outer ones are compatible with those found 
in previous works 
\citep{keady_1993,cernicharo_1999,cernicharo_1996b,wiedemann_1991,dayal_1995,lindqvist_2000}.
The abundance of HCN between Regions I and II increases by
a factor 3.7 
while for C$_2$H$_2$ the increase is by a factor of 10, 
incompatible in this case with the predictions of chemical 
LTE models (M. Ag\'undez, private communication)
and non-LTE stellar pulsating ones
\citep{cherchneff_2006}, where a diminishing in the HCN abundance with growing radii is found
(see \S\ref{sec:sens.disc} for a discussion on this topic). 
The decrease by a factor of $\simeq 2$
in the abundance 
between Regions II and III seems to be real (see Table~\ref{tab:errors}).

\section{Sensitivity to Diverse Parameters and Discussion}
\label{sec:sens.disc}

To obtain complete information about the CSE, it is necessary to study the 
behavior of the model while varying the parameters and their uncertainties.
For this purpose, we have selected lines of all the molecules
studied in the present work. Each line is radiatively active over different 
regions of the CSE. For example, the H$^{13}$CCH line $2\nu_5(\delta)
\leftrightarrow{}\nu_5(\pi)$R$_e(7)$ extends until 22~R$_*$ and has an optical 
depth of $8.50\times 10^{-2}$ in Region I, 1.41 in Region II, and 
$9.94\times 10^{-3}$ in Region III. On the other hand, the C$_2$H$_2$ line 
$\nu_5(\pi_u)\leftrightarrow{}$G.S.$(\sigma_g^+)$R$_e(30)$ extends until 27.5~R$_*$ and has 
an optical depth of 3.39 in Region I, 39.0 in Region II, and 1.35 in Region III.
The first line is useful to study Region II while the second is useful for 
Region III because they are optically thin. 
Table~\ref{tab:errors} shows the calculated errors and the
lines used for this purpose. The chosen lines appear
to be the best for a determination of the errors in the physical parameters. The
velocity gradients introduce different spatial contributions
to the line profiles. Hence the impact of each region of the CSE becomes 
measurable through the analysis of these lines. Moreover, a line that is globally 
optically thick can also carry information on specific
regions of the envelope. However, 
there are some parameters such as \tdin{} or \rdin{} whose errors must be calculated 
by fitting only the continuum because the changes they produce on the lines, when
varied, can be overcome by modifying other parameters, e.g., vibrational temperatures.

\begin{deluxetable*}{ccc|ccc}
\tabletypesize{\footnotesize}
\tablecolumns{6}
\tablewidth{0.9\textwidth}
\tablecaption{Errors\label{tab:errors}}
\tablehead{\colhead{Abundance} & \colhead{Region} & \colhead{Value} & \colhead{Parameter} & \colhead{Unit} & \colhead{Value}}
\startdata
                &     &                                               &  \rdin{}                                       & R$_*$      & $5.2^{+0.6}_{-0.5}\tablenotemark{i}$\\
$x$(C$_2$H$_2$)                & I   & $\left( 8^{+6}_{-4}\tablenotemark{a}\right) \times 10^{-6}$   & \rdout{}                                       & R$_*$      & $21\pm 3\tablenotemark{b}$\\
                & II  & $\left( 8.0^{+1.2}_{-1.1}\tablenotemark{a}\right) \times 10^{-5}$   & \tk{}(\rdin{})                              & K           & $900^{+300}_{-200}\tablenotemark{e}$ \\
                & III & $\left( 8.0^{+4.0}_{-2.3}\tablenotemark{b}\right) \times 10^{-5}$   & \tk{}(\rdout{})                            & K          & $400^{+27}_{-25}\tablenotemark{b}$\\
$x$(H$^{13}$CCH)                & I   & $\left( 3.7^{+1.6}_{-1.5}\tablenotemark{d}\right) \times 10^{-7}$   & $\alpha_\subscript{I}$                     &            & $0.58\pm 0.16\tablenotemark{e}$\\
                & II  & $\left( 3.9^{+0.5}_{-0.4}\tablenotemark{d}\right) \times 10^{-6}$   & $\alpha_\subscript{II}$                     &            & $0.58^{+0.05}_{-0.04}\tablenotemark{b}$\\
                & III & $\left( 3.9^{+1.0}_{-0.9}\tablenotemark{f}\right) \times 10^{-6}$   & \tvib($\nu_5$R$_e$,\rdin)                 & K          & $175^{+22}_{-20}\tablenotemark{b}$\\
$x$(HCN)                & I   & $\left( 1.2\pm 0.4\tablenotemark{h}\right) \times 10^{-5}$   & \tvib($\nu_5$R$_e$,\rdout)                 & K          & $82^{+10}_{-9}\tablenotemark{b}$\\
                & II  & $\left( 4.5^{+0.7}_{-0.6}\tablenotemark{h}\right) \times 10^{-5}$   & \tdin{}                                    & K          & $850\pm 50\tablenotemark{i}$\\ 
                & III & $\left( 2.0^{+0.8}_{-0.6}\tablenotemark{g}\right) \times 10^{-5}$   & $v_\subscript{e,\textnormal{I}}$              & km~s$^{-1}$ & $5.0^{+1.6}_{-0.9}\tablenotemark{e}$\\
$x$(H$^{13}$CN)                & I   & $\left( 3.0^{+1.1}_{-0.9}\tablenotemark{e}\right) \times 10^{-7}$   & $v_\subscript{e,\textnormal{II}}$             & km~s$^{-1}$ & $11.0^{+1.5}_{-1.4}\tablenotemark{c}$\\
                & II  & $\left( 1.10^{+0.18}_{-0.16}\tablenotemark{e}\right) \times 10^{-6}$  & $v_\subscript{e,\textnormal{III}}$            & km~s$^{-1}$ & $14.5\pm 0.4\tablenotemark{c}$\\
                & III & $\left( 5.0^{+2.0}_{-1.5}\tablenotemark{h}\right) \times 10^{-7}$       & $\alpha_\textnormal{d}$                       &            & $0.39^{+0.08}_{-0.06}\tablenotemark{i}$\\
                &     &                                                   & $\tau(11~\mu\textnormal{m})$                &            & $0.70^{+0.13}_{-0.11}\tablenotemark{i}$
\enddata
\tablecomments{The parameters derived by the model with their errors,
estimated through a sensitivity study. All the errors translate
into an interval 
of the normalized flux with a maximum width of 20\%, containing the best fit.
The uncertainty of a given parameter has been calculated by comparing the synthetic
spectrum with the observed one and by forcing the former to be within that
interval, while the rest of the parameters
are modified until getting the maximum/minimum value for the considered one
(see \S\ref{sec:modeldesc} for more information about the method).
Some of the values in the upper Table could be slightly different
from those appearing along the text, since the latter produce the best fits and
the former produce just acceptable fits and are complemented with the uncertainties
of the parameters.
The $x$'s are the abundances in the 
three regions of the envelope. \rdin{} and \rdout{} are the positions of the 
inner and outer dust formation shells, respectively. The kinetic temperature, 
\tk{}, has been evaluated at \rdin{} and \rdout{}. $\alpha_\subscript{I}$ and
$\alpha_\subscript{II}$ are the \tk{} exponents in Regions I and II.
\tvib($\nu_5$R$_e$,\rdin) and 
\tvib($\nu_5$R$_e$,\rdout) are the vibrational temperatures of the transition 
$\nu_5-$G.S.R$_e$ evaluated at \rdin{} and \rdout{} respectively (shown in the 
Table as an example of the \tvib{} uncertainties). \tdin{} is the dust temperature 
at \rdin{}. The expansion velocity of the gas, $v_e$, is shown over the three regions 
of the CSE. $\alpha_\textnormal{d}$ is the exponent of the dust temperature law. 
Finally, the last parameter is the dust optical depth, $\tau$, at 11~$\mu$m.\\
A list of the lines used to calculate the error for each parameter is shown below. 
Each line is accompanied by the set of parameters 
$P=[r_{1\%}(\textnormal{R$_*$}),\tau_\subscript{Total},\tau_\subscript{I},\tau_\subscript{II},
\tau_\subscript{III}]$, where $r_{1\%}$ is the minimal radius at which the opacity of the 
line is $1\%$ of the maximum, $\tau_\subscript{Total}$ is the total optical depth, and 
$\tau_\subscript{I}$, $\tau_\subscript{II}$, and $\tau_\subscript{III}$ are the optical depths in 
Regions I, II, and III, respectively.}
\tablenotetext{a}{C$_2$H$_2$~$2\nu_5(\delta_g)\leftrightarrow\nu_5(\pi_u)$R$_e(30)$ with 
$P=[20.1, 3.57, 0.549, 2.83, 0.00664]$}
\tablenotetext{b}{C$_2$H$_2$~$\nu_5(\pi_u)\leftrightarrow~$G.S.$(\sigma_g^+)$R$_e(30)$ with 
$P=[28.4, 42.0, 1.83, 37.2, 0.988]$}
\tablenotetext{c}{C$_2$H$_2$~$\nu_5(\pi_u)\leftrightarrow~$G.S.$(\sigma_g^+)$R$_e(6)$ with 
$P=[206, 241, 1.60, 103.0, 127.4]$}
\tablenotetext{d}{H$^{13}$CCH~$2\nu_5(\delta)\leftrightarrow\nu_5(\pi)$R$_e(7)$ with 
$P=[22.6, 1.47, 0.0492, 1.38, 0.00660]$}
\tablenotetext{e}{H$^{13}$CN~$\nu_2(\pi)\leftrightarrow~$G.S.$(\sigma^+)$R$_e(14)$ with 
$P=[46.4, 3.14, 0.132, 2.64, 0.234]$}
\tablenotetext{f}{H$^{13}$CCH~$\nu_5(\pi)\leftrightarrow~$G.S.$(\sigma^+)$R$_e(26)$ with 
$P=[37.8, 2.98, 0.0629, 2.59, 0.188]$}
\tablenotetext{g}{HCN~$\nu_2(\pi)\leftrightarrow~$G.S.$(\sigma^+)$R$_e(22)$ with 
$P=[32.8, 62.6, 5.16, 53.9, 1.07]$}
\tablenotetext{h}{The absolute error of the abundance in the given Region for the 
isotopologue considered
has been calculated with the relative error for the other isotopologue.
Taking the ratio [$^{12}$C]/[$^{13}$C] 
as constant throughout the CSE 
the relative errors of both isotopologues must be equal.}
\tablenotetext{i}{The errors have been calculated by fitting the continuum with different
values of the parameter.}
\end{deluxetable*}

The results obtained in this paper have been based on several hypotheses, 
most of them
supported by observational data. However, the assumption for which the
observations of IRC+10216 by ISO/SWS and IRTF/TEXES are compatible
(see \S\ref{sec:continuum}) needs discussion.
As we pointed out in \S\ref{sec:obs}, the difference in the IR phase
between both observations is $\Delta\varphi_\subscript{IR}\simeq 0.3$. 
Hence, the physical properties
in the innermost CSE are different for observations of gas 
(IRTF/TEXES) and dust (ISO/SWS). Each pulsation of the star is followed
by an increment in the emitted radiation and, consequently, \rdin{} and
\tdin{} are magnified because of the new input of optical and IR 
photons from the star and the ejected gas.
Nevertheless, the large optical depth of dust shells near \rdin{} 
diminishes the number of short wavelength photons from the stellar surface as 
the radius increases. The rest of the high frequency stellar emission is largely
diluted and does not significantly affect
the middle and outer envelope.
The result is that only the inner shells of the dusty CSE are significantly heated.
The dust grains near \rdin{} reemit the absorbed energy at longer wavelengths
increasing the observed NIR intensity of the continuum. Since FIR and radio
emission from the source arise almost entirely from the outer dust shells
(the black-body emission at T$_\subscript{bb}=150$~K peaks at 19.3~$\mu$m and the
dust shell at that temperature at $\simeq 450$~R$_*$),
stellar pulsation has little effect on the long wavelength range
of the continuum. However, it does affect the MIR contribution
\citep{monnier_1998}. By modifying \tdin{} from 850 to 950~K and the exponent 
$\alpha_\subscript{d}$
from 0.39 to 0.42 to model the increment of \tdust{} in the 
inner dusty CSE during a stellar pulsation, the continuum changes less
than 15\% over the frequency range considered in this paper. Any variations
introduced in the lineshapes can be overcome by changing the \tev{} profiles.
Despite these changes, the rough behavior of \tev{} remains. The abundance
profile does not seem to be affected by these modifications of \tdust{}. 
The increase in the abundance detected between Regions I and II seems to be
real and not an effect of an incompatibility between the ISO/SWS and
IRTF/TEXES observations.

The expansion velocity in Region I derived by us is larger than that suggested
by \citet{keady_1988} and \citet{keady_1993} by a factor of 1.75. 
The main effect produced by this difference on the P-Cygni profiles is found
at the red wing of the line emission, at positive velocities.
Increasing the expansion velocity of the gas in Region I 
expands the red wing of P-Cygni profiles. A reduction to 3~km~s$^{-1}$ 
does not improve the fits. Expansion velocities larger than 5~km~s$^{-1}$ seem 
to give better results, but are improbable. The synthetic profiles with 5~km~s$^{-1}$
fit the observed lines quite well, although a slight lack
of emission
at velocities near terminal in C$_2$H$_2$ $\nu_5(\pi_u)-$G.S.$(\sigma_g^+)$R$_e$ 
and HCN $\nu_2(\pi)-$G.S.$(\sigma^+)$R$_e$ lines can be seen.

In order to detect any effect produced by the difference between the pulsation 
phase of the IRTF/TEXES and ISO/SWS observations, we have varied the parameters 
\rdin{} and \tdust{}. They do not cause any modification of the width of the red wing.
The fact that the width of the red part of the observed P-Cygni profiles is
larger than that of the synthetic ones, considering the accepted microturbulence, 
i.e., $\Delta v_1=5$~km~s$^{-1}$, suggests either more emission from
the outer CSE only at the back of the star or larger linewidths in the innermost
envelope. The former explanation is in disagreement with the spherical symmetry
established by a large number of observations. The latter seems to
be more realistic. A possible explanation could be an onion-structured Region I with
shells expanding at positive and negative velocities corresponding to expansion and 
collapse \citep{bowen_1988}. 
However, we could suggest another scenario explaining large linewidths in the innermost
CSE where it would be a clumpy region.
The clumps would be composed of hot gas and
would move inwards and outwards at high velocities along different radial directions.
Unfortunately, little information on the spatial distribution of these clumps 
in Region I could be derived through the observed spectrum
due to heavy dilution.
Both scenarios have been invoked by \citet{fonfria_2006} in order to explain the observed
SiS masers towards IRC+10216, where each maser feature has widths of $\simeq 2-3$~km~s$^{-1}$
providing valuable information about the physical conditions 
of the emitting regions, i.e., shells or clumps. 
Modeling these structures as in \S\ref{sec:linewidth} 
with $\Delta v_1=30$~km~s$^{-1}$ and $\ell=1.5$~R$_*$ produces synthetic profiles having
an emission wing wider than with $\Delta v_1=5$~km~s$^{-1}$.  
Related to the derived column densities for C$_2$H$_2$ and HCN, 
they are beam averaged and thus
their actual values for the clumps in Region I could be much larger if we take
into account dilution.
More spatial resolution is needed to study the complex, small scale
structure of the innermost CSE.

The possible condensation of C$_2$H$_2$ (and isotopologues)
onto the dust grains in
the outer acceleration zone is somewhat constrained by our derived 
uncertainties in the abundances.  In Regions II and III, these allow for 
$20-30$\% of the C$_2$H$_2$ to be deposited onto grains. 
We can estimate how much the dust grains should increase in radius from this 
deposition. Using the density of C$_2$H$_2$ and the density of dust grains at 
\rdout{} derived from the fits ($\simeq 1200$~cm$^{-3}$ and 
$\simeq 3\times 10^{-4}$~cm$^{-3}$ respectively), assuming that 30\% of
the C$_2$H$_2$ turns into solid state with density $\simeq 1$~g~cm$^{-3}$ and 
that the dust grains are spheres of radius 0.05~$\mu$m,
we calculate that $1.2\times 10^{6}$ molecules of C$_2$H$_2$
condense onto a single dust grain. The contribution to the mass of the dust grain is 
$10\%$ and the radius of the grain grows by $3.2\%$. For a larger initial 
dust grain radius, the 
increase is even smaller. The small increase in the size of the dust grains suggests 
that the condensation of C$_2$H$_2$ onto the dust grains does not contribute 
significantly to the grain growth and so other carbon molecules are more likely to
be responsible for it. However, it might be possible that C$_2$H$_2$ reacts
with other molecules on the grain surface giving rise to
more complex organic molecules.

Regarding the decrease of the HCN abundance in \rdout{}, it might be 
thought that HCN suffers a depletion onto dust grains. 
In that case, it would be reasonable to expect
some emission/absorption in the observed 
continuum at wavelengths between 4 and 5~$\mu$m due to the stretching mode of
the group C$\equiv$N \citep*[see, for example,][]{lacy_1984,pendleton_1999}.
Nevertheless, 
no significant emission/absorption can be seen in the ISO/SWS spectrum 
of IRC+10216 at these wavelengths
although it is possible that some HCN molecules condense onto dust grains.
However, their contribution to the continuum could be masked by the CO 
vibrational band at 4.67~$\mu$m.
Even in that case, we think that 
this deposition process does not account for the abundance
decay observed at \rdout{}, 
which could actually be due to chemical processes. The most important HCN decrease
is produced by photodissociation in the external layers of the CSE 
\citep*[][and references therein]{agundez_2006}.

\section{Conclusions}
\label{sec:conclusions}

IRC+10216 has been observed from 11 to 14~$\mu$m with the high resolution 
spectrograph TEXES at the 3~m IRTF. We complemented this
data with information on 
the continuum observed with ISO/SWS. We have identified 462 ro-vibrational lines 
of C$_2$H$_2$ (involving vibrational levels up to $3\nu_5$ at $\simeq 2185$~cm$^{-1}$), 
95 of HCN (involving levels up to 
$2\nu_2$ at $\simeq 1425$~cm$^{-1}$), 106 of
H$^{13}$CCH (involving levels up to $2\nu_5$ at $\simeq 1454$~cm$^{-1}$)
and 7 of H$^{13}$CN (the fundamental band $\nu_2$ at $\simeq 707$~cm$^{-1}$).
By means of a model of an AGB star developed by us, we have fitted over 300 lines. 
The results can be summarized as follows:

1.--~The geometrical structure and physical properties of dust and gas over the 
whole envelope are compatible in many cases with those already 
proposed. However, the values of certain parameters are quite different. In 
particular, we find the position of the outer acceleration zone 
($\simeq 21$~R$_*\simeq 0\arcsecond4$) to be farther 
out than found by previous work. 

2.--~The abundances of C$_2$H$_2$ 
and HCN in the innermost CSE, Region I
($8\times 10^{-6}$ for C$_2$H$_2$ and $1.2\times 10^{-5}$ for HCN with
[$^{13}$C]/[$^{12}$C]$\simeq 41$) are lower than those over the outer envelope,
Regions II 
($8.0\times 10^{-5}$ for C$_2$H$_2$ and $4.5\times 10^{-5}$ 
for HCN) and III ($8.0\times 10^{-5}$ for C$_2$H$_2$ and 
$2.0\times 10^{-5}$ for HCN). 
For HCN, the ratio 
of the abundance in Region II to that in Region I is only a factor 
$\simeq 4$ but,
in the case of C$_2$H$_2$, it is a factor $\simeq 10$. 
The latter result is in accordance with the increase predicted by the
chemical models. Contrarily, the derived abundance of HCN grows as radius increases
while the chemical models suggest a decay.
In addition, the determined
abundances in Regions II and III allow the condensation of $\le 20-30\%$ of
molecules of C$_2$H$_2$ onto the dust grains.
On the other hand, the decrease of the HCN abundance when reaching the outer
dust formation zone is probably due to chemical processes.

3.--~The vibrational temperatures determined by fitting the lines suggest
the existence of a complex pumping mechanism driven by near-IR radiation 
($3.5-7$~$\mu$m) emitted by the star and the inner dusty CSE. 
High-energy vibrational levels play an important role in the non-LTE pumping of 
low- and mid-energy vibrational states throughout
Region II and the edge
of Region III through radiative cascades. Reliable modeling using 
spectroscopic methods of any molecular species in the innermost CSE requires 
an analysis of high-energy vibrational levels at temperatures as high as 
$\simeq 3000$~K.

4.--~Most of the rotational levels behave as expected: low-$J$ rotational levels 
can be considered to be in LTE, while high-$J$ ones are not thermalized. Interestingly, 
we have found it necessary to add \textit{ad-hoc} rotational temperatures to fit 
several ro-vibrational lines. These \trot{} are quite different from those of 
adjacent levels. The involved ro-vibrational transitions do not seem to be affected by 
instrumental or telluric effects so we assume these variations to be
real and produced by molecular processes such as overlaps with lines of other 
molecular species.

Infrared spectroscopic data provide us with extensive information about physical 
conditions of warm sources with good spatial resolution and without 
resorting to interferometric 
methods. However, physical processes and the time-dependent
chemistry in the 
innermost CSE remain unknown. Future high angular resolution observations with 
TEXES in the infrared domain will provide us with high quality data to 
study the dynamics, chemistry, and physical conditions in the warm asymmetric 
innermost envelope. On the other hand, ALMA will supply us in several years
with the interferometric observations needed to delve
more deeply
into the dynamics of the near environment of the central star and 
to further our understanding of dust
formation and growth.

\acknowledgments

J. Cernicharo and J. P. Fonfr\'{\i}a would like to thank the Spanish 
Ministerio de Educaci\'on y Ciencia for funding support through grant 
ESP2004-665, AYA2003-2785, and the ``Comunidad de Madrid''
government under PRICIT project S-0505/ESP-0237 (ASTROCAM). 
During this study, J. P. Fonfr\'{\i}a was supported by the CSIC and the
``Fondo Social Europeo'' under internship grant from the I3P Programme.
This study is supported in part by the European Community's human potential
Programme under contract MCRTN-CT-2004-51230, ``Molecular Universe''.
M. J. Richter is supported by grant AST-0307497. Development of TEXES was
supported by grants from the NSF and USRA.  Observations with TEXES were
supported by NSF grant AST-0205518 and AST-0607312.
M. J. Richter, J. H. Lacy and others at the Infrared Telescope
Facility, which is 
operated by the University of Hawaii under Cooperative Agreement no. 
NCC 5-538 with the National Aeronautics and Space Administration, 
Office of Space Science, Planetary Astronomy Program.
We also thank J. R. 
Pardo, M. Taylor and the referee 
for suggestions and valuable corrections to this manuscript, M. Ag\'undez 
for the very interesting talks about chemistry in the inner envelope of IRC+10216
and
H. Mutschke for providing us with SiC opacities and his comments 
on dust properties.

\appendix

\section{Line Frequencies}
\label{sec:LineFrequencies}

The spectroscopic data for all the molecular species have been taken mainly 
from the \citet{HITRAN_database} \citep{HITRAN}.  Further information came from
data published by \citet{herman_2003}, \citet{kabbadj_1991}, and
\citet{dilonardo_1993,dilonardo_2002} for C$_2$H$_2$ and its
isotopologues, and from \citet{maki_1996,maki_2000} 
and \citet{malathy_devi_2005} for HCN and H$^{13}$CN.
The spectroscopic data relative to the modeled lines can be found
in Table~\ref{tab:spectroscopy}.

\begin{deluxetable}{cc|cccccc}
\tabletypesize{\footnotesize}
\tablecolumns{8}
\tablewidth{0pt}
\tablecaption{Spectroscopic Data of the Modeled Lines\label{tab:spectroscopy}}
\tablehead{\colhead{Mol.} & \colhead{Transition} & \colhead{Freq. (cm$^{-1}$)} & 
\colhead{$R^2$ ($10^{-2}$~D$^2$)} & \colhead{$E_l$ (cm$^{-1}$)} & \colhead{$S_{ul}$} &
\colhead{$g_u$} & \colhead{$g_l$}}
\startdata
HCN    & $2\nu_2(\delta)\leftrightarrow{}\nu_2(\pi)$R$_f$(3) & 726.423637 & 7.460 &  729.8054 &  37.500 &   9 &   7\\
HCN    & $2\nu_2(\sigma^+) \leftrightarrow{}\nu_2(\pi)$R$_e$(8) & 726.700075 & 3.740 &  818.3829 &  40.000 &  19 &  17\\
H$^{13}$CCH & $\nu_5(\pi) \leftrightarrow{}$G.S.$(\sigma^+)$R$_e$(1) & 732.820280 & 8.744 &    2.2969 &  15.000 &   5 &   3\\
H$^{13}$CCH & $\nu_4+\nu_5(\sigma^+)\leftrightarrow{}\nu_4(\pi)$R$_e$(7) & 734.177200 & 5.118 &  671.4353 &  35.000 &  17 &  15\\
H$^{13}$CCH & $\nu_5(\pi)\leftrightarrow{}$G.S.$(\sigma^+)$R$_e$(2) & 735.115630 & 6.995 &    6.8907 &  20.000 &   7 &   5\\
HCN    & $2\nu_2(\delta)\leftrightarrow{}\nu_2(\pi)$R$_f$(6) & 735.315339 & 7.480 &  774.3662 &  51.429 &  15 &  13\\
C$_2$H$_2$   & $\nu_4+\nu_5(\sigma_u^+)\leftrightarrow{}\nu_4(\pi_g)$R$_e$(7) & 735.543410 & 5.118 &  677.5067 &  35.000 &  51 &  45\\
HCN    & $\nu_2(\pi)\leftrightarrow{}$G.S.$(\sigma^+)$R$_e$(7) & 735.611573 & 3.740 &   82.7713 &  45.000 &  17 &  15\\
H$^{13}$CCH & $\nu_5(\pi)\leftrightarrow{}$G.S.$(\sigma^+)$R$_e$(3) & 737.410380 & 6.247 &   13.7813 &  25.000 &   9 &   7\\
C$_2$H$_2$   & $\nu_4+\nu_5(\sigma_u^+)\leftrightarrow{}\nu_4(\pi_g)$R$_e$(8) & 737.980170 & 5.176 &  696.3085 &  40.000 &  19 &  17
\enddata
\tablecomments{Spectroscopic data relative to the modeled lines.
The transitions are labeled in Columns 1 and 2, while in Columns 3 to 8 
are shown the frequency in cm$^{-1}$, the square dipole moment in $10^{-2}$~D$^{2}$,
the energy of the lower level in cm$^{-1}$ and the line strength of each transition,
and the degeneracy of the upper and lower levels respectively.
See the text for details on the notation relative to the transitions.
\textit{[The complete version of this table is in the electronic edition of
the Journal. The printed edition contains only a sample.]}}
\end{deluxetable}

The molecules studied in the present work, acetylene and hydrogen cyanide, 
are linear. Consequently, the notation we have used is the same for both species. 
The notation adopted to refer to the vibrational normal modes is 
$v_i\nu_i(\ell_S^p)$, where $i$ is the mode number ($i=1,2,\ldots,3N-5-d$, where
$N$ is the number of atoms of the molecule and $d$ is the number of 
degenerate
modes), $v_i$ is the vibrational quantum number related to the $i^\subscript{th}$ 
normal mode, $\ell$ is a Greek letter corresponding to the quantum number of
the total vibrational angular momentum of the vibrational state ($\ell=0\equiv\sigma$,
$\ell=1\equiv\pi$, $\ell=2\equiv\delta$, etc.), $S$ is the symmetry of the
ro-vibrational state with respect to the molecular midplane ($S=g$ for 
\textit{gerade} or even states
and $S=u$ for \textit{ungerade} or odd ones), and 
$p$ is the parity of the total molecular wavefunction ($+,-$). If no parity indicator 
appears, the label refers to the $+$ and $-$ levels simultaneously.
If no symmetry indicator is present, the molecule is not symmetric
with respect to the molecular midplane, as is the case of H$^{13}$CCH and HCN.
For the combination bands, $\ell$, $S$, and $p$ refer to the mixture
of all vibrational modes which participate in the combined state. Combination
bands are sometimes split in different vibrational levels 
having
the same quantum number $\ell\ne 0$, each one with both parities $+$ and $-$.
In these cases, Roman numbers are added to label these states from
higher to lower energies (see Figure~\ref{fig:vlc2h2}).
In addition, the interaction between the vibrational and molecular 
angular momenta splits the rotational levels of vibrational states with 
$\ell\ne 0$ into two sub-levels with opposite parities denoted 
$e$ and $f$. The energy gap between those sub-levels for a given rotational
state depends on $J$ and on the vibrational state. For low-$J$ levels it is 
between 200 and 300 times lower than the rotational constant for acetylene
and hydrogen cyanide. 
The gaps are large enough to be observed in high-resolution 
spectra, as in our case. In the label for each transition, the parity $e$ or $f$ of 
the lower ro-vibrational level is included. Deriving the parity of the upper level 
can be done with the selection rules governing transitions in linear molecules:
\begin{eqnarray}
& \Delta v_i=\pm 1 \quad & i=\textnormal{bending modes,}\nonumber\\
& \Delta v_i=\pm 1,\pm 2,\pm 3,\ldots\quad & i=\textnormal{stretching modes,}\nonumber\\
& \Delta\ell=0,\pm 1\quad & \Delta\ell=0~\textnormal{allowed only for}~\ell_\subscript{low}=0,\nonumber\\
& \sigma^+\leftrightarrow\sigma^- & \textnormal{forbidden,}\nonumber\\
& g\leftrightarrow{}u\quad & \textnormal{allowed (C$_2$H$_2$, $^{13}$C$_2$H$_2$),}\nonumber\\
& g\leftrightarrow{}g~~~u\leftrightarrow{}u\quad & \textnormal{forbidden (C$_2$H$_2$, $^{13}$C$_2$H$_2$),}\nonumber\\
& \Delta J=0,\pm 1\quad & \textnormal{$\Delta J=0$ forbidden for $J=0\leftrightarrow J=0$, when $\Delta\ell=0$ and $\ell_\subscript{low}=0$, and}\nonumber\\
&                       & \hspace{35pt}\textnormal{for parallel transitions; allowed otherwise,}\nonumber\\
& e\leftrightarrow{}e~~~f\leftrightarrow{}f\quad & \textnormal{allowed for R and P branches; forbidden for Q branch,}\nonumber\\
& e\leftrightarrow{}f\quad & \textnormal{allowed for Q branch; forbidden for R and P branches,}\nonumber
\end{eqnarray}
where the transitions with $\Delta\ell=0$ are named \textit{parallel transitions}
(with the change in the dipole moment parallel to the molecular axis)
and those with $\Delta\ell=\pm 1$ are named \textit{perpendicular transitions}
(the dipole moment changing in a 
perpendicular direction to the molecular axis).

The three most abundant isotopologues of acetylene are $^{12}$C$_2$H$_2$ (C$_2$H$_2$), 
H$^{13}$C$^{12}$CH (H$^{13}$CCH) and $^{13}$C$_2$H$_2$. Acetylene is a linear molecule 
with five fundamental vibrational modes (see Figure~\ref{fig:vlc2h2}).
The bending modes $\nu_4$ and $\nu_5$ are doubly-degenerate to a
first approximation
with energies of 612.871 and 730.332~cm$^{-1}$ respectively \citep{herman_2003}. 
The $\nu_4$ mode is infrared inactive for C$_2$H$_2$ and $^{13}$C$_2$H$_2$, but active 
for H$^{13}$CCH. The $\nu_1$, $\nu_2$, and $\nu_3$ vibrational modes correspond to 
the different stretching modes of acetylene and have energies of 3372.849, 1974.316, 
and 3294.839~cm$^{-1}$ respectively for C$_2$H$_2$, and 
differ slightly for the 
isotopologues \citep{herman_2003}. Both C$_2$H$_2$ and $^{13}$C$_2$H$_2$ are symmetric 
and show a degeneracy, $g_s$, in the rotational levels due to nuclear spin 
statistics. The \textit{para-} levels ($J=\textnormal{even}$) have $g_s($C$_2$H$_2)=1$ 
and $g_s(^{13}$C$_2$H$_2)=6$, while the \textit{ortho-} levels ($J=\textnormal{odd}$) have
$g_s($C$_2$H$_2)=3$ and $g_s(^{13}$C$_2$H$_2)=10$. These two isotopologues have a null 
permanent dipole moment due to symmetry with respect to the molecular midplane. 
Therefore, they do not have purely rotational transitions. However, H$^{13}$CCH is not 
symmetric with respect the molecular midplane and has a very small dipole moment along 
the molecular axis. The bands of C$_2$H$_2$ and H$^{13}$CCH identified
are shown in the Table~\ref{tab:lines}.

HCN (and H$^{13}$CN) has a permanent dipole moment along the molecular axis allowing 
purely rotational radiative transitions. 
HCN has three vibrational modes: $\nu_2$, is a doubly-degenerated
bending mode with an energy of 713.461~cm$^{-1}$, and 
the other two, $\nu_1$ and $\nu_3$, are stretching modes with energies of 3311.480 and 
2096.846~cm$^{-1}$, respectively \citep{maki_2000}. The vibrational energy
pattern for HCN and H$^{13}$CN is shown in Figure~\ref{fig:tvhcn}.
The detected bands are given in Table~\ref{tab:lines}.

\section{Line Intensities}
\label{sec:LineIntensities}

The line intensity as a function of temperature is given by \citep{jacquemart_2001}:
\begin{equation}
\label{eq:intensity}
S(T) = \frac{1}{4\pi\varepsilon_0} \frac{8\pi^3}{3hc}
\frac{g_\subscript{low}^\subscript{s}\nu_0}{Z(T)} \frac{1}{g_\subscript{v}} \left| R \right|^2
L(J,\ell) e^{-hcE_\subscript{low}/k_B T} \left[ 1 - e^{-hc\nu_0/k_B T} \right] 
\end{equation}
where the dipole moment squared, $\left|R\right|^2$ is expressed in Debye$^2$,
$\nu_0$ is the centre of the line in cm$^{-1}$, $E_\subscript{low}$ is the energy of the
lower level in cm$^{-1}$, $g_\subscript{low}^\subscript{s}$ is the spin degeneracy of the lower level, 
$L(J,\ell)$ is the rotational line strength (see below) and $Z(T)$ is the partition 
function. $g_\subscript{v}=2$ when the upper and lower vibrational levels involved in a 
ro-vibrational transition present $\ell$-type doubling, i.e., 
$\ell_\subscript{up},\ell_\subscript{low}\ne 0$; otherwise, 
$g_\subscript{v}=1$. In the case of $g_\subscript{v}=2$, 
the given value of $\left|R\right|^2$ considers the transitions 
having parities $e$ and $f$ at the same time.

The line strength of a ro-vibrational transition, $L(J,\ell)$ for a linear molecule 
is given by \citep{rothman_1992,herzberg_i}:
\begin{equation}
L(J,\ell)= \left\{
\begin{array}{ll}
(J+1+\ell)(J+1-\ell)/(J+1) & \textnormal{R branch,}~\Delta\ell=0, \\
(2J+1)\ell^2/J(J+1) & \textnormal{Q branch,}~\Delta\ell=0,\\
(J+\ell)(J-\ell)/J & \textnormal{P branch,}~\Delta\ell=0,\\[8pt]
(J+2+\ell\Delta\ell)(J+1+\ell\Delta\ell)/2(J+1) & \textnormal{R branch,}~\Delta\ell=\pm 1, \\
(J+1+\ell\Delta\ell)(J-\ell\Delta\ell)(2J+1)/2J(J+1) & \textnormal{Q branch,}~\Delta\ell=\pm 1,\\
(J-1-\ell\Delta\ell)(J-\ell\Delta\ell)/2J & \textnormal{P branch,}~\Delta\ell=\pm 1.
\end{array}
\right.
\end{equation}

The dipole moments used to calculate the spectrum of C$_2$H$_2$ and the $\nu_5$ 
ro-vibrational transitions of H$^{13}$CCH, have been obtained from the 
\citet{HITRAN_database}, while the hot bands of H$^{13}$CCH have been assumed 
to be identical to those of C$_2$H$_2$. The corresponding values for the $\nu_2$ 
transition of HCN have been taken from \citet{malathy_devi_2005}. The adopted 
dipole moment for HCN $\Delta\nu_2$=1 hot bands and for H$^{13}$CN $\nu_2$
transitions are similar to that of the $\nu_2$ transition of HCN.
All dipole moments are shown in Table~\ref{tab:dm}.

\begin{deluxetable}{ccc|ccc}
\tabletypesize{\footnotesize}
\tablecolumns{4}
\tablewidth{0pt}
\tablecaption{Dipole moments for C$_2$H$_2$ and HCN vibrational transitions\label{tab:dm}}
\tablehead{\multicolumn{3}{c}{C$_2$H$_2$} & \multicolumn{3}{c}{HCN}\\
\colhead{Transition} & $g_\subscript{v}$ & \colhead{$\left|R_0\right|^2 (\times 10^{-2}\textnormal{D}^2)$} & \colhead{Transition} & $g_\subscript{v}$ &  \colhead{$\left|R_0\right|^2 (\times 10^{-2}\textnormal{D}^2)$}}
\startdata
$\nu_5(\pi_u)\leftrightarrow~$G.S.$(\sigma_g^+)$ & 1 & $9.81 \pm 0.11$\tablenotemark{a} & $\nu_2(\pi)\leftrightarrow~$G.S.$(\sigma^+)$ & 1 & $3.8925 \pm 0.0016$\tablenotemark{b}\\
$\nu_4+\nu_5(\sigma_u^+)\leftrightarrow\nu_4(\pi_g)$ & 1 & $4.904 \pm 0.063$\tablenotemark{a} & $2\nu_2(\sigma^+)\leftrightarrow\nu_2(\pi)$ & 1 & 3.39\tablenotemark{c}\\
$\nu_4+\nu_5(\sigma_u^-)\leftrightarrow\nu_4(\pi_g)$ & 1 & $5.006 \pm 0.062$\tablenotemark{a} & $2\nu_2(\delta)\leftrightarrow\nu_2(\pi)$ & 2 & 7.14\tablenotemark{c}\\
$\nu_4+\nu_5(\delta)\leftrightarrow\nu_4(\pi_g)$ & 2 & $9.599 \pm 0.084$\tablenotemark{a} & & & \\
$2\nu_5(\sigma_g^+)\leftrightarrow\nu_5(\pi_u)$ & 1 & $9.79 \pm 0.19$\tablenotemark{a} & & & \\
$2\nu_5(\delta)\leftrightarrow\nu_5(\pi_u)$ & 2 & $18.86 \pm 0.14$\tablenotemark{a} & & &
\enddata
\tablenotetext{a}{\citet{jacquemart_2001}}
\tablenotetext{b}{\citet{malathy_devi_2005}}
\tablenotetext{c}{The vibrational dipole moment has been obtained from
fits to the available data in the \citet*[][2004]{HITRAN_database}.
\citet{malathy_devi_2005} established that the values of this database
for the $\nu_2\leftrightarrow{}$G.S. transition are lower
than they should be. Hence we could expect the same for the other transitions.
Therefore, we will use the same values for $2\nu_2(\sigma^+)
\leftrightarrow\nu_2(\pi)$ and $2\nu_2(\delta)
\leftrightarrow\nu_2(\pi)$ rather than $\nu_2(\pi)
\leftrightarrow{}$G.S.$(\sigma^+)$ \citep{malathy_devi_2005} as an approximation.}
\tablecomments{See the text for a definition of $g_\subscript{v}$ and the ro-vibrational selection rules.}
\end{deluxetable}

The partition function has been calculated by directly summing ro-vibrational
levels under these conditions:
\begin{enumerate}
\item For each vibrational state, at least all the rotational levels with
$J\le J_\subscript{max}=65$ are summed,
\item the relative contribution to the rotational partition function 
of the last rotational level of each
vibrational state must be less than $10^{-3}$, which 
may require a sum over
rotational levels having $J>J_\subscript{max}= 65$,
\item the highest considered vibrational state, $v_\subscript{max}$ must be higher
than that of the upper ro-vibrational level involved in the considered transition,
\item the relative contribution to the molecular partition function of
$v_\subscript{max}$ (including the rotational
partition function) must be less than $10^{-3}$ implying the
possibility of summing vibrational levels with $v>v_\subscript{max}$.
\end{enumerate}
For the high rotational temperatures prevailing near the stellar
photosphere, condition 2 implies that the highest rotational level needed
to be included for C$_2$H$_2$ in the models is between $J=116-130$. 
Conditions 3 and 4 lead to $v_\subscript{max}=110$.
On the other hand, for low and intermediate \tk{},
the highest rotational level included to fulfill condition 2
is well below $J_\subscript{max}=65$, e.g., for $T_\subscript{rot}\simeq 400$~K
the error on the partition function
will be $10^{-3}$ for $J\simeq 32$, 
although we have to sum up to $J_\subscript{max}=65$ to
fulfill condition 1, and $v_\subscript{max}=9$ 
(conditions 3 and 4).


\begin{thebibliography}{}
\bibitem[Ag\'undez \& Cernicharo(2006)]{agundez_2006} Ag\'undez, M. \& Cernicharo, J., 2006, \apj, 650, 374
\bibitem[Bergeat et al.(2001)]{bergeat_2001} Bergeat, J., Knapik, A., \& Rutily, B., 2001, \aap, 369, 178
\bibitem[Betz(1981)]{betz_1981} Betz, A. L., 1981, \apj, 244, L103
\bibitem[Bowen(1988)]{bowen_1988} Bowen, G. H., 1988, \apj, 329, 299
\bibitem[Cernicharo \& Gu\'elin(1987)]{cernicharo_1987} Cernicharo, J. \& Gu\'elin, M., 1987, \aap, 183, L10
\bibitem[Cernicharo \& Gu\'elin(1996a)]{cernicharo_1996a} Cernicharo, J. \& Gu\'elin, M., 1996, \aap, 309, L27
\bibitem[Cernicharo et al.(1996b)]{cernicharo_1996b} Cernicharo, J., Barlow, M. J., Gonz\'alez-Alfonso, E. et al., 1996, \aap, 315, L201
\bibitem[Cernicharo et al.(1999)]{cernicharo_1999} Cernicharo, J., Yamamura, I., Gonz\'alez-Alfonso, E. et al., 1999, \apj, 526, L41
\bibitem[Cernicharo et al.(2000)]{cernicharo_2000} Cernicharo, J., Gu\'elin, M., \& Kahane, C., 2000, \aas, 142, 181
\bibitem[Cernicharo et al.(2004)]{cernicharo_2004} Cernicharo, J., Gu\'elin, M., \& Pardo, J. R., 2004, \apjl, 615, L145
\bibitem[Cherchneff et al.(1992)]{cherchneff_1992} Cherchneff, I., Barker, J. R., \& Tielens, A. G. G. M., \apj, 1992, 401, 269
\bibitem[Cherchneff(2006)]{cherchneff_2006} Cherchneff, I., 2006, \aap, 456, 1001
\bibitem[Cox(2000)]{cox_2000} Cox, A. N., ed. 2000, Allen's Astrophysical Quantities (4th ed.; New York: AIP)
\bibitem[Crosas \& Menten(1997)]{crosas_1997} Crosas, M. \& Menten, K. M., 1997, \apj, 483, 913
\bibitem[Dayal \& Bieging(1993)]{dayal_1993} Dayal, A. \& Bieging, J. H., 1993, \apj, 407, L37
\bibitem[Dayal \& Bieging(1995)]{dayal_1995} Dayal, A. \& Bieging, J. H., 1995, \apj, 439, 996
\bibitem[Di Lonardo et al.(1993)]{dilonardo_1993} Di Lonardo, G., Ferracuti, P., Fusina, L. et al., 1993, \jms, 161, 466
\bibitem[Di Lonardo et al.(2002)]{dilonardo_2002} Di Lonardo, G., Baldan, A., Bramati, G. et al., 2002, \jms, 213, 57
\bibitem[Doty \& Leung(1997)]{doty_1997} Doty, S. D. \& Leung, C. M., 1997, \memras, 286, 1003
\bibitem[Dyck et al.(1991)]{dyck_1991} Dyck, H. M., Benson, J. A., Howell, R. R. et al., 1991, \apj, 102, 200
\bibitem[Fonfr\'{\i}a Exp\'osito et al.(2006)]{fonfria_2006} Fonfr\'{\i}a Exp\'osito, J. P., Ag\'undez, M. et al., J., 2006, \apj, submitted
\bibitem[Gilman(1972)]{gilman_1972} Gilman, R. C., 1972, \apj, 178, 423
\bibitem[Gonz\'alez-Alfonso \& Cernicharo(1997)]{gonzalezalfonso_1997} Gonz\'alez-Alfonso, E. \& Cernicharo, J., 1997, \aap, 322, 938
\bibitem[Groenewegen(1997)]{groenewegen_1997} Groenewegen, M. A. T., 1997, \aap, 317, 503
\bibitem[Gu\'elin \& Cernicharo(1997)]{guelin_1997} Gu\'elin, M., Cernicharo, J. et al., 1997, \aap, 317, L1
\bibitem[Herbig \& Zappala(1970)]{herbig_1970} Herbig, G. H. \& Zappala, R. R., 1970, \apj, 162, L15
\bibitem[Herman et al.(2003)]{herman_2003} Herman, M., Campargue, A., Idrissi, M. I. et al., 2003, \jpcrd, 32, 921
\bibitem[Herzberg I()]{herzberg_i} Herzberg, G., ``Molecular Spectra and Molecular Structure: I.Spectra of Diatomic Molecules'', Krieger Publishing Company, 1989, ISBN: \mbox{0-89464-268-5}
\bibitem[Herzberg II()]{herzberg_ii} Herzberg, G., ``Molecular Spectra and Molecular Structure: II.Infrared and Raman Spectra of Polyatomic Molecules'', Krieger Publishing Company, 1989, ISBN: \mbox{0-89464-269-3}
\bibitem[Hoyle \& Wickramasinghe(1991)]{hoyle} Hoyle, F. \& Wickramasinghe, N. C., 1991, ``The theory of cosmic grains'', Kluwer Academic Publishers, ISBN: \mbox{0-7923-1189-2}
\bibitem[Huggins \& Healy(1986)]{huggins_1986} Huggins, P. J. \& Healy, A. P., 1986, \apj, 304, 418
\bibitem[Ivezi\'c \& Elitzur(1996)]{ivezic_and_elitzur_1996} Ivezi\'c, \v Z. \& Elitzur, M., 1996, \memras, 279, 1019
\bibitem[Jacquemart et al.(2001)]{jacquemart_2001} Jacquemart, D., Claveau, C., Mandin, J.-Y. et al., 2001, \jqsrt, 69, 81
\bibitem[Jones et al.(1989)]{jones_1989} Jones, T. J., Bryja, C. O., Gehrz, R. D. et al., 1989, \apjs, 74, 785
\bibitem[Justtanont et al.(2005)]{justtanont_2005} Justtanont, K., Bergman, P., Larsson, B., et al., 2005, \aap, 439, 627
\bibitem[Kabbadj et al.(1991)]{kabbadj_1991} Kabbadj, Y., Herman, M., Di Lonardo, G. et al., 1991, \jms, 150, 535
\bibitem[Keady et al.(1988)]{keady_1988} Keady, J. J., Hall Donald, N. B. \& Ridgway, S. T., 1988, \apj, 326, 832
\bibitem[Keady \& Ridgway(1993)]{keady_1993} Keady, J. J. \& Ridgway, S. T., 1993, \apj, 406, 199
\bibitem[Knapp \& Morris(1985)]{knapp_1985} Knapp, G. R. \& Morris, M., 1985, \apj, 292, 640
\bibitem[Kwok(1975)]{kwok_1975} Kowk, S., 1975, \apj, 198, 583
\bibitem[Lacy et al.(1984)]{lacy_1984} Lacy, J. H., Baas, F., Allamandola, L. J., et al., 1984, \apj, 276, 533
\bibitem[Lacy et al.(2002)]{lacy_2002} Lacy, J. H., Richter, M. J., Greathouse, T. K. et al., 2002, \pasp, 114, 153
\bibitem[Lafont et al.(1982)]{lafont_1982} Lafont, S., Lucas, R., \& Omont, A., 1982, \aap, 106, 201
\bibitem[Lindqvist et al.(2000)]{lindqvist_2000} Lindqvist, M., Sch\"oier, F. L., Lucas, R., et al., 2000, \aap, 361, 1036
\bibitem[Loup et al.(1993)]{loup_1993} Loup, C., Forveille, T., Omont, A. et al., 1993, \aap, 99, 291
\bibitem[Lucas \& Cernicharo(1989)]{lucas_1989} Lucas, R. \& Cernicharo, J., 1989, \aap, 218, L20
\bibitem[Lucas et al.(1995)]{lucas_1995} Lucas, R., Gu\'elin, M., Kahane, C., et al., 1995, \apss, 224, 293
\bibitem[Maki et al.(1996)]{maki_1996} Maki, A., Quapp, W., Klee, S. et al., 1996, \jms, 180, 323
\bibitem[Maki et al.(2000)]{maki_2000} Maki, A., Mellau, G. Ch., Klee, S. et al., 2000, \jms, 202, 67
\bibitem[Malathy Devi et al.(2005)]{malathy_devi_2005} Malathy Devi, V., Chris Benner, D., Smith, M. A. H. et al., 2005, \jms, 231, 66
\bibitem[Mandin et al.(2005)]{mandin_2005} Mandin, J.-Y., Jacquemart, D., Dana, V. et al., 2005, \jqsrt, 92, 23
\bibitem[Mauron \& Huggins(1999)]{mauron_and_huggins_1999} Mauron, N. \& Huggins, P. J., 1999, 349, 203
\bibitem[Mauron \& Huggins(2000)]{mauron_and_huggins_2000} Mauron, N. \& Huggins, P. J., 2000, \aap, 359, 707
\bibitem[Men'shchikov et al.(2001)]{menshchikov_2001} Men'shchikov, A. B., Balega, Y., Bl\"ocker, T. et al., 2001, \aap, 368, 497
\bibitem[Men'shchikov et al.(2002)]{menshchikov_2002} Men'shchikov, A. B., Hofmann, K.-H., \& Weigelt, G., 2002, \aap, 392, 921
\bibitem[Monnier et al.(1998)]{monnier_1998} Monnier, J. D., Geballe, T. R., \& Danchi, W. C., 1998, \apj, 502, 833
\bibitem[Monnier et al.(2000a)]{monnier_2000a} Monnier, J. D., Danchi, W. C., Hale, D. S. et al., 2000, \apj, 543, 861
\bibitem[Morris et al.(1975)]{morris_1975} Morris, M., Gilmore, W., Palmer, P., et al., \apj, 199, L47
\bibitem[Murakawa et al.(2002)]{murakawa_2002} Murakawa, K., Tamura, M., Suto, H. et al., 2002, \aap, 395, L9
\bibitem[Mutschke et al.(1999)]{mutschke_1999} Mutschke, H., Andersen, A. C., Cl\'ement, D. et al., 1999, \aap, 345, 187
\bibitem[Neugebauer \& Leighton(1969)]{neugebauer_1969} Neugebauer, G. \& Leighton, R. B., 1969, ``Two-Micron Sky Survey -- a Preliminary Catalog'' (NASA SP-3047 [Washington D.C.: Government Printing Office])
\bibitem[Pendleton et al.(1999)]{pendleton_1999} Pendleton, Y. J., Tielens, A. G. G. M., Tokunaga, A. T., 1999, \apj, 513, 294
\bibitem[Pijpers \& Hearn(1989a)]{pijpers_1989a} Pijpers, F. P. \& Hearn, A. G., 1989, \aap, 209, 198
\bibitem[Pijpers \& Habing(1989b)]{pijpers_1989b} Pijpers, F. P. \& Habing, H. J., 1989, \aap, 215, 334
\bibitem[Ridgway \& Keady(1988)]{rigdway_1988} Ridgway, S. T. \& Keady, J. J., 1988, \apj, 326, 843
\bibitem[Rothman et al.(1992)]{rothman_1992} Rothman, L. S., Hawkins, R. L., Wattson, R. B., \& Gamache, R. R., 1992, \jqsrt, 48, 5/6, 537
\bibitem[Rothman et al.(2003)]{HITRAN} Rothman, L. S., Barbe, A., Benner, D. C. et al., 2003, \jqsrt, 82, 5
\bibitem[Rouleau \& Martin(1991)]{rouleau_1991} Rouleau, F. \& Martin, P. G., 1991, \apj, 377, 526
\bibitem[Schilke \& Menten(2003)]{schilke_2003} Schilke, P. \& Menten, K. M., 2003, \apj, 583, 446
\bibitem[Sch\"oier \& Olofsson(2006)]{schoier_2001} Sch\"oier, F. L. \& Olofsson, H., 2001, \aap, 368, 969
\bibitem[Sch\"oier et al.(2006)]{schoier_2006} Sch\"oier, F. L., Fong, D., Olofsson, H., et al., 2006, \apj, 649, 965
\bibitem[Skinner et al.(1999)]{skinner_1999} Skinner, C. J., Justtanont, K., Tielens A. G. G. M., Betz, A. L., Boreiko, R. T., \& Baas, F., 1999, \mnras, 302, 293
\bibitem[Tejero \& Cernicharo(1991)]{tejero_1991} Tejero, J. \& Cernicharo, J., 1991, ``Modelos de equilibrio termodin\'amico aplicados a envolturas circunestelares de estrellas evolucionadas'', Ministerio de obras p\'ublicas y transportes, Instituto geogr\'afico nacional
\bibitem[Tuthill et al.(2000)]{tuthill_2000} Tuthill, P. G., Monnier, J. D., Danchi, W. C., et al, 2000, \apj, 543, 284
\bibitem[Tuthill et al.(2005)]{tuthill_2005} Tuthill, P. G., Monnier, J. D., \& Danchi, W. C., 2005, \apj, 624, 352
\bibitem[Weigelt et al. (1998)]{weigelt_1998} Weigelt, G., Balega, Y., Bl\"ocker, T. et al., 1998, \aap, 333, L51
\bibitem[Weigelt et al.(2002)]{weigelt_2002} Weigelt, G., Balega, Y. Y., Bl\"ocker, T. et al., 2002, \aap, 392, 131
\bibitem[Wiedemann et al.(1991)]{wiedemann_1991} Wiedemann, G. R., Hinkle, K. H., Keady, J. J. et al., 1991, 382, 321
\bibitem[HITRAN Database()]{HITRAN_database} Rothman, L. S. et al., The HITRAN Database, (USA: Atomic and Molecular Physics Division, Harvard-Smithsonian Center for Astrophysics), \url{http://cfa-www.harvard.edu/hitran/}
\end{thebibliography}
\end{document}